\documentclass{aa}  
\usepackage{graphicx}
\usepackage[varg]{txfonts} 
\usepackage{longtable} 
\usepackage{pdflscape} 
\usepackage{booktabs}
\usepackage{multirow} 
\usepackage{array}
\usepackage{ulem}
\usepackage{enumitem}
\usepackage[colorlinks=true,linkcolor=cyan,urlcolor=cyan,citecolor=cyan]{hyperref}
\usepackage{natbib}
\usepackage{comment}
\usepackage{pdfpages}
\usepackage{xcolor}

\newcommand{\tabref}[1]{\hyperref[#1]{\textcolor{cyan}{Table~\ref*{#1}}}}
\newcommand{\secref}[1]{\hyperref[#1]{\textcolor{cyan}{Section~\ref*{#1}}}}
\newcommand{\figref}[1]{\hyperref[#1]{\textcolor{cyan}{Figure~\ref*{#1}}}}
\newcommand{\eqnref}[1]{\hyperref[#1]{\textcolor{cyan}{Equation~\ref*{#1}}}}

\begin{document}
\title{Kilonova and progenitor properties of merger-driven gamma-ray bursts}

\author{P. Singh\inst{1,2}
\fnmsep\thanks{Corresponding author: \href{mailto:psingh@itp.uni-frankfurt.de}{psingh@itp.uni-frankfurt.de}}
\and
G. Stratta\inst{2,3} 
\and
A. Rossi\inst{2}
\and
P. T. H. Pang\inst{4,5}
\and
M. Bulla\inst{6,7,8}
\and
F. Ragosta\inst{9,10}
\and
A. De Rosa\inst{2}
\and
D. A. Kann\inst{11,12}
\and
F. Cogato\inst{13,2}
}

\institute{Institut für Theoretische Physik, Goethe Universität Frankfurt, Max-von-Laue-Str 1, 60438, Frankfurt am Main, Germany
\and
INAF – Osservatorio di Astrofisica e Scienza dello Spazio, Via Piero Gobetti 101, I-40129, Bologna, Italy
\and
INFN – Sezione di Bologna Viale C. Berti Pichat 6/2 - 40126 Bologna, Italy
\and
Nikhef, Science Park 105, 1098 XG Amsterdam, The Netherlands
\and
Institute for Gravitational and Subatomic Physics (GRASP), Utrecht University, Princetonplein 1, 3584 CC Utrecht, Netherland
\and 
Department of Physics and Earth Science, University of Ferrara, Via Saragat 1, Ferrara, 44122, Ferrara, Italy
\and
INFN, Sezione di Ferrara, Via Saragat 1, Ferrara, 44122, Ferrara, Italy
\and
INAF – Osservatorio Astronomico d’Abruzzo, Via Mentore Maggini snc, Teramo, 64100, Teramo, Italy
\and
Dipartimento di Fisica ‘Ettore Pancini’, Universita` di Napoli Federico II, Via Cinthia 9, 80126 Naples, Italy
\and
INAF – Osservatorio Astronomico di Capodimonte, Via Moiariello 16, I-80131 Naples, Italy
\and
Giersch Science Center, Goethe Universität Frankfurt, Max-von-Laue-Strasse 12, 60438 Frankfurt am Main, Germany
\and
Instituto de Astrofísica de Andalucía (IAA-CSIC), Glorieta de la Astronomía s/n, 18008 Granada, Spain
\and 
Dipartimento di Fisica e Astronomia “Augusto Righi” – Alma Mater Studiorum, Università di Bologna, via Piero Gobetti 93/2, I-40129, Bologna, Italy
}

\date{Received 23 December 2025 / Accepted 11 April 2026}
\abstract 
{
Gamma-ray burst (GRB) prompt and afterglow emission, as well as kilonova (KN) emission, are the expected electromagnetic (EM) counterparts of binary neutron star (BNS) and neutron star-black hole (NSBH) mergers. Only one BNS merger (GW170817) detected on the basis of gravitational waves (GWs) has a confirmed association with EM counterparts. Without a GW signal, the ejecta parameters from a KN can be used to infer the progenitor properties.   
}
{
We aim to infer the KN ejecta parameters and the progenitor properties by modeling merger-driven GRBs with a claim of KN, good data, and robust redshift measurements.
}
{ 
We modeled the afterglow and KN, performing a Bayesian analysis, within the Nuclear physics and Multi-Messenger Astrophysics (\texttt{NMMA}) framework. The KN emission is modeled with the radiative transfer code \texttt{POSSIS} and for the afterglow, we used the \texttt{afterglowpy} library. In contrast to previous approaches, our methodology simultaneously models both the afterglow and KN.
} 
{
We find that all GRBs in our sample have a KN, but we were unable to confirm or exclude its presence in GRB\,150101B. A BNS progenitor is favored for GRB\,160821B, GRB\,170817A/AT2017gfo, GRB\,211211A, and GRB\,230307A. For GRB\,150101B and GRB\,191019A, we found a slight preference for the NSBH scenario; however, a BNS is also a viable possibility. For the KN emission, we see that the median wind mass $\langle M_{\rm wind}\rangle=0.027^{+0.046}_{-0.019}$ $M_{\odot}$ is higher than the dynamical $\langle M_{\rm dyn}\rangle = 0.012^{+0.007}_{-0.006}$ $M_{\odot}$. Finally, we find that $M_{\rm wind}$ and the collimation-corrected energy of the jet can be expressed as $log(M_{\rm wind})=-20.23+0.38\,log(E_{0,J})$. We confirm the results of numerical simulations, demonstrating that the mass-weighted tidal deformability, $\tilde\Lambda$, increases along with a decrease in the chirp mass, $\mathcal{M}_{\rm \,Chirp}$. 
}
{
Our results exemplify and reinforce that EM modeling is the only feasible pathway to investigate binary mergers in absence of GW observations. In this work, we present for the first time the binary progenitor properties of a sizable sample of merger-driven GRBs.
}


\maketitle
\section{Introduction}\label{Introduction}
A kilonova (KN) is the thermal emission powered by the radioactive decay of newly synthesized r-process elements. The neutron-rich matter produced in the coalescence of a binary neutron star (BNS) or neutron star-black hole (NSBH) has long been predicted to host heavy element production \citep{Burbidge1957RvMP, LattimerSchramm1974ApJ,LiPaczynski1998ApJ,Eichler1989Nature,Freiburghaus1999ApJ,Metzger2010MNRAS}. See also \citet{BaiottiRezzolla2017RPPh,Shibata2019,Metzger2020LRR,Barnes2020FrP,Lattimer2021ARNPS}, and \citet{Cowan2021RvMP} for recent reviews. The discovery of AT2017gfo \citep{Coulter2017Sci,Pian2017Nature,Smartt2017Nature}, succeeding the gravitational wave (GW) event GW170817 \citep{Abbott2017ApJ}, confirmed that a merger of BNS resulted in a KN and the production of heavy elements. Simultaneously, the detection of a gamma-ray burst (GRB), designated GRB\,170817A linked the same merger event to the formation of a relativistic jet \citep{Goldstein2017ApJ,Savchenko2017ApJ}. This observation represented the long-awaited, first direct confirmation that at least some GRBs can indeed originate from such mergers. The high-energy emission was followed by an afterglow, detected first in radio and X-rays about a week after theGRBs GRB, indicating an off-axis inclination of the jet axis with the observer's line of sight \citep{Troja2017Nature,Hallinan2017Sci}. \citep{Goldstein2017ApJ,Savchenko2017ApJ}
 
The nonthermal afterglow emission in GRBs arises as the relativistic jet propagates outwards and interacts with the surrounding medium \citep{Sari1998ApJ,Granot2002ApJ,Granot2002ApJ.570L,Piran2004RvMP,Zhang2006ApJ}. The synchrotron radiation from the jet comes from the forward shock, but it can also have contributions from the reverse and refreshed shocks \citep{Lemoine2011CRPhy,Lamb2019ApJ,Fong2021ApJ200522A,Abdikamalov2025MNRAS}. Furthermore, the angle of inclination, the jet opening angle, and the jet structure also play a critical role in shaping the nonthermal emission \citep{Ryan2020ApJ,Beniamini2020MNRAS.493,Lamb2021MNRAS,Beniamini2022MNRAS.515,Eerten2024MNRAS,Ryan2024ApJ}. In the context of early observational epochs of KN, afterglow emission can be seen as contamination, where the jet interacts with post-merger ejecta and causes the KN to appear dimmer and bluer \citep{Klion2021MNRAS,Nativi2021MNRAS,Shrestha2023MNRAS,Combi2023PhRvL}. 

Overall, KN modeling is governed by several aspects, such as the ejected mass, the expansion velocity, the composition of the ejecta, and the angle of inclination relative to the observer \citep{Korobkin2012MNRAS,Barnes2016ApJ,Radice2016MNRAS,Bovard2017PhRvD,Perego2017ApJ,Wollaeger2018MNRAS,Radice2018ApJ,Kawaguchi2018ApJ,Bulla2019MNRAS,Hotokezaka2020ApJ,Zhu2021ApJ,Bulla2023MNRAS,Fujibayashi2023ApJ}. From simulations of BNS mergers, it has been suggested that the total ejected mass is composed of two components: the dynamical and the secular (also referred to as wind) \citep[see][for a review]{Nakar2020,Metzger2020LRR}. The dynamical mass ($\lesssim10^{-2} M_{\odot}$) is expelled on millisecond timescales \citep{Radice2018ApJ,Shibata2019,Kruger2020PhRvD,Han2025PhRvD,Cook2025arXiv2508,Gutierrez2025arXiv2506}, while a massive secular (wind) ejecta ($\gtrsim10^{-2} M_{\odot}$) is emitted over longer timescales \citep{Siegel2018ApJ,Fujibayashi2018ApJ,Gill2019ApJ,Foucart2023LRCA}. The computation of KN emission further requires accurate knowledge of local heating rates, thermalization, and wavelength-dependent opacities of heavy elements, along with their ionization states \citep{Tanaka2020MNRAS,Banerjee2024ApJ,Brethauer2024ApJ,Kato2024MNRAS}. 

Understanding the KN emission and its connection to the merging progenitors has been at the forefront of theoretical and numerical modeling. The GW observations allow us to estimate parameters such as the binary mass ratio ($q$), chirp mass ($\mathcal{M}_{\rm \,Chirp}$), mass-weighted tidal deformability ($\tilde\Lambda$), effective spin ($\chi_{\rm eff}$), and put constraints on the neutron star's equation of state (EOS) \citep[see, e.g.,][]{Abbott2017PhRvL,Koppel2019ApJ,Bauswein2017ApJ}. The matter ejected during the binary merger and the resulting KN would be imprinted with the effect of the EOS and can be linked to the post-merger dynamics and the accretion disk \citep{Dietrich2017CQGra,Radice2018ApJ,Coughlin2019MNRAS,Kruger2020PhRvD, Dietrich2020Sci,Barbieri2020EPJA,Huth2022Nature,Pang2023NatCo,Kunert2024,Lund2025ApJ}. The peak brightness of the KN emission can be used further to investigate the compactness of neutron star, the estimation of the ejected mass, the tidal deformability, and the identification of BNS or NSBH \citep{Radice2019EPJA,Kawaguchi2020ApJ,Raaijmakers2021ApJ,PerezGarcia2022A&A, Topolski2024b}. Furthermore, the successful formation and detection of a relativistic jet after a merger and its behavior could provide constraints on the nature of the remnant \citep{Siegel2014, Rezzolla2014b, Gill2019ApJ,Ascenzi2019ApJ877,Salafia2022A&A}.

The KN observations of the AT2017gfo were exceptional because of the proximity and large off-axis inclination, which caused the afterglow from GRB\,170817A to emerge $\gtrsim160$ days, with the thermal contribution dominating at earlier epochs ($\lesssim10$ days) over all optical and near-infrared (NIR) wavelengths. As a result, its KN emission could be modeled without any stringent afterglow contamination \citep{Arcavi2017Nature,Tanvir2017ApJ,Valenti2017ApJ,Kasen2017Nature,Chornock2017ApJ,Kilpatrick2017Sci,Villar2017ApJ,Cowperthwaite2017ApJ,Drout2017Sci}. However, such a significant observational bifurcation among thermal and nonthermal emissions has not been a commonplace scenario with other observed GRBs, which are typically bright, on-axis events taking place at much greater distances. In reality, it is more common for both afterglow and KN emissions to be temporally intertwined \citep{Gompertz2018ApJ,Ascenzi2019,Rossi2020MNRAS,Rastinejad2021ApJ}. A conventionally popular procedure for isolating a KN relies on the extrapolation of the afterglow flux from the X-ray/radio to the optical/NIR wavelengths. After subtracting the afterglow component, the residual optical/NIR emission is generally considered to be a ``pure'' KN \citep{Troja2019MNRAS,OConnor2021MNRAS,Fong2021ApJ,Rastinejad2022Nature,Troja2022Natur211211A,Yang2022Nature,Rastinejad2025}. Although this approach has been successful for the identification of a KN, modeling cannot feasibly account for the combined thermal and nonthermal emission mechanisms simultaneously. Such procedures have been shown to introduce uncertainties in the inferred physical parameters while producing reasonable fits to the observations \citep{Wallace2025MNRAS}. 

In the first four duty cycles of \texttt{LIGO/Virgo/KAGRA} (\texttt{LVK}) \citep{Acernese2015CQGra,LIGOScientificCollaboration2015CQGra}, GW170817 remains the ``first" (and ``only") ever-observed BNS merger with an observed EM counterpart. Hence, our understanding of the binary mergers and their subsequent EM counterparts is scarce. Therefore, it is imperative to study previously detected GRBs with evidence (or claims) of a KN and to systematically explore their properties. Recent advances in inference methods, combining GW information, afterglow, and KN have shown promise in the characterization of observational properties, merger ejecta, and binary properties. For example, \texttt{bajes} incorporates GW data with KN, but not afterglow \citep{Breschi2021MNRAS,Breschi2024A&A}; \texttt{MOSFiT} combines GW data with analytical fits for ejecta masses derived from BNS simulations and multicomponent geometry, but it requires an independent modeling of afterglows \citep{Guillochon2018ApJS,Nicholl2021MNRAS}; furthermore, the \texttt{Redback} module can model the combination of GW, afterglow and thermal emission \citep{Sarin2024MNRAS}. In our analysis, we utilized the Nuclear physics and Multi-Messenger Astrophysics (\texttt{NMMA}\footnote{\url{https://nuclear-multimessenger-astronomy.github.io/nmma/index.html}}) framework, which simultaneously models the afterglow, KN, and GW, while allowing the binary parameters to be inferred \citep{Dietrich2020Sci,Almualla2021arXiv211215470A,Pang2021ApJ,Pang2023NatCo}. In the absence of a GW detection, we used EM data from merger-driven GRB to analyze the progenitor properties. This allowed us to present, for the first time, the binary properties for a sample of events other than GW170817. We describe our sample in \secref{Sample}, along with the modeling of the electromagnetic counterpart and the properties of the binary progenitor in \secref{ModelingEM}. Our results are presented in \secref{Results}, followed by discussions in \secref{Discussion} and our conclusions in \secref{ConclusionsSummary}. Throughout this work, we have adopted a flat standard cosmological model with $H_{\rm 0} = 67.4$\,km\,s$^{-1}$\,Mpc$^{-1}$, $\Omega_{\rm M} =0.315$, and $\Omega_{\rm \Lambda}=0.685$ \citep{PlanckCollaboration2020}. 

\section{Sample of the GRBs}\label{Sample}
In this work, we restricted our analysis solely to GRBs for which evidence or claims of KN emission have been established. We compiled our GRB sample from previous studies \citep{Ascenzi2019,Rossi2020MNRAS,OConnor2022,Fong2022,Troja2023EighteenYears} by imposing the selection conditions detailed below.  

\subsection{Selection criteria}\label{SelectionCriteria}
\begin{itemize}[itemsep=0.35em]
    \item[{\it a})] We evaluated the reported redshift and the association of GRB with its host galaxy; namely, the probability of chance coincidence ($pcc$) \citep{Bloom2002AJ}. We only chose cases with accurate redshift measurements and robust host association $pcc \lesssim 10^{-3}$ (with the exception of GRB\,230307A, see \secref{230307A_Sample}).
    \item[{\it b})] We further limited our sample to events with a redshift below $z\lesssim0.5$. This criterion is imposed because of the poor detectability of KN emission at larger distances. For example, an event such as AT2017gfo, peaking in the optical/NIR bands with absolute magnitudes $\simeq-15$ to $-16$ mag, would appear fainter than $\sim 26$ to $27$ AB mag if it were at $z>0.5$. Thus, a KN at a larger redshift would be above the limiting magnitude achievable by the vast majority of the current generation of ground-based telescopes \citep[e.g,][]{Gompertz2018ApJ,Rossi2020MNRAS}.
    \item[{\it c})] We imposed the condition that each GRB should have sufficiently good optical/NIR spectral coverage. Specifically, we only consider events that have more than four optical/NIR bands (both observations\footnote{For all GRBs in our sample, we individually evaluated each observation and omitted the data that have an observed magnitude error larger than $0.3$ mag, to avoid values dominated by systematics and/or unsure detections.} and upper limits\footnote{In our inference methodology, the upper limits are also included to constrain the models.}) and long monitoring in the optical/NIR bands, from early afterglow phase to several of days after the burst onset.
    \item[{\it d})] For our multiwavelength fitting procedure, the nonthermal afterglow emission is primarily constrained by X-ray observations (and radio when available). Hence, we selected only those GRBs that have well-sampled X-ray data\footnote{For all GRBs, except where otherwise stated, the X-ray data were obtained from the \texttt{SWIFT-XRT} repository (\url{https://www.swift.ac.uk/xrt_curves/}).}. Furthermore, we did not consider GRBs that have any flaring activity in the afterglow, since flares are believed to originate from a mechanism different from the afterglow.
\end{itemize} 

As a result, six GRBs satisfied our selection criteria, GRB\,150101B ($z=0.134$), GRB\,160821B ($z=0.1619$), GRB\,170817A ($z=0.009787$), GRB\,191019A ($z=0.248$), GRB\,211211A ($z=0.0763$), and GRB\,230307A ($z=0.0646$). The observational data are available online (see \secref{DataAvailability} and \tabref{tab:ObservationsDataTable}). Further details of the selected GRBs are summarized in the following sections. 

\subsection{GRB\,150101B}\label{150101B_Sample}
On January 1, 2015, at 15:23 UT, the GRB\,150101B was detected by the Burst Alert Telescope (\texttt{BAT}) on board the \texttt{Swift} satellite \citep{Barthelmy2005,Cummings2015GCN}. It was located at a projected distance of $7.35\pm0.07$ kpc from the host and did not show any evidence of a supernova, suggesting a BNS or NSBH merger as the source \citep{Fong2016ApJ}. The prompt burst duration, $T_{90} = 0.018$\,s \citep{Fong2016ApJ,Burns2018ApJ} makes it a short GRB and it has a redshift $z=0.134$ \citep{Levan2015GCN,Fong2016ApJ}. The \texttt{Swift} X-ray telescope (\texttt{XRT}) observations are contaminated with background emission from an active galaxy, whereas the \texttt{Chandra} data are unaffected and used in our analysis \citep{Gompertz2018ApJ}. All observational data have been collected from \citet{Fong2015ApJ} and an upper limit in X-rays at $\sim1.5$ days comes from \citet{Troja2018GRB150101B}. There is no intrinsic extinction from the host galaxy (i.e., $A_{\rm V}^{\rm Host} = 0$) and the foreground MW extinction in the direction of the burst is $E (B-V) = 0.036$ mag \citep{Fong2016ApJ}. The analysis of \citet{Troja2018GRB150101B} favors the presence of an early KN at optical wavelengths. 

\subsection{GRB\,160821B}\label{160821B_Sample}
The Fermi Space Telescope's Gamma-ray Burst Monitor (\texttt{GBM}) \citep{Meegan2009ApJ} and the \texttt{Swift-BAT} detected GRB\,160821B on August 21, 2016, at 22:29:13 UT. GRB\,160821B has redshift, $z = 0.1619$, with its host galaxy at an offset of $15.74 \,\rm{kpc}$ and with $T_{90} = 0.48$\,s, making it a short GRB \citep{Fong2022}. The intrinsic absorption due to the host galaxy is negligible \citep[ $\rm{A_{V}^{Host}} \approx 0$,][]{Sbarufatti2016GCN}. For the galactic foreground extinction, we took $A_{\rm V} = 0.118$ mag \citep{Schlafly2011ApJ}. The observations of GRB\,160821B revealed the presence of a KN \citep{Kasliwal2017ApJ,Jin2018ApJ160821B,Lamb2019ApJ160821B,Troja2019MNRAS}. \cite{Lamb2019ApJ160821B}, showed that at early times $\leq0.9$ days (especially at 0.1 days), a reverse shock is required to explain the X-ray/radio behavior and after $\sim1$ day, a forward shock is consistent with the observations. In our afterglow model that is based only on the forward shock (see \secref{ModelingEM}), we did not find evidence of energy injection. Hence, we converted data $t\leq 0.9$ days to upper limits. We have used X-ray observations from \texttt{Swift-XRT}, \texttt{XMM-Newton}, and the K band by \citet{Troja2019MNRAS}. The rest of the optical/NIR and radio data were adopted from \citet{Lamb2019ApJ160821B}.

\subsection{GRB\,170817A and AT2017gfo}\label{170817A_Sample}
The GRB\,170817A signal \citep{Goldstein2017ApJ,Savchenko2017ApJ} was detected by both the \texttt{Fermi-GBM} and \texttt{INTEGRAL}, 1.7 seconds after the Advanced Laser Interferometer Gravitational wave Observatory \texttt{LIGO} \citep{LIGOScientific:2014pky} detection of the BNS merger GW170817 \citep{Abbott2017ApJ}. Subsequent optical and spectroscopic observations confirmed the detection of an electromagnetic counterpart AT2017gfo \citep{Arcavi2017Nature,Coulter2017Sci,Lipunov2017ApJ,Tanvir2017ApJ,Soares-Santos2017ApJ,Valenti2017ApJ,Chornock2017ApJ,Pian2017Nature,Kilpatrick2017Sci,Smartt2017Nature,Villar2017ApJ,Cowperthwaite2017ApJ,Drout2017Sci}. The redshift for GRB\,170817A is $z=0.009787$ and the distance, inferred from its host-galaxy, is $\sim41$ Mpc \citep{Hjorth2017ApJ,Cantiello2018ApJ}. The electromagnetic counterpart of GW170817 has been extensively investigated in the literature. Therefore, we can use it as a calibrator for our method, focusing on the KN properties and binary progenitor inference. The optical/NIR data used in this work were corrected for extinction and adapted from \citealt{Coughlin2018MNRAS} (and references therein), while the X-ray data were taken from \citet{Troja2022MNRAS}. Our resulting data cover the temporal range from $\sim0.2$ to $\sim$1300 days. 

\subsection{GRB\,191019A}\label{191019A_Sample}
The GRB\,191019A was discovered by \texttt{Swift-BAT} \citep{Simpson2019GCN,Barthelmy2005}. Its burst duration of $T_{90} = 64.6 \pm 4.5$\,s places it among the class of long GRBs. At the location of this burst, \texttt{Swift-XRT} detected a transient that was identified as the X-ray afterglow  \citep{Reva2019GCN.26036}. The spectroscopic absorption lines from the host galaxy allowed inferring its redshift as $z=0.248$, along with a small amount of dust extinction ($A_{\rm V} = 0.19 \pm 0.08$ mag). There was no evidence of an associated supernova, even in the deep limits of the Nordic Optical Telescope (\texttt{NOT}) and the Hubble Space Telescope (\texttt{HST}) \citep{Levan2023NatAs}. The GRB\,191019A was further evaluated against the prompt emission minimum variability timescale criterion and its high variability (low FWHM$_{\rm min} = 0.196^{+0.068}_{-0.050}$) suggested a compact merger origin \citep{Camisasca2023A&A}. GRB\,191019A was located at a projected distance of $\leq$ 100 pc from its host galaxy, thought to be an active galactic nucleus (AGN) that is relatively close to the host center. In the hypothesis of an AGN disk environment, \citet{Lazzati2023ApJ} showed that the intrinsic prompt emission duration of this GRB could be altered if the surrounding interstellar matter density was shown to be on the order of $10^{7}$ to $10^{8}$ cm$^{-3}$. However, if the circumburst density is assumed to be high, as suggested by \citet{Lazzati2023ApJ}, the interaction of the KN ejecta with the circumstellar medium would result in bright late-time emission; however, this was not observed for GRB\,191019A \citep{Wang2024arXiv240114418W}. Recently, \citet{Stratta2025ApJ} used Gamma-ray Burst Optical/Near-infrared Detector (\texttt{GROND}) observations \citep{Greiner2008PASP,NicuesaGuelbenzu2019GCN.26042}, lasting from $\sim10.3$ hours to $\sim15$ days post-burst, and identified a KN emission peaking at about $\sim1$ day.
    
\subsection{GRB\,211211A}\label{211211A_Sample}
The GRB\,211211A signal was first detected by \texttt{Swift-BAT} and \texttt{Fermi-GBM}, on 11 December 2021 at 13:09:59 UT \citep{Mangan2021GCN.31210,Stamatikos2021GCN.31209}. GRB\,211211A was a peculiar event, with a long burst duration of $T_{90} = 50.7\pm{0.9}$\,s, demonstrating that it is not a classically short GRB; instead, it is deemed a hybrid case with properties that are also akin to those of long GRBs (see Table 2 in \citet{Troja2023EighteenYears}, Table 4 from \citet{Levan2023NatAs}, and the discussion in \citet{Zhang2025JHEAp}). From the follow-up observational campaign in the optical/NIR wavelengths, this burst was located at the redshift $z=0.0763$. A KN was identified in the optical/NIR while no associated supernova was found even $17.7$ days post-burst \citep{Rastinejad2022Nature,Troja2022Natur211211A}. In our analysis, we began by gathering the data available in \citet{Rastinejad2022Nature} and supplemented it with \texttt{UVOT}, b band (similar to the B band) observation at $0.9799$ days, in the Rc band at $0.4500$ days, and with an upper limit of $3.39$ days in the g band from \citet{Troja2022Natur211211A}. An additional observation from \citet{Kumar2021GCN.31227} in the Rc band at $0.4506$ days was also included \footnote{We tested the combined afterglow and KN inference by considering uvw1, uvw2, uvm2, r, and z bands. These observations are aptly described by our modeling and we omitted them from the light curves shown in the left panel of \figref{Fig:211211AAnd230307ALightcurves} for visualization purposes.}. The Galactic extinction was $A_{\rm V} = 0.048$ mag \citep{Rastinejad2022Nature}, whereas the contribution from the host was negligible \citep{Troja2022Natur211211A}.

\subsection{GRB\,230307A}\label{230307A_Sample}
\texttt{Fermi-GBM} detected a long and bright burst, GRB\,230307A on 7 March 2023 at 15:44:06.67 UT \citep{FermiGBMTeam2023GCN,Dalessi2023GCN.33407,Dalessi2023GCN.33411}. It was localized by further \texttt{Konus-Wind} \citep{Svinkin2023GCN.33427} and \texttt{Swift} observations \citep{Evans2023GCN.33419}. Based solely on its prompt burst duration of $T_{\rm 90}\approx35$\,s, it belongs to long GRBs \citep{Dalessi2023GCN.33411}. For GRB\,230307A, two candidate host galaxies were initially identified at $z\sim3.87$ ($pcc=0.05$) and $z=0.0646$ ($pcc=0.09$), while the projected physical offset of the burst location was $\sim1-2$ kpc and $\sim40$ kpc, respectively \citep{Levan2024Nature}. GRB\,230307A did not have a GeV emission, implying a low circumburst density \citep{Dai2024ApJ}. Furthermore, if the host was the galaxy at $z\sim3.87$, the isotropic kinetic energy would have been greater than $10^{55}$ erg, which is unrealistically high with respect to any other GRBs previously observed \citep[see][for an extensive discussion]{Atteia2017ApJ,Burns2023ApJ,Levan2024Nature,Dai2024ApJ}. Hence, the host was established as the galaxy at $z=0.0646$ \citep{Levan2024Nature}. The optical/NIR observations of GRB\,230307A contain a prominent bright thermal component at $\sim10$ days, especially seen in the K band and with James Webb Space Telescope (\texttt{JWST}) around $\sim28$ days, which is brighter than the afterglow flux \citep{Levan2024Nature}. The early X-ray and optical data before $\lesssim2-4$ days are consistent with a power-law decay of afterglow, but after a few days, emission from radioactive material is predominant \citep{Yang2024Nature}. While AT2017gfo was the first KN to be observed spectroscopically \citep[see][]{Pian2017Nature,Smartt2017Nature}, the follow-up carried out by \texttt{JWST} made GRB\,230307A the second KN characterized by spectroscopic observations \citep{Levan2024Nature}. At $2.4$ days, it contains a featureless thermal continuum, and at $28.9$ days the spectrum exhibits excess red emission that cannot be explained solely based on the assumption of afterglow emission \citep{Gillanders2023,Gillanders2024}. Our GRB\,230307A dataset is accumulated from \citet{Yang2024Nature}, as the main source, and it includes unique observations from \citet{Levan2024Nature}. We omitted the r band observation from \citet{Yang2024Nature} at $0.43$ days from our final data because of its error $\geq0.3\%$ ($r=19.4 \pm 0.6$). We combined the K band observations from \citet{Yang2024Nature} and \citet{Levan2024Nature}. The Transiting Exoplanet Survey Satellite (\texttt{TESS}) observations were adopted from \citet{Levan2024Nature}, where the measured flux within $600-1000$ nm was converted to the Ic band. Due to a large offset among the host galaxy and the GRB location, as seen from the \texttt{JWST-NIRCAM} images, the intrinsic extinction is negligible \citep{Levan2024Nature}. The correction for Milky Way extinction $E(B-V)=0.0758$ mag was required \citep{Schlafly2011ApJ,Yang2024Nature}. In our data, we combine the z, z$^{\prime}$, and Z bands. 

\subsection{Excluded GRBs}\label{ExcludedGRBs}
Based on the sample selection criteria (\secref{SelectionCriteria}), we can go on to discuss the GRBs omitted from our analysis. GRB\,050709 (z=0.1607), is not considered due to poor X-ray monitoring and flaring activity observed at $\sim16$ days \citep{Fox2005Nature,Watson2006A&A,Jin2016GRB050709}. The X-ray light curve of GRB\,050724A (z=0.254) shows a strong flaring activity $\sim 0.2-3.0$ days after the burst and, thus, it was also rejected \citep{Berger2005NatureGRB050724,Grupe2006ApJ,Malesani2007A&A,Panaitescu2006MNRAS,Gao2017ApJ}. The GRB\,060505 (z=0.089) has only three X-ray data points and no radio data, while its optical/NIR counterpart is limited to a single epoch $\sim1$ day \citep{Ofek2007ApJ,Xu2009ApJ060505060614,Jin2021arXiv}. Hence, we excluded GRB\,060505 from our sample. In the case of GRB\,060614A (z=0.125), it has been reported that it might be harboring a KN given the minor flux excess seen in the F814W band at $\sim13$ days \citep{Jin2015ApJ060614,Yang2015NatCo060614,Tanaka2016AdAst2016E}. A weak ``plateau-like” emission that (potentially due to an AGN) is also present towards the end of the X-ray light curve \citep{Mangano2007A&A}. Furthermore, the optical emission of GRB\,060614A shows complex behavior, exhibiting achromatic breaks, a fast decay, and then a long shallow phase \citep[see Fig. 5 in ][]{Mangano2007A&A}. Our afterglow model described in \secref{ModelingEM} cannot take such complex behavior into account. Therefore, we omitted GRB\,060614A from our current analysis and we plan to address it in a future work. Furthermore, a KN was suggested for GRB\,070809 by \citet{Jin2020Nat070809}, although its redshift is not secure, with two possible values $z=0.2187$ \citep{Perley2008GCN,OConnor2022} and $z=0.473$ \citep{Berger2010ApJ,Fong2022,Nugent2024}. The optical/NIR observations were poorly monitored with no data at $\lesssim1$ days and also did not meet our spectral coverage criterion. The GRB\,130603B (z=0.3568) X-ray light curve and F606W band fade rapidly, while the F160W band at $\sim7$ days requires additional thermal emission \citep{Berger2013ApJ,Tanvir2013Nature}. Although a KN emission is predominant for GRB\,130603B, it has only been constrained by a single observational data point and it has been rejected due to poor multiwavelength coverage. The GRB\,200522A has only two observed data points at $\approx3.55$ days, in the F160W and F125W bands, in addition to X-ray and radio, and a KN has been suggested in this case \citep{OConnor2021MNRAS,Rastinejad2025}. However, it can be modeled equally well by thermal emission or only by the combination of forward plus reverse shock \citep{Fong2021ApJ}. Hence, given the limited data and the absence of substantial KN claim/evidence, GRB\,200522A was not considered in our work. 

\section{Analysis of electromagnetic counterparts}\label{ModelingEM}
Characterizing the thermal KN emission as an extra component buried within the nonthermal GRB optical/NIR observations is challenging, because both types of emissions are temporally overlapped. Consequently, in the presence of a KN, any effective modeling requires an approach that would simultaneously account for both the thermal and nonthermal contributions to the multiwavelength observations. 

In our analysis, we performed a simultaneous inference of afterglow and KN, using the \texttt{NMMA} framework, version 0.2.0 \citep{Dietrich2020Sci,Almualla2021arXiv211215470A,Pang2021ApJ,Pang2023NatCo}. \texttt{NMMA} is trained via a neural network on theoretical KN models and incorporates Bayesian inference. Thus, it allows us to test and account for the high-dimensionality of the combined KN and afterglow parameter space. 

\subsection{Afterglow modeling}\label{AfterglowModeling}
The nonthermal afterglow emission model used in \texttt{NMMA} is based on \texttt{afterglowpy}\footnote{\url{https://afterglowpy.readthedocs.io/en/stable/}. We note that a newer version of \texttt{afterglowpy} \citep{Ryan2024ApJ}, is currently not used in \texttt{NMMA}.} module developed by \citet{vanEerten2010MNRAS,Ryan2020ApJ}. This allows us to investigate the effects of inclination and various complex jet structures shaping the GRB afterglow light curves. In this work, we considered Gaussian (GS) and top-hat (TH) jet structures. A GRB jet structure is described by the energy profile $E(\theta)\propto f(\theta)$, where $\theta$ is the half-opening angle from the jet axis. A TH jet structure is marked by a sharp cutoff of energy outside the core, given as
\begin{equation}
E ({\theta}) \propto \left\{ 
\begin{array}{ll} 
\displaystyle
const. & \theta \leq \theta_{\rm c},\\
\displaystyle
0 & \theta > \theta_{\rm c},
\end{array}\right. 
\end{equation}
where $\theta_{\rm c}$ denotes the half-opening angle of the jet core. A Gaussian jet structure can be parametrized as
\begin{equation}
E ({\theta}) \propto \left\{ 
\begin{array}{ll} 
\displaystyle
\text{exp} \left[ -\frac{1}{2} \left[ \frac{\theta}{\theta_{\rm c}}\, \right]^2 \, \right] & \theta \leq \theta_{\rm w},\\
\displaystyle
0 & \theta> \theta_{\rm w},
\end{array}\right. 
\end{equation}
where $\theta_{\rm w}$ is the half-opening angle of the jet truncated-wings. Following the analysis of \citet{Ryan2024ApJ}, we consider that $\theta_{\rm w}$, is at least larger than $4\times\theta_{\rm c}$. In modeling the afterglow, we limited our analysis to the synchrotron radiation originating from the forward shock, which fairly describes the afterglow emission in the temporal window of interest for KN observations. 

The posterior parameters for the afterglow modeling are: $E_{0}$, kinetic isotropic equivalent energy;  $\iota$, viewing angle with respect to the jet axis; $n_{0}$, particle number density in the circumburst environment; $p$, electron energy distribution power-law index; $\epsilon_{e}$, shock energy fraction that goes into the electrons; and $\epsilon_{B}$, shock energy fraction that goes into the magnetic energy density. Finally, we assume that the bulk $\xi_{\rm N}=1$ of the electron population is shock-accelerated \citep{Cunningham2020ApJ,Ryan2020ApJ,Urrutia2021MNRAS,Hayes2023ApJ,Ryan2024ApJ}. 

We assigned uniform priors to all afterglow parameters, except the viewing angle, $\iota$, which is uniform on a sphere, see \tabref{tab:All_KN_Priors}. In the case of GRB\,170817A, we adopted prior values for $\iota = Sine (0.20, 0.60)$, $\theta_{\rm c} = \mathcal{U} (0.02, 0.15)$, and $\theta_{\rm w} = \mathcal{U} (0.6, 0.99)$ in radians, from \citet{Gianfagna2024MNRAS}. For GRB\,191019A, to address the high density suggested by \citet{Lazzati2023ApJ}, we chose a wide prior for $log\,n_{0}= \mathcal{U} (-3, 7)$ (cm$^{-3}$), where $\log(\epsilon_{\rm e})$ and $\log(\epsilon_{\rm B})$ were initially fixed to $-0.3$ and $-2.0$, respectively; see \tabref{tab:191019A_AllModelsResults}. Once we were able to consistently find that $\log(n_0)$ does not attain the high values obtained by \citet{Lazzati2023ApJ}, we allowed $\log(\epsilon_{\rm e})$ and $\log(\epsilon_{\rm B})$ to vary freely. We report the details of the model in \tabref{tab:All_KN_Results}. 

\subsection{Kilonova modeling}\label{KilonovaModeling}
For KN modeling in \texttt{NMMA}, we utilized predictions from the time-dependent 3D Monte Carlo radiative transfer code \texttt{POSSIS}\footnote{\url{https://github.com/mbulla/kilonova_models?tab=readme-ov-file}}, which generates spectra and light curves from BNS and NSBH progenitors \citep{Bulla2019MNRAS}. In our work, we relied on simulations carried out with the assumption of two distinct ejecta components \citep{Nakar2020,Dietrich2020Sci}: 1) a dynamical ejecta component with a mass, $M_{\rm dyn}$, emitted at high velocities ($0.08 < v_{\rm dyn}/c < 0.3$) as the two compact objects merge; and (2) a spherical component with a mass, $M_{\rm wind}$, emitted as a slower wind ($0.025 < v_{\rm wind}/c < 0.08$) through various mechanisms from a post-merger disk. 

The composition of the ejecta differs between the BNS and NSBH models. In the BNS scenario \citep{Dietrich2020Sci}, the dynamical ejecta are assumed to be lanthanide-rich in regions that are within the latitude $\pm\Phi$ from the merger plane, while they would be lanthanide-poor at higher latitudes. In the NSBH scenario \citep{Anand2021}, instead, the lanthanide-poor ejecta at high latitudes are not present and the lanthanide-rich component is fixed within $\Phi=30^\circ$. The wind component of both grids has compositions (and, hence, opacities) that are intermediate between lanthanide-poor and lanthanide-rich. The KN grids are constructed by varying $M_{\rm dyn}$, $M_{\rm wind}$ and $\Phi$ in the BNS case, while $M_{\rm dyn}$ and $M_{\rm wind}$ in the NSBH case.
 
\subsection{Bayesian analysis}\label{BayesianAnalysis}
In \texttt{NMMA}, an inference is carried out for a specific filter, $j$, in the AB magnitude, $m_i^j$, over observational time series, $t_{i}$, for given model parameters, $\Omega$, and providing the estimated AB mag as $m_i^{j,\text{est}}({\Omega})$ \citep{Pang2021ApJ,Pang2023NatCo}. To account for underlying uncertainties in the afterglow, KN models, and systematic errors, an additional error $\sigma_{\rm sys}$ is included that is free to vary in the range 0-2 mag \citep{Kunert2024,Heinzel2021MNRAS}. The likelihood is described as
\begin{equation}
	\label{equation:likelihood_f}
    \begin{aligned}
	\ln\mathcal{L}({\Omega}) &=  \sum_{i j} -\frac{1}{2}\frac{\left(m_i^j-m_i^{j, \text { est }}({\Omega})\right)^2}{\left(\sigma_i^j\right)^2+\left(\sigma_{\rm sys}\right)^2} + \ln2\pi\left[\left(\sigma_i^j\right)^2+\left(\sigma_{\rm sys}\right)^2\right],
    \end{aligned}
\end{equation}
where $\sigma_i^j= \sigma^{j}(t_i)$ is the statistical error. Lastly, for each GRB, we fixed the luminosity distance ($D_{\rm L}$ in Mpc) corresponding to the observed redshift value (see \tabref{tab:All_KN_Results}) assuming a flat standard cosmological model. 

Bayesian inference is performed with \texttt{PyMultiNest}\footnote{\url{https://johannesbuchner.github.io/PyMultiNest/}} \citep{Buchner2014}. We can evaluate different scenarios and compute the Bayes factor, which allows us to select the most plausible model that satisfies the electromagnetic observations. In Bayesian analysis, given a set of data, $d$, prior beliefs, $p(\Omega \mid \mathcal{H})$, likelihood, $p(d \mid \Omega,\mathcal{H})$, and evidence, $p(d\mid\mathcal{H})$, the posterior can be written as
\begin{equation}
    p({\Omega} \mid d,\mathcal{H}) =  \frac{p(d \mid \Omega,\mathcal{H}) \, p(\Omega \mid \mathcal{H})} {p(d\mid\mathcal{H})},
\end{equation}
where $\Omega$ describes the model parameters and $\mathcal{H}$ accounts for the model hypothesis. Rewriting the above as
\begin{equation}
\mathcal{P}({\Omega}) =  \frac{\mathcal{L}({\Omega}) \, \pi({\Omega})}{\mathcal{Z}},
\end{equation}
where the posterior, $\mathcal{P}({\Omega})$, is evaluated from its likelihood, $\mathcal{L}({\Omega})$, to describe the observations based on evidence, $\mathcal{Z}$, and priors, $\pi({\Omega})$, which essentially reflect the model choice. Assuming that we have two models, $\mathcal{H}_1$ and $\mathcal{H}_2$, their odds ratio, $\mathcal{O}^1_2$, can be compared as 
\begin{equation}
	\label{equation:bayesfactor}
	\mathcal{O}^1_2 \equiv \frac{p(d|\mathcal{H}_1)}{p(d|\mathcal{H}_2)}\frac{p(\mathcal{H}_1)}{p(\mathcal{H}_2)} \equiv \mathcal{B}^1_2\Pi^1_2,
\end{equation}
where $\mathcal{B}^1_2$ is the Bayes factor and $\Pi^1_2$ is the prior odds. In this way, if $\mathcal{H}_1$ is more plausible than $\mathcal{H}_2$, we would have $\mathcal{O}^1_2 > 1$ and vice versa. We can further assume that the prior odds $\Pi^1_2=1$, such that we consider both models to be equally plausible; thus, we can simply compare the Bayes factor assuming the model with highest likelihood to be our reference. Hence, we can perform a model selection by comparing the evidence of a test model relative to our reference model, $\ln \mathcal{B}^{\rm Test}_{\rm{Ref}}$, as

\begin{tabular}{m{4.0cm} m{3.7cm}}
    $\ln [\mathcal{B}^{\rm Test}_{\rm{Ref}}] > 0$,  & test model is favored \\[0.6em]
    $ -1.10 < \ln [\mathcal{B}^{\rm Test}_{\rm{Ref}}] \leq 0$,  & both models are equally-likely \\[0.6em]
    $-2.30 < \ln [\mathcal{B}^{\rm Test}_{\rm{Ref}}] \leq -1.10$,  & substantial evidence, \\[0.5em]
    $-4.61 < \ln [\mathcal{B}^{\rm Test}_{\rm{Ref}}] \leq -2.30$,  & strong evidence, \\[0.5em]
    $ \ln [\mathcal{B}^{\rm Test}_{\rm{Ref}}] \leq -4.61$, & decisive evidence against a test model, \\[0.6em]
\end{tabular}
where higher values imply that the reference is unfavorable in comparison to a test model. We chose the model with the highest level of evidence ($\mathcal{Z}$) as our reference model. The decrease in the Bayes factor to lower values indicates the level of disfavor for a given test model \citep{Jeffreys1939,Kass1995,Trotta2007MNRAS,Robert2008arXiv0804,Nesseris2013JCAP}. We chose the model with the highest Bayes factor ($\ln [\mathcal{B}^{\rm Test}_{\rm{Ref}}]$) as the most-favored model.

\subsection{Inference of progenitor properties}\label{ProgenitorProperties}
The origin of GW170817 was well-established due to the coalescence of two neutron stars and allowed measurement of the binary properties \citep{Abbott2017ApJ,Abbott2017PhRvL}. However, because of the rarity of other coincident EM observations from GW detected binary mergers, performing a complete (GW+EM) inference for any other case is not possible. Therefore, in the absence of GW information, the inferred best-fit KN properties from the EM counterpart modeling can be linked to likely progenitor properties. We selected the most-favored model for each GRB (see \secref{Results}) and performed an inference of the binary progenitor properties with \texttt{NMMA}. 

Our methodology is identical to \citet{Pang2023NatCo} (see also \citep{Coughlin2019MNRAS,Dietrich2020Sci}). Hence, we only report a brief overview here. We started by constructing GW priors, where we utilized the accurately measured redshift (see \tabref{tab:All_KN_Results}) and provide a small uniform prior for the luminosity distance ($D_{\rm L}$) as $\mathcal{U} \in[D_{\rm L}^{\,-}, D_{\rm L}^{\,+}]$ (see \tabref{tab:All_Binary_Results}). The priors for individual neutron star mass are $M_{\rm NS} \in [1.0,M_{\rm TOV}\footnote{$M_{\rm TOV}$ is the Tolman-Oppenheimer-Volkoff mass \citep{Oppenheimer1939PhRv}. }]\,M_{\odot}$ that depend on a given EOS \citep[see][]{Kruger2020PhRvD,Dietrich2020Sci} and the prior black hole mass span $M_{\rm BH}\in[3.0,20.0]\,M_{\odot}$, where the dependence on $R_{\rm ISCO}(\chi_{\rm eff})$\footnote{The spin-dependent radius of the innermost stable circular orbit of the black hole.}, is included through a prior on the effective spin $\chi_{\rm eff}\in[-0.9,0.9]$. Thus, these priors are directly derived from binary masses and spins for a given set of EOSs. For the EOS, the prior weights are based on astrophysical, theoretical, and experimental constraints, as also noted \citep[see][]{Huth2022Nature}.

In \texttt{NMMA}, an inverse sampling of the EM posteriors ($\mathcal{P}_{\rm EM}$), for the KN ejecta $(M_{\rm dyn},M_{\rm wind})$ is performed and these are mapped to their predicted expectations as
\begin{align}
    M_{\rm dyn} \equiv M^{\rm pred}_{\rm dyn}  + \alpha \,; \quad M_{\rm wind} \equiv \zeta\,M_{\rm disk}^{\rm pred} 
    \label{eq:NMMAMassesAlphaZeta1},
\end{align}
where $\alpha$ is the dynamical ejecta mass fitting error and $\zeta$\footnote{The parameter $\zeta$ in our formalism is denoted as $\xi$ in \cite{Pang2023NatCo}.} is the fraction of the disk mass ejected as wind \citep{Pang2023NatCo}. In the BNS merger scenario, the phenomenological relations for $M^{\rm pred}_{\rm dyn}$ \citep{Kruger2020PhRvD} and $M_{\rm disk}^{\rm pred}$ \citep{Dietrich2020Sci}, are evaluated against $\mathcal{P}_{\rm EM}$ of $M_{\rm dyn}$ and $M_{\rm wind}$, marginalizing over the uncertainty on $\alpha$ and $\zeta$ (\eqnref{eq:NMMAMassesAlphaZeta1}). Here, the neutron star spins are assumed to be zero \citep[see][]{Kruger2020PhRvD,Dietrich2020Sci}. Similarly, for the NSBH scenario, the phenomenological relations for $M_{\rm dyn}^{\rm pred}$ \citep{Kawaguchi2016ApJ,Kruger2020PhRvD} and $M_{\rm disk}^{\rm pred}$ \citep{Foucart2012PhRvD,Barbieri2020EPJA} are connected to the posteriors $\mathcal{P}_{\rm EM}$. The likelihood ($\mathcal{L}$) for the binary parameters for given EM observations ($d_{\rm EM}$) and EM posteriors ($\mathcal{P}_{\rm EM}$) is expressed as
\begin{equation}
\begin{aligned}
    \mathcal{L} (M_{\rm i}, \ \dots \ , {\rm EOS} \,\vert\, d_{\rm EM} )
&\equiv
\mathcal{P}_{\rm EM}(M_{\rm dyn}, M_{\rm wind})\\
&\times\delta(M_{\rm dyn} - M^{\rm pred}_{\rm dyn}(M_{\rm i}, \ \dots \ , {\rm EOS}) - \alpha)\\
&\times\delta(M_{\rm wind} - \zeta M^{\rm pred}_{\rm disk}(M_{\rm i}, \ \dots \ , {\rm EOS})),
\end{aligned}
\end{equation}
where $i \in [1,2]$, indicates individual binaries and $\delta(x)$ is the Dirac function. Hence, from the resulting posteriors we obtain the following binary parameters: (I) $M_{\rm TOV}$; (II) the ratio of binary masses, $q$; (III) the chirp mass $M_{\rm Chirp}$ calculated from the reparameterization of binary masses; and (IV) the mass-weighted tidal deformability, $\tilde{\Lambda}$\footnote{In the NSBH case, $\Lambda_{\rm BH} = 0$, while $\Lambda_{\rm NS}$ is nonzero.}, computed from individual, $\Lambda_{\rm i}$, which is a function of mass of a given EOS \citep[see ][]{Flanagan2008PhRvD,Hinderer2008ApJ,Hinderer2010PhRvD,Damour2010PhRvD,Favata2014PhRvL,Raithel2018ApJ,Rezzolla2018ASSL,Chatziioannou2020GReGr,Kyutoku2021LRR}. 

\begin{figure*}[pt]
   \centering
   \includegraphics[width=\hsize]{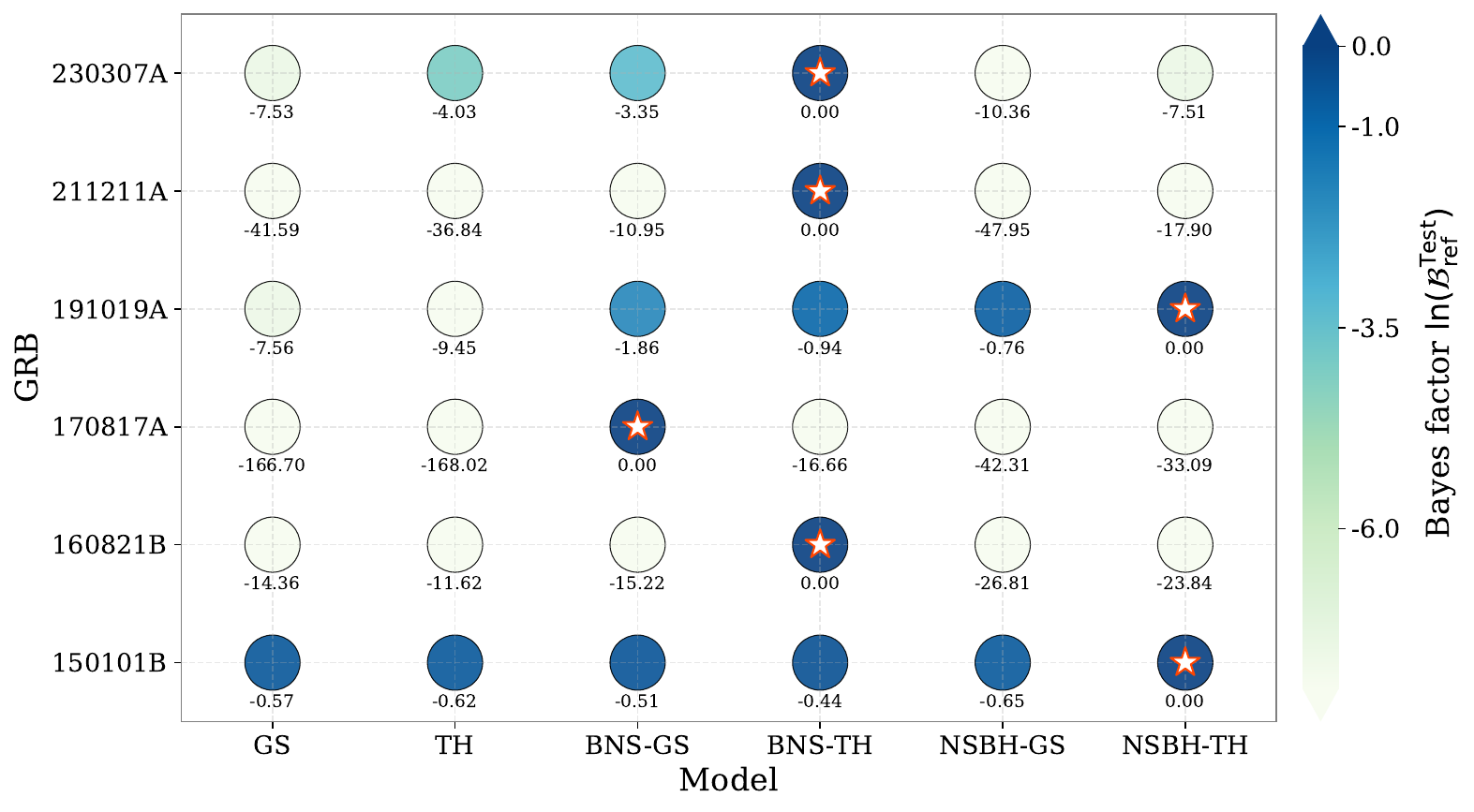}
    \caption{Bayes factor $\ln [\mathcal{B}^{\rm Test}_{\rm{Ref}}]$, quantifying the relative disfavor of test models compared the reference model. Afterglow-only models are characterized by a GS and TH jet structure. Combined afterglow and KN models correspond to a binary neutron star (BNS-GS \& BNS-TH) and a neutron star-black hole (NSBH-GS \& NSBN-TH). The star highlights the most-plausible (favored) model.
    } 
    \label{Fig:ModelSelectionAndBayes}
\end{figure*}

\begin{figure*}
	\centering
	\includegraphics[width=0.46\linewidth]{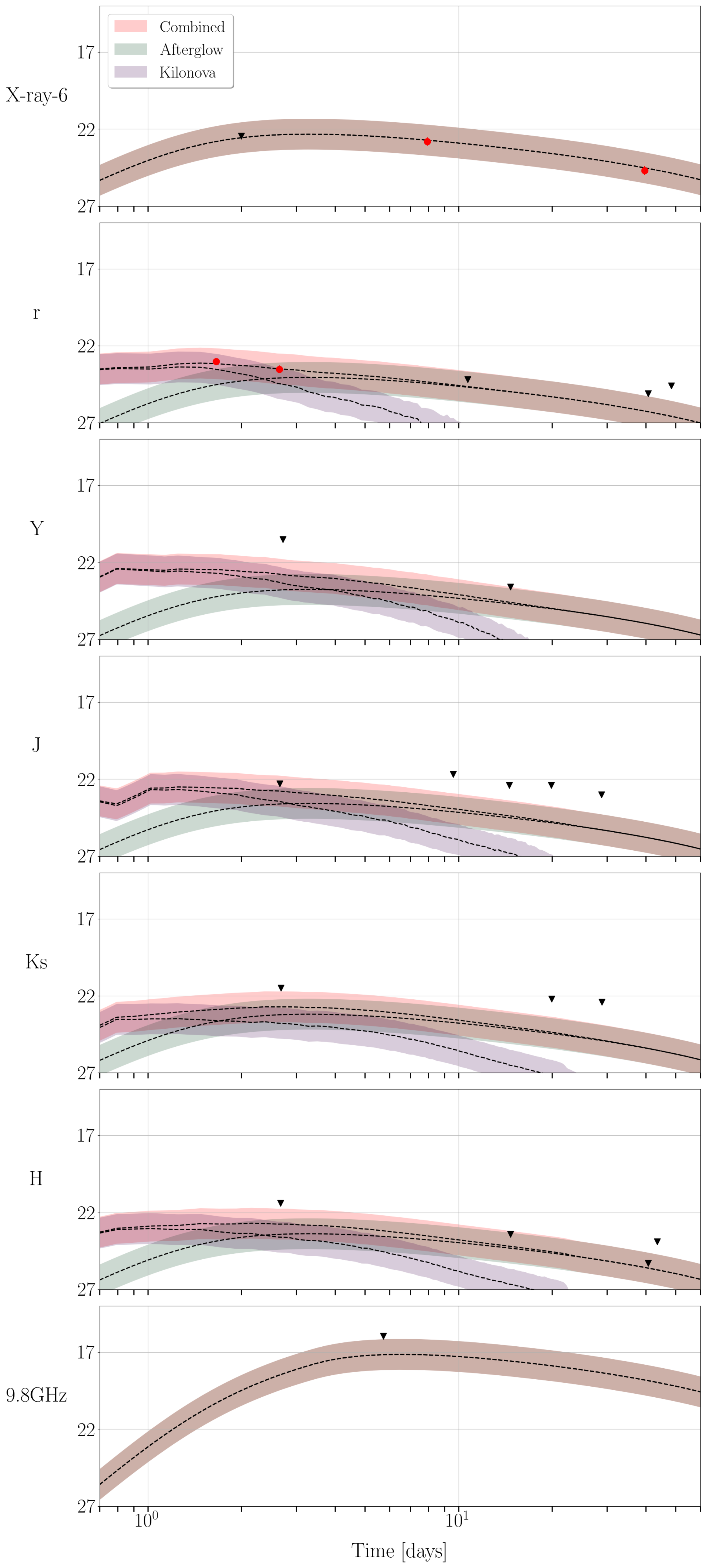}
	\includegraphics[width=0.46\linewidth]{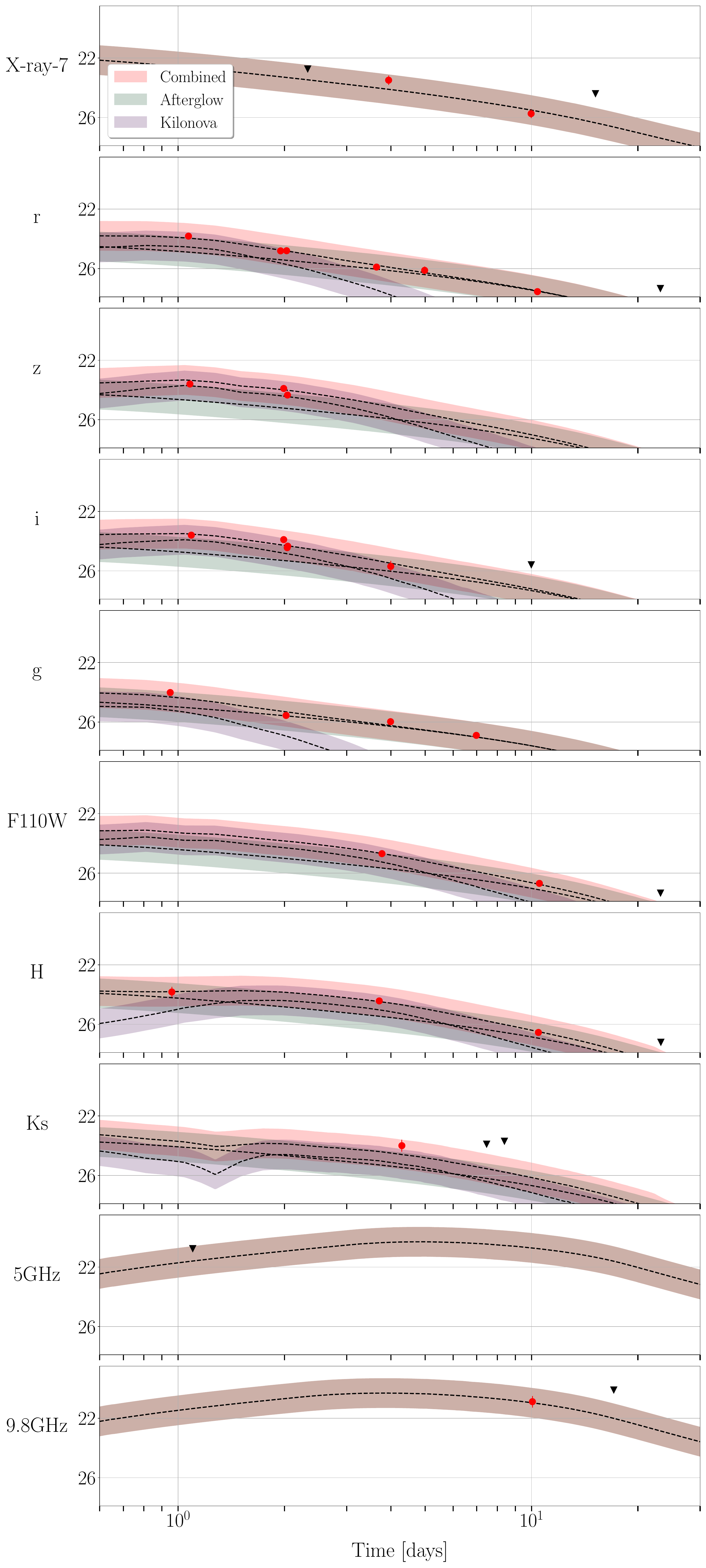}
	\caption{The best-fitting light curves corresponding to the parameters in \tabref{tab:All_KN_Results}, obtained from the joint Bayesian inference of afterglow and KN for GRB\,150101B (left) and GRB\,160821B (right). The dashed lines show the median light curve, while the shaded bands correspond to the 95\% interval. The observations are indicated in red circles and the upper limits are marked with black triangles, in AB mag. The X-ray data is shifted by the indicated magnitude for better visualizations. 
    }
	\label{Fig:150101BAnd160821BLightcurves}
\end{figure*}

\section{Results}\label{Results}
We present the results of the afterglow and KN properties for our GRB sample in \secref{KilonovaProperties} and on the binary progenitor properties in \secref{ProgenitorParameters}. We highlight that this is the first time that the latter have been obtained for a sizable sample of merger-driven GRBs. Our model comparison is reported in \figref{Fig:ModelSelectionAndBayes}, where the model with the highest Bayes factor represents the most plausible model and is highlighted with a star. The best-fit parameters for each GRB are summarized in \tabref{tab:All_KN_Results}, and the corresponding priors are provided in \tabref{tab:All_KN_Priors}. In Appendix (\tabref{tab:150101B_AllModelsResults} to \tabref{tab:230307A_AllModelsResults}), we provide the complete analysis of all models investigated for each GRB. The posterior distribution corresponding to the most-plausible model (\tabref{tab:All_KN_Results} and \tabref{tab:All_Binary_Results}) is shown in \figref{Fig:AllKNPosterior_CornerSmall1} to \figref{Fig:AllGWPosterior_CornerSmall3}.

\begin{table*}[t]
\centering
\caption{Results of the afterglow and KN parameters. }
\renewcommand{\arraystretch}{1.9}
\begin{tabular}{@{}>{\raggedright\arraybackslash}m{2.1cm}*{6}{>{\centering\arraybackslash}m{2.4cm}}@{}}
\toprule
& GRB\,150101B  & GRB\,160821B & GRB\,170817A & GRB\,191019A & GRB\,211211A &  GRB\,230307A   \\ 
\hline 
Redshift & $z=0.134$  & $z=0.1619$ & $z=0.0098$ & $z=0.248$ & $z=0.0763$ &  $z=0.0646$   \\ 
$D_{\rm L}$ (Mpc)	& 651.8 & 781.7 & 41.0 & 1289.3 & 350.0 & 300.1   \\ 
\midrule
$T_{90}$ (seconds)& 0.018 & 0.48 & 2.64 & 64.6 & 50.7  &  35.0  \\
\hline 
Best fit model	& NSBH-TH & BNS-TH & BNS-GS & NSBH-TH & BNS-TH & BNS-TH   \\  
\midrule
$\log(M_{\rm dyn})$  $M_{\odot}$ &  $-1.95^{+0.56}_{-0.51}$ & $-2.29^{+0.12}_{-0.14}$  & $-2.21^{+0.02}_{-0.02}$ &  $-1.42^{+0.26}_{-0.28}$ &  $-1.88^{+0.07}_{-0.10}$ &  $-1.85^{+0.17}_{-0.21}$ \\ 
$\log(M_{\rm wind})$   $M_{\odot}$& $-1.56^{+0.36}_{-0.38}$ &  $-2.06^{+0.11}_{-0.13}$  & $-1.17^{+0.01}_{-0.01}$ &  $-1.01^{+0.27}_{-0.39}$ &  $-2.20^{+0.07}_{-0.09}$ & $-1.57^{+0.19}_{-0.30}$ \\ 
$\Phi$  (deg)& 30 & $70^{+7}_{-7}$ & $70^{+1}_{-1}$ &  30 & $72^{+2}_{-5}$ &  $64^{+7}_{-10}$\\
\midrule
$\iota$ (rad)        & $0.20^{+0.11}_{-0.13}$   & $0.30^{+0.05}_{-0.07}$ & $0.57^{+0.01}_{-0.01}$ & $0.06^{+0.09}_{-0.03}$ & $0.005^{+0.001}_{-0.001}$    & $0.06^{+0.03}_{-0.02}$  \\	
$\log(E_0)$  (erg)  &  $52.36^{+0.84}_{-0.71}$ & $50.52^{+0.50}_{-0.25}$ & $52.23^{+0.10}_{-0.09}$ & $52.43^{+0.30}_{-0.40}$ & $51.54^{+0.22}_{-0.23}$  & $51.22^{+0.56}_{-0.53}$  \\ 
$\log(n_0)$  (cm$^{-3}$)  &  $-2.83^{+1.13}_{-0.67}$   & $-1.67^{+0.79}_{-1.00}$ & $-2.96^{+0.10}_{-0.08}$ & $0.53^{+1.01}_{-1.22}$ & $-6.25^{+0.33}_{-0.34}$  & $-4.25^{+0.96}_{-0.77}$\\
$\theta_{\rm c}$ (rad)&  $0.22^{+0.05}_{-0.07}$ &  $0.22^{+0.05}_{-0.09}$ & $0.138^{+0.002}_{-0.002}$ & $0.24^{+0.04}_{-0.08}$ &   $0.017^{+0.003}_{-0.002}$  & $0.09^{+0.04}_{-0.03}$ \\ 
$\theta_{\rm w}$ (rad) &  -  & - & $0.64^{+0.01}_{-0.01}$ & - &   - & - \\ 
$p$   &  $2.19^{+0.08}_{-0.05}$  & $2.34^{+0.15}_{-0.15}$ & $2.14^{+0.01}_{-0.01}$   & $2.74^{+0.05}_{-0.05}$ &  $2.39^{+0.02}_{-0.03}$  & $2.50^{+0.05}_{-0.05}$\\
$\log(\epsilon_e)$ &  $-0.93^{+0.54}_{-1.02}$  & $-0.33^{+0.15}_{-0.18}$ & $-1.85^{+0.10}_{-0.10}$ & $-0.86^{+0.47}_{-0.53}$ &  $-0.10^{+0.05}_{-0.07}$ & $-0.33^{+0.20}_{-0.14}$\\ 
$\log(\epsilon_B)$ &  $-3.28^{+1.12}_{-0.99}$   & $-2.23^{+0.51}_{-0.65}$ & $-2.25^{+0.08}_{-0.12}$ & $-5.12^{+0.94}_{-1.06}$ &  $-1.40^{+0.37}_{-0.35}$ & $-1.92^{+0.57}_{-0.68}$\\ 
\bottomrule
\end{tabular}
\tablefoot{
The parameters are: $M_{\rm dyn}$ = dynamical ejecta mass; $M_{\rm wind}$ = wind ejecta mass; $\Phi$ = half-opening angle of lanthanide-rich equatorial ejecta; $\iota$ = inclination angle; $E_{0}$ = kinetic isotropic equivalent energy; $n_{0}$ = particle number density in the circumburst environment; $\theta_{\rm c}$ = half-opening angle of the jet core; $\theta_{\rm w}$ = half-opening angle of the jet truncated-wings;  $p$ = electron energy distribution power-law index; $\epsilon_{e}$ = shock energy fraction that goes into the electrons and $\epsilon_{B}$ = shock energy fraction that goes into the magnetic energy density. In the case of the TH jet structure, $\theta_{\rm w}$ is not a model parameter and $\Phi$ is fixed = 30 (deg) for all NSBH models (see \secref{ModelingEM}). }
\label{tab:All_KN_Results}
\end{table*}

\begin{figure}
	\centering
	\includegraphics[width=\linewidth]{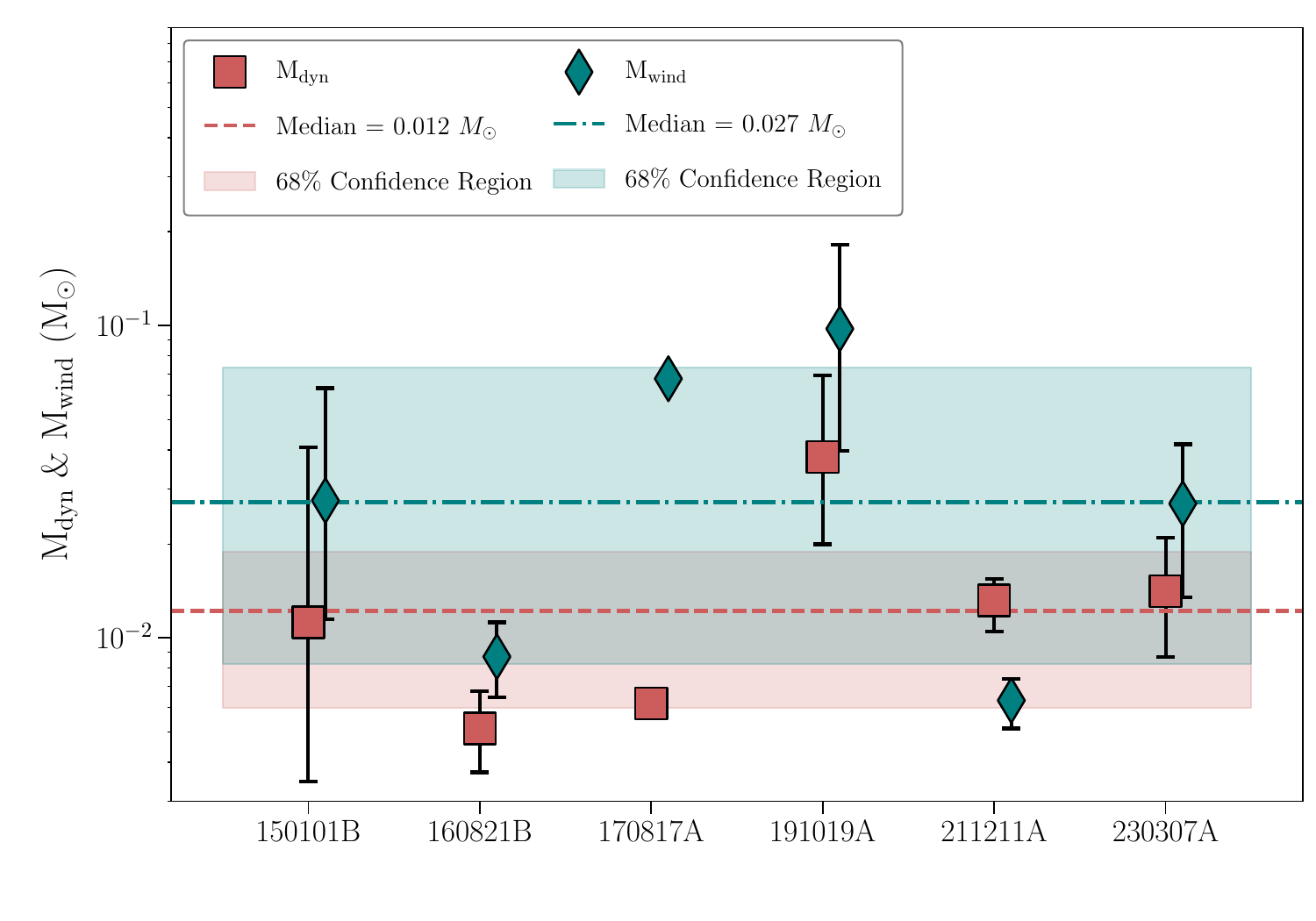}
	\caption{Dynamical mass, $M_{\rm Dyn}$ ($M_{\odot}$), and the wind mass, $M_{\rm Wind}$ ($M_{\odot}$), of KN ejecta. The shaded region highlights the 68\% confidence region and the dotted lines show the median. }
	\label{Fig:EjectaMassesMWindDyn}
\end{figure}

\begin{figure}
   \centering
   \includegraphics[width=\hsize]{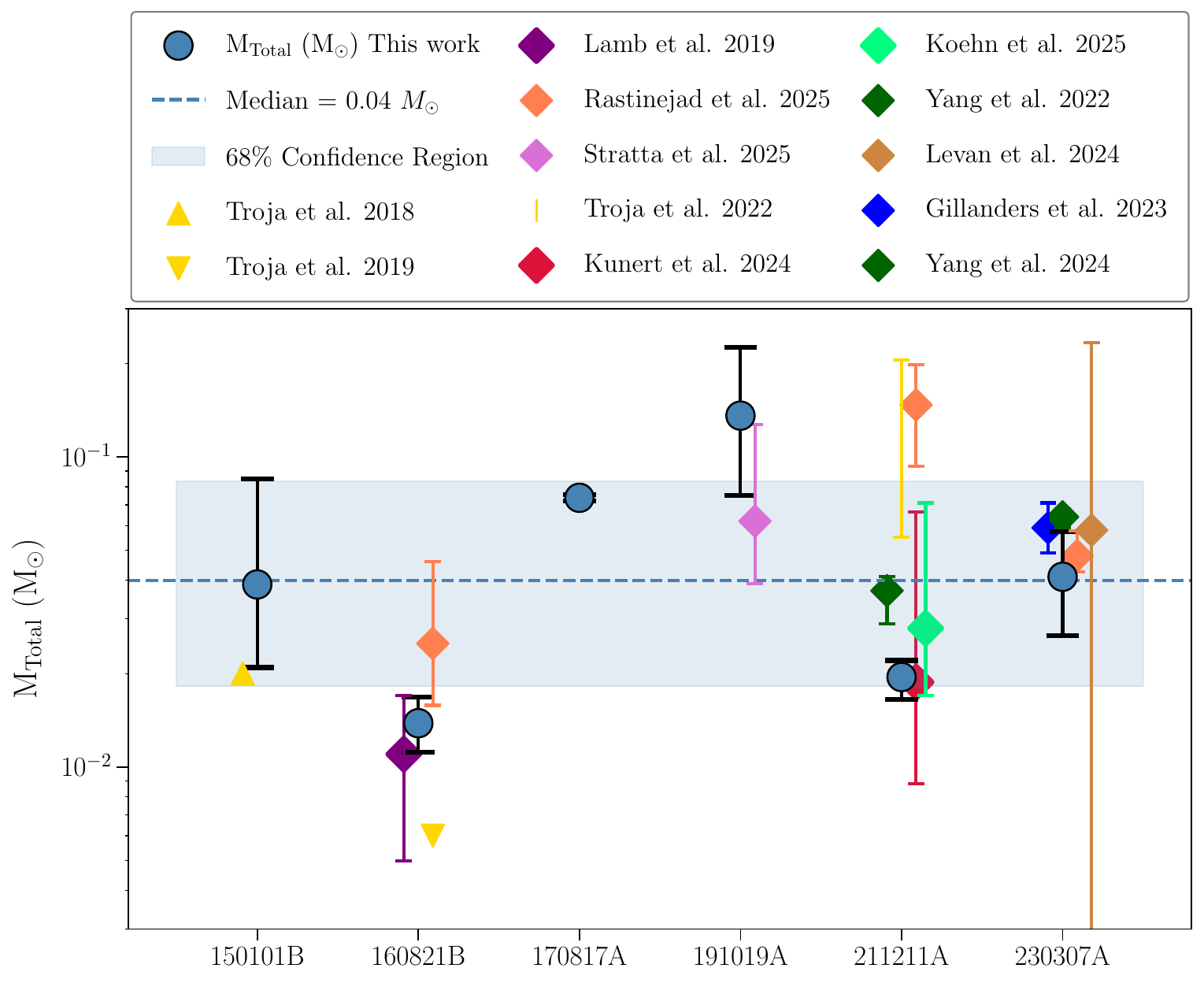}
    \caption{Distribution of the total ejected mass, $M_{\rm Total}$, $M_{\odot}$, and comparison with pervious studies. The dotted line shows the median and the shaded part highlights the 68\% confidence region. The measurements for each GRB have been slightly shifted along the x-axis for clarity. The comparison of AT2017gfo is shown in a separate plot (see \figref{Fig:170817MTotalComparison}). The lower limit \citep{Troja2018GRB150101B} for For GRB\,150101B and the upper limit for GRB\,160821B \citep{Troja2019MNRAS} are shown. For GRB\,211211A, vertical bar indicates the upper and lower limits from \citet{Troja2022Natur211211A}. } 
    \label{Fig:EjectaMassesMtotal}
\end{figure}

\begin{figure}
	\centering
	\includegraphics[width=\linewidth]{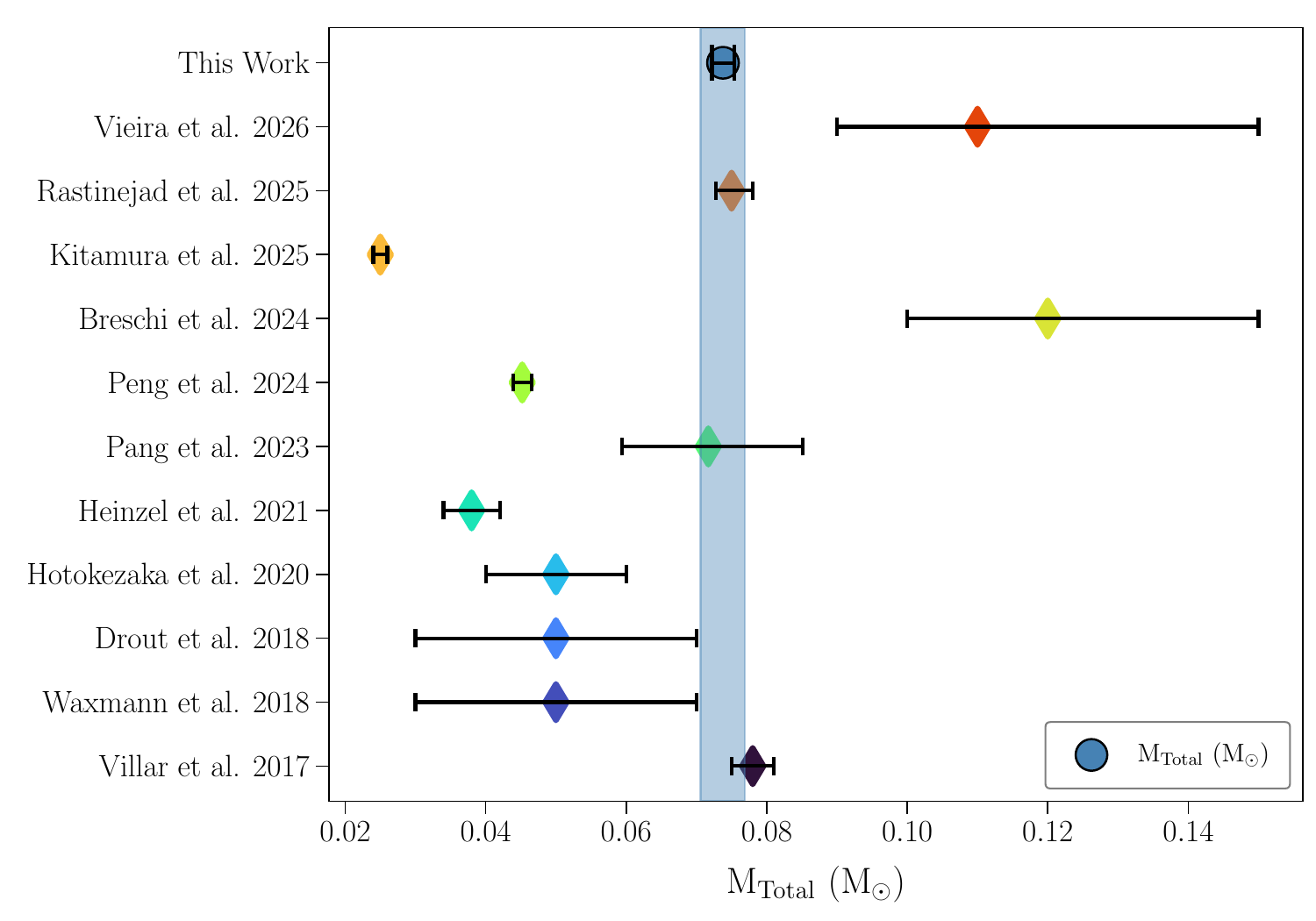}
	\caption{Comparison of the $M_{\rm Total}$ and $M_{\odot}$, for AT2017gfo. The shaded region corresponds to $2\sigma$ region of $M_{\rm Total}$ from our analysis.}
	\label{Fig:170817MTotalComparison}
\end{figure}

\subsection{Afterglow and kilonova properties}\label{KilonovaProperties}
GRB\,150101B Bayes factor (\figref{Fig:ModelSelectionAndBayes}) comparison shows a preference for a KN originating from an NSBH progenitor with a TH jet structure, but other scenarios cannot be conclusively ruled out. The light curves in different bands are shown in the left panel of \figref{Fig:150101BAnd160821BLightcurves}, where the r band observations $\sim0.5$ days after the burst onset clearly show an excess with respect to the afterglow flux. The afterglow emission is compatible with a slightly off-axis inclination with $\iota=11.46^{+6.30}_{-7.45}$ (deg), in line with previous results \citep{Troja2018GRB150101B}. Its high isotropic equivalent kinetic energy, $\log(E_0)=52.36^{+0.84}_{-0.71}$ (erg), results in a dimmer KN emission with a total ejected mass of $M_{\rm Total}=0.039^{+0.046}_{-0.018} M_{\odot}$.

GRB\,160821B is best modeled with a BNS merger and a TH jet structure (\figref{Fig:ModelSelectionAndBayes}), favoring a KN while all other models are strongly disfavored. The multiband light curves in the right panel of \figref{Fig:150101BAnd160821BLightcurves}, show an excess emission that starts to rise above the afterglow after $\sim1$ day in the optical bands and further peaks in the NIR wavelengths in the F110W, Ks, and H bands around $\sim3$ days. Subsequently (at $\gtrsim4$ days) the excess emission subsides and the afterglow again becomes dominant at optical wavelengths, as seen in r and g band observations. We note that the total ejected mass of GRB\,160821B is the lowest of the entire sample, with $M_{\rm Total}=0.014\pm{0.003} M_{\odot}$ \citep[compatible with ][]{Troja2019MNRAS,Lamb2019ApJ160821B,Rastinejad2025} and the faint afterglow is described by a low $\log(E_0)=50.52^{+0.50}_{-0.25}$ (erg), resulting in a brighter KN in contrast to GRB\,150101B. The $\Phi=70\pm{7}$ (deg) is indicative of large amounts of lanthanide-rich ejecta. We find that GRB\,160821B has an off-axis inclination of $\iota=17.19^{+2.86}_{-4.01}$ (deg), consistent with the literature \citep{Troja2019MNRAS}.

GRB\,170817A is the well known GRB associated with the GW-detected BNS GW170817. We specifically included this event in our sample to test the robustness and consistency of our methodology. Our analysis shows a strong preference for a KN emission originating from a BNS merger with a GS jet structure for the afterglow (\figref{Fig:ModelSelectionAndBayes}), aptly confirming past results and validating our methodology. From the light curves (see left panel in \figref{Fig:170817And191019ALightcurves}), at all epochs prior to $\sim10$ days, the KN emission arising from $M_{\rm Total}=0.073\pm{0.001} M_{\odot}$ and high $\Phi=70.41\pm{1}$ (deg) completely outshines the off-axis afterglow emission described by $\log(E_0)=52.23^{+0.10}_{-0.09}$ (erg). Specifically, our inferred values for the inclination angle $\iota=32.73^{\circ}\pm0.57^{\circ}$ (deg) and the jet-core $\theta_{\rm core}=7.91^{\circ}\pm0.11^{\circ}$ (deg) are consistent with the widely reported values $\iota \approx 20^{\circ}-35^{\circ}$ and $\theta_{\rm core}\lesssim 5^{\circ}-10^{\circ}$, respectively \citep{Troja2017Nature,Lazzati2018PhRvL,Lyman2018NatAs,Finstad2018ApJ,Lamb2019ApJ,Troja2019MNRASGW170817,Ryan2020ApJ,Dhawan2020ApJ,Takahashi2021MNRAS,Gill2019ApJ,Nathanail2020MNRAS,Mpisketzis2024MNRAS}. We note that a slightly smaller inclination of $\iota \approx 15^{\circ}-22^{\circ}$ was inferred from radio observations that account for the measurement of the source size, superluminal motion, and displacement of the centroid \citep{Mooley2018Natur,Ghirlanda2019Sci,Mooley2022Nature}. Recent works have included such a centroid motion in the modeling of the GRB\,170817A afterglow \citep{Ryan2024ApJ,Gianfagna2024MNRAS} and we took into account their analysis, while selecting our priors (see \secref{AfterglowModeling}). However, they require an additional model-independent constant luminosity in the X-ray, radio, and F606W bands to be fully consistent with $\iota \approx 15^{\circ}-22^{\circ}$, which we have not considered in our analysis. 

GRB\,191019A is the first of the three merger-driven long GRBs in our sample. Although KN is strongly favored compared to an afterglow-only solution, both BNS and NSBH are viable as progenitors (\figref{Fig:ModelSelectionAndBayes}). From our analysis, the highest Bayes factor is obtained for an NSBH-TH model. The optical observations show an excess emission $\sim1$ day in the optical bands, which is incompatible with the forward shock afterglow emission (see right panel in \figref{Fig:170817And191019ALightcurves}). Our result of excess emission is compatible with the presence of a KN \citep{Stratta2025ApJ}. The KN emission is described by $M_{\rm Total}=0.136^{+0.089}_{-0.061} M_{\odot}$, while the afterglow is shaped by $\iota=3.44^{+5.16}_{-1.72}$ (deg), $\log(E_0)=52.43^{+0.30}_{-0.40}$ (erg), and $\log(n_0) = 0.53^{+1.01}_{-1.22}$ (cm$^{-3}$), which are the highest values in our sample. 

GRB\,211211A is another long-duration event in our sample that provides strong evidence of a KN stemming from a BNS as a progenitor and its afterglow explained by a TH jet structure (\figref{Fig:ModelSelectionAndBayes}). This GRB accounts for the most extensive dataset within our sample, gathered by cross-checking all available literature (see \secref{211211A_Sample}). The early observations $\lesssim0.9$ days across all wavelengths are well described by a dominant afterglow emission with $\log(E_0)=51.54^{+0.22}_{-0.23}$ (erg), whereas at later epochs ($>1$ days) a shallow decay phase compatible with a KN becomes apparent (see left panel in \figref{Fig:211211AAnd230307ALightcurves}). Specifically, the KN arising from $M_{\rm Total}=0.019\pm{0.002}$ $M_{\odot}$ \citep[consistent with][]{Yang2022Nature,Kunert2024,Koehn2025arXiv} and $\Phi=72^{+2}_{-5}$ (deg) outshines the afterglow up to about $\sim10$ days in the NIR bands. The late deep upper limits ($>10$ days) rule out any further prolonged KN emission. The jet aperture of this GRB, $\theta_{\rm c}=1.0^{+0.2}_{-0.1}$ (deg), is among the narrowest, while the viewing angle $\iota=0.29\pm0.07$ (deg) was found well within the jet cone, while it has the lowest circumburst density, with $\log(n_0) = -6.25^{+0.33}_{-0.34}$ (cm$^{-3}$). 

GRB\,230307A is the third binary-driven long GRB of our sample and is well modeled by a BNS with a TH jet structure, while clearly favoring a KN emission (\figref{Fig:ModelSelectionAndBayes}). Its light curve shows a typical decaying behavior in X-rays constrained by early ($<1$ days) \texttt{TESS} (Ic band) observations. The total ejected mass of $M_{\rm Total}=0.041^{+0.016}_{-0.014} M_{\odot}$ \citep[consistent with][]{Levan2024Nature,Yang2024Nature,Gillanders2024,Rastinejad2025} and $\Phi=64^{+7}_{-10}$ (deg) powers the KN emission, which appears brighter than the afterglow in the Ks band at $\sim10$ days (see right panel in \figref{Fig:211211AAnd230307ALightcurves}). Also for this event, like GRB\,211211A, we found a very low circumburst density $\log(n_0) = -4.25^{+0.96}_{-0.97}$ (cm$^{-3}$) and the lowest $\log(E_0)=51.22^{+0.56}_{-0.53}$ (erg), while its inclination $\iota=3.43^{+1.71}_{-1.14}$ (deg) is almost on-axis and constrained inside the jet cone with $\theta_{\rm c}=5.2^{+2.3}_{-1.7}$ (deg).

When considering the entire GRB sample, we see that the median value of the dynamical ejecta mass is $\langle M_{\rm dyn}\rangle= 0.012^{+0.007}_{-0.006}$ $M_{\odot}$, is nearly half of the wind mass median $\langle M_{\rm wind}\rangle= 0.027^{+0.046}_{-0.019}$ $M_{\odot}$ (\figref{Fig:EjectaMassesMWindDyn}). The median value of the total ejected mass reported in \tabref{tab:All_TotalMass} was obtained from the summation of $M_{\rm dyn}$ and $M_{\rm wind}$ is $\langle M_{\rm Total}\rangle=0.039^{+0.043}_{-0.033} $ $M_{\odot}$ (\figref{Fig:EjectaMassesMtotal}). 

\begin{table*}[t]
\centering
\caption{Values of the total ejected mass $M_{\rm Total}$ ($M_{\odot}$) responsible for the KN emission.  }
\renewcommand{\arraystretch}{1.9}
\begin{tabular}{@{}>{\raggedright\arraybackslash}m{2.1cm}*{6}{>{\centering\arraybackslash}m{2.4cm}}@{}}
\toprule
& GRB\,150101B  & GRB\,160821B & GRB\,170817A & GRB\,191019A & GRB\,211211A &  GRB\,230307A   \\ 
\hline
$M_{\rm Total}$ ($M_{\odot}$) & $0.039^{+0.046}_{-0.018}$  & $0.014^{+0.003}_{-0.003}$ & $0.074^{+0.002}_{-0.001}$ & $0.136^{+0.089}_{-0.061}$ & $0.019^{+0.003}_{-0.003}$ & $0.041^{+0.016}_{-0.014}$   \\ 
\bottomrule
\end{tabular}
\tablefoot{The $M_{\rm Total}$ is computed by summing the $M_{\rm dyn}$ and $M_{\rm wind}$ values in \tabref{tab:All_KN_Results}.}
\label{tab:All_TotalMass}
\end{table*} 

\subsection{Progenitor properties}\label{ProgenitorParameters}
In this section, we present the properties inferred for the progenitor systems. We highlight once again that this is the first time these properties have been obtained for a sample of merger-driven GRBs. Given the absence of the GW observations of other GRBs, in the case of GW170817, we used only its EM counterpart, similar to the other events in our sample. In doing so, we can validate our methodology by comparing our results with the current estimates of GW170817 and buttress the binary properties of other GRBs. In the appendix (\figref{Fig:AllGWPosterior_CornerSmall1} to \figref{Fig:AllGWPosterior_CornerSmall3}), we show the posterior distribution of binary properties for individual GRBs.

Our results for GRB\,170817A/AT2017gfo are quoted in \tabref{tab:All_Binary_Results}, where we find the mass-weighted tidal deformability, $\tilde{\Lambda}= 595^{+226}_{-171}$, the chirp mass $M_{\rm Chirp}=1.13^{+0.07}_{-0.07}$ $M_{\odot}$, and the binary mass ratio, $q = 0.93^{+0.04}_{-0.06} $, consistent with $\tilde{\Lambda} < 800$,  $M_{\rm Chirp}=1.186$ $M_{\odot}$, and $q\approx0.7-1.0$, respectively, measured by \texttt{LVK} \citep{Abbott2017PhRvL}. Furthermore, our value of tidal deformability is compatible with $323\lesssim \tilde{\Lambda} \lesssim776$ \citep{Radice2019EPJA}, $200\lesssim \tilde{\Lambda} \lesssim800$ \citep{Altiparmak2022ApJ} as well as with other studies \citep{Coughlin2019MNRAS,Bauswein2020PhRvL,Breschi2024A&A}. The compatibility of our results with previous estimates validates the consistency of our analysis. Next, we focus on the binary properties of the other GRBs in our sample summarized in \tabref{tab:All_Binary_Results} and also displayed in \figref{Fig:TidalMassRatio}.

Focusing on the NSBH cases of our sample, GRB\,150101B has $\tilde{\Lambda}=14^{+31}_{-9}$, ${M}_{\rm \,Chirp}=2.05^{+0.43}_{-0.34} M_{\odot}$ and the binary mass ratio, $q=0.24^{+0.13}_{-0.08}$. Similarly, GRB\,191019A also shows a low $\tilde{\Lambda}=27^{+67}_{-20}$, high ${M}_{\rm \,Chirp}=1.86^{+0.45}_{-0.36} M_{\odot}$, and low $q=0.27^{+0.11}_{-0.09}$. These values are attributed to a heavier primary component. 

For the BNS progenitor cases, GRB\,160821B has $\tilde{\Lambda}=309^{+290}_{-108}$ and ${M}_{\rm \,Chirp}=1.25^{+0.10}_{-0.13} M_{\odot}$, while $q=0.89^{+0.06}_{-0.07}$. GRB\,211211A also has large $\tilde{\Lambda}=300^{+235}_{-112}$ and ${M}_{\rm \,Chirp}=1.27^{+0.12}_{-0.11} M_{\odot}$, while the binary mass ratio $q=0.81^{+0.06}_{-0.06}$. This suggests that for both GRBs, the binaries are close to having equal mass. Finally, GRB\,230307A is explained by $\tilde{\Lambda}=294^{+284}_{-116}$ and ${M}_{\rm \,Chirp}=1.27^{+0.12}_{-0.14} M_{\odot}$. The result of $q=0.67^{+0.08}_{-0.06}$ indicates that the primary is more massive than the secondary compact object. 

\section{Discussion}\label{Discussion}
In the following, we discuss our results on the electromagnetic counterparts, binary properties, and trends obtained from our analysis. We also include a comparison with previous studies. 

\begin{figure}
	\centering
	\includegraphics[width=\linewidth]{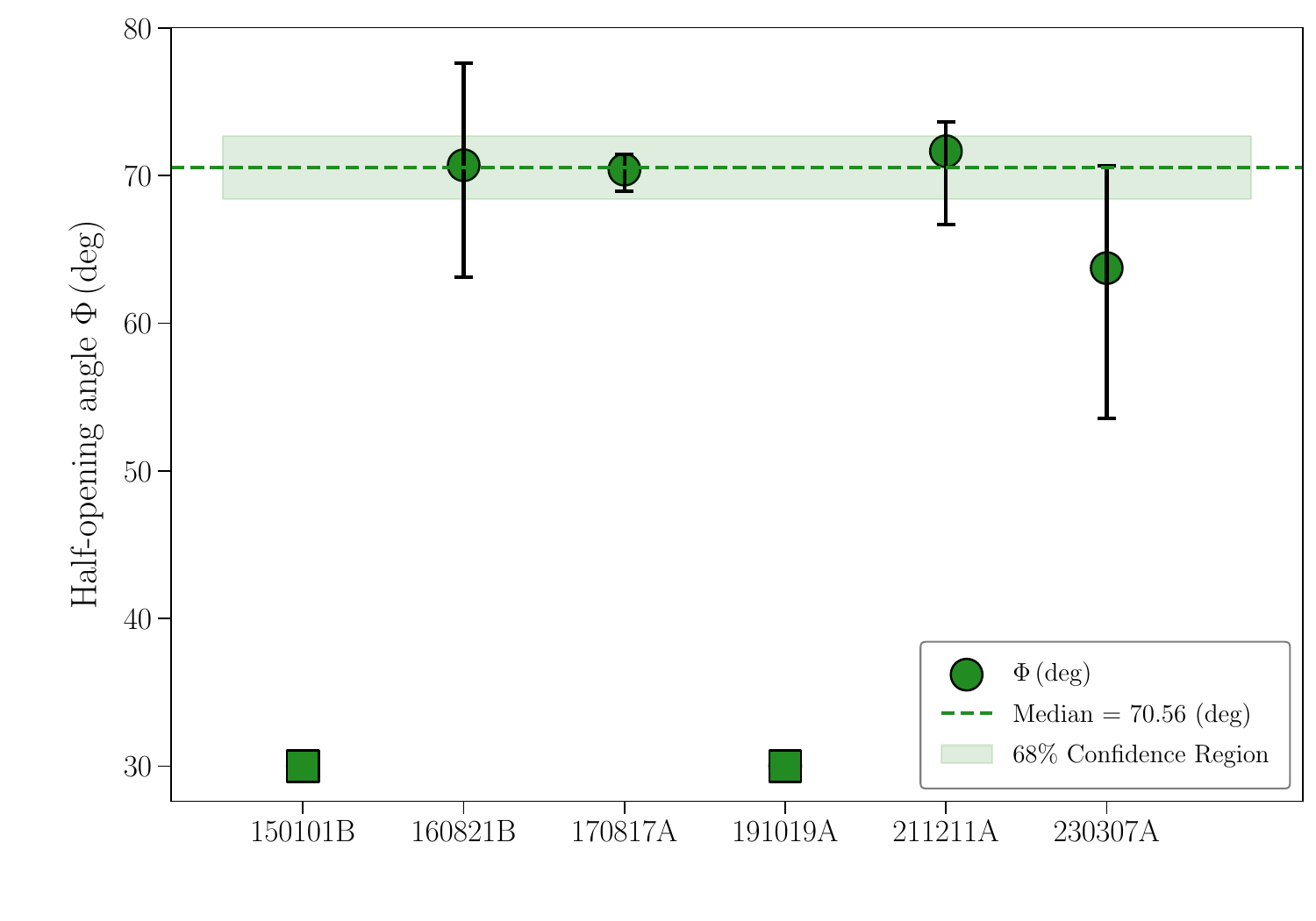}
	\caption{Distribution of the half-opening angle, $\Phi$ (deg), of lanthanide-rich ejecta. For GRB\,150101B and GRB\,191019A, highlighted with square, the $\Phi$ value is fixed = 30 (deg) corresponding to the NSBH models. The median value is computed by excluding the former two GRBs.}
	\label{Fig:PhiAngel}
\end{figure}

\begin{figure}
	\centering
    \includegraphics[width=\linewidth]{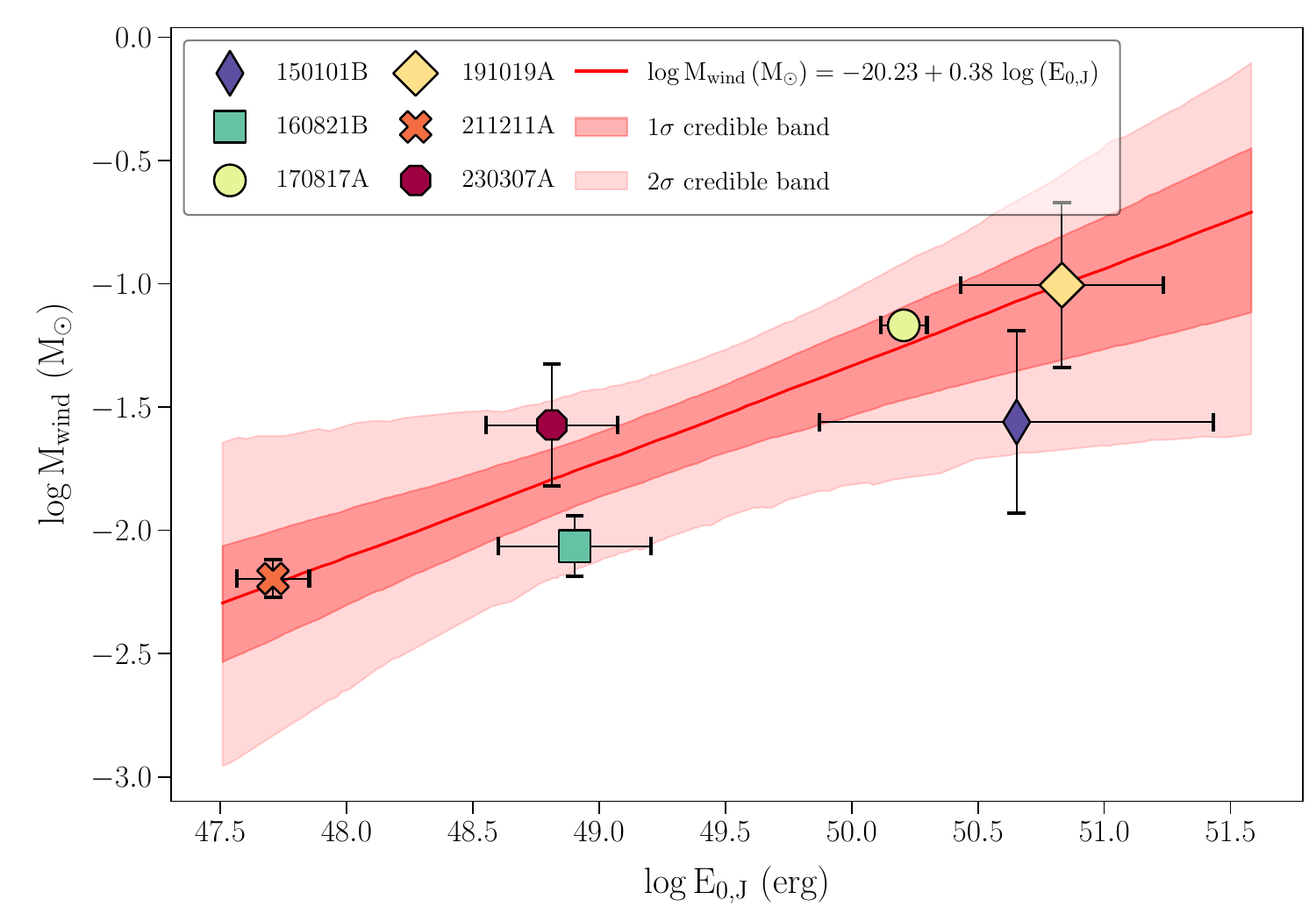}
	\caption{Relation between the collimation-corrected isotropic equivalent kinetic energy of the jet $E_{0,J}$ (erg) and the ejected wind mass, $M_{\rm Wind}$ ($M_{\odot}$). 
    }
	\label{Fig:EnergyMass}
\end{figure}

\subsection{Dynamical and wind mass in KNe}\label{DynAndWindMassComparison}
In \secref{KilonovaProperties} and \figref{Fig:EjectaMassesMWindDyn}, we show that the post-merger wind mass is the leading term that contributes to the total ejected mass, where we find that $M_{\rm wind}$ is almost twice as large as $M_{\rm dyn}$. This result confirms simulations and past findings on compact binary mergers, where the dynamical mass is expected to be smaller $\lesssim10^{-2} M_{\odot}$ \citep{Radice2018ApJ,Shibata2019,Kruger2020PhRvD,Han2025PhRvD,Cook2025arXiv2508,Gutierrez2025arXiv2506}, while more massive wind ejecta ($\gtrsim10^{-2} M_{\odot}$) have been found \citep{Siegel2018ApJ,Fujibayashi2018ApJ,Shibata2019,Foucart2023LRCA}. Focusing on the mass ejection timescales, results from numerical merger simulations suggest that less mass is ejected dynamically after merger onset ($\lesssim10$ ms) and more mass is expelled in wind (secular) outflows on a much longer viscous timescale ($\gtrsim10$ ms), persisting to much later post-merger epochs \citep[see, e.g.,][]{Rosswog2015IJMPD,Fujibayashi2018ApJ,Shibata2019,Kawaguchi2020ApJ,Nakar2020}, consistent with our findings.

Furthermore, the recent work by \cite{Kitamura2025ApJ} showed that radiative transfer models with dynamical and wind components and photon reprocessing (as also applied in our analysis) are in closer agreement with numerical simulations. In contrast, models utilizing a blue or red decomposition of the dynamical and wind components can yield less consistent interpretations of the KN ejecta \citep{Kitamura2025ApJ}; see also \citep{Bulla2019MNRAS,Kawaguchi2020ApJ,Bulla2023MNRAS}. Hence, our result is consistent with both numerical simulations and detailed radiative transfer modeling.

\subsection{Comparison of the total ejected mass}\label{DiscussionEjectaMass}
A one-to-one comparison for the properties of KN ejecta from the previously published literature, where various components and morphologies of the ejecta have been utilized, is challenging \citep[see, e.g.,][]{Ascenzi2019,Kawaguchi2020ApJ,Nicholl2021MNRAS,Heinzel2021MNRAS,King2025arXiv2505,Vieira2026ApJ}. Therefore, for completeness, we show the comparison of the total ejected mass obtained from our analysis with previous studies (\figref{Fig:EjectaMassesMtotal}). We found compatible values for the majority of the GRBs and we discuss only the two cases where different values have been inferred below. 

For AT2017gfo, our results within $2\,\sigma$ are compatible with the vast majority of past findings (see \figref{Fig:170817MTotalComparison}). However, we note that considerably smaller values of the total ejected mass have been obtained by some studies. \citet{Heinzel2021MNRAS} employed \texttt{POSSIS} \citep{Bulla2019MNRAS}, similarly to the approach taken in our analysis, but they considered simulations from \citet{Coughlin2020NatCo}, finding $\Phi$ angle $=45^{+9}_{-8}$ (deg) and $M_{\rm Total}=0.038^{+0.004}_{-0.004} M_{\odot}$. These values  are lower than the results we obtained. \citet{Peng2024PhRvR} used \texttt{SuperNu} \citep{Wollaeger2014ApJS} for the radiative transfer and a different prescription of the ejecta geometry \citep{Wollaeger2021ApJ}, while incorporating a rapid iterative fitting \citep{Wofford2023PhRvD} to perform the Monte Carlo analysis. They found $M_{\rm Total}=0.045^{+0.001}_{-0.001} M_{\odot}$, while the viewing angle in their analysis $\iota\approx6$ (deg) was considerably smaller than current estimates (see \secref{KilonovaProperties}). \citet{Kitamura2025ApJ} obtained $M_{\rm Total}=0.025^{+0.001}_{-0.001} M_{\odot}$ using a different model for the KN emission \citep{Villar2017ApJ,Metzger2017LRR}, included in \texttt{MOSFiT} \citep{Guillochon2018ApJS}, while also considering the polar angle for the inclination. 

In the case of GRB\,211211A, a higher value $M_{\rm Total}=0.048^{+0.021}_{-0.009}$ $M_{\odot}$ was found by \citet{Rastinejad2025}, while $M_{\rm Total}\approx 0.02\,(0.13)$ and $M_{\odot}$ lower (upper) limits have been suggested by \citet{Troja2022Natur211211A}. Both cited works modeled the KN and afterglow components independently, unlike the method used in the present work. 

\subsection{On the $\Phi$ angle}\label{DiscussionPhiAngle}
The \texttt{POSSIS} code (see \secref{KilonovaModeling}) we utilized allowed us to constrain not only the ejecta masses, but also the lanthanide-rich component in the ejecta by constraining the half-opening angle ($\Phi$). Thermal KN emission from this component is expected to be predominant at NIR wavelengths at late times, and the high values of $\Phi$ correspond to the large amount of the lanthanide-rich ejecta. In \figref{Fig:PhiAngel}, we show the values of $\Phi$, for individual GRB cases,  finding that cases with BNS progenitors display similar values, with a  median of $\langle\Phi\rangle=70\pm2$ (deg), suggesting a similar fraction of lanthanide-free vs lanthanide-rich masses. In the case of GRB 150101B and GRB 191019A, where the best fit progenitor is a NSBH, the $\Phi$ parameter in \texttt{POSSIS} is fixed at $30$ (deg). 

\subsection{Afterglow and kilonova properties}\label{DiscussionAfterglow}
A GRB jet originates via mass accretion from the post-merger disk onto a compact object left behind as a remnant in the aftermath of a binary merger \citep[see, e.g.,][]{Eichler1989Nature,Narayan1992ApJ,Rezzolla2011ApJ,Ascenzi2019ApJ877}. Therefore, it is natural to expect that the ejected wind mass, which is a fraction of the accretion disk, would be correlated with the energy output of the GRB jet. This expectation has been reported for GRB\,170817A \citep{Salafia2021}. Another analysis focusing on short GRBs (but not KNe) suggested a similar behavior based on the predicted disk mass responsible for the jet energy \citep{Mpisketzis2024A&A}. Recently, \citet{Rastinejad2025} (see their Figure 5) noticed a possible hint of such a trend, but for different parameters: the $M_{\rm Total}$ and $E_{\gamma}$ (beaming-corrected $\gamma-$ray energy); also $E_{\gamma,\,\rm{iso}}$ (isotropic-equivalent $\gamma-$ray energy). We confirm that this expectation has been reproduced in our analysis (see \figref{Fig:EnergyMass}) with a positive correlation between the wind mass and the collimation-corrected kinetic energy of the jet in logarithmic scale, with $\log\, M_{\rm Wind}\,(M_{\odot}) = -20^{+6}_{-5} + 0.38^{+0.11}_{-0.12}\, log\, (E_{\rm 0,J})$ (erg), where we accounted for the collimation-correction as $(E_{\rm 0,J}) = (1-\cos(\theta_{\rm core}))E_{\rm 0}$. The Pearson correlation coefficient ($r = 0.85$, $p = 0.034$) and Spearman's rank correlation coefficient ($\rho = 0.89$, $p = 0.019$) both indicate a statistically significant correlation between $M_{\rm wind}$ and $E_{\rm 0,J}$. A larger sample of GRBs with KNe in future observations would further improve the statistical significance of this correlation, ultimately providing valuable insights into the physics of the system, such as the fraction of disk mass contributing to wind ejecta, the accretion-to-jet energy conversion efficiency, and the conversion efficiency of kinetic energy into observed gamma-ray radiation \citep[e.g.,][]{Salafia2021}.

\begin{table*}[t]
\centering
\caption{Results of the binary progenitor properties, corresponding to the models \tabref{tab:All_KN_Results}.}
\renewcommand{\arraystretch}{1.9}
\begin{tabular}{@{}>{\raggedright\arraybackslash}m{2.1cm}*{6}{>{\centering\arraybackslash}m{2.4cm}}@{}}
\toprule
& GRB\,150101B  & GRB\,160821B & GRB\,170817A & GRB\,191019A & GRB\,211211A &  GRB\,230307A  \\ 
\hline
Best-fit model	& NSBH-TH & BNS-TH & BNS-GS & NSBH-TH & BNS-TH & BNS-TH   \\  
\midrule
$\mathcal{U} D_{\rm L}$ (Mpc) & $\mathcal{U} (400-800)$ & $\mathcal{U} (600-900)$ & $\mathcal{U} (35-50)$ & $\mathcal{U} (1000-1300)$ & $\mathcal{U} (200-400)$ &  $\mathcal{U} (200-400)$   \\
\midrule
$M_{\rm Chirp}$  ($\rm M_{\odot}$) &  $2.05^{+0.43}_{-0.34}$ & $1.25^{+0.10}_{-0.13}$  & $1.13^{+0.07}_{-0.07}$ &  $1.86^{+0.45}_{-0.36}$ &  $1.27^{+0.12}_{-0.11}$ &  $1.27^{+0.12}_{-0.14}$ \\ 

$\tilde\Lambda$ & $14^{+31}_{-9}$ & $309^{+290}_{-108}$ & $595^{+226}_{-171}$ &  $27^{+67}_{-20}$ & $300^{+235}_{-112}$ &  $294^{+284}_{-116}$\\

$q$ & $0.24^{+0.13}_{-0.08}$ & $0.89^{+0.06}_{-0.07}$ & $0.93^{+0.04}_{-0.06}$ &  $0.27^{+0.11}_{-0.09}$ & $0.81^{+0.06}_{-0.06}$ &  $0.67^{+0.08}_{-0.06}$\\

$M_{\rm TOV}$ ($\rm M_{\odot}$) & $2.20^{+0.14}_{-0.11}$ & $2.19^{+0.13}_{-0.10}$ & $2.21^{+0.13}_{-0.10}$ &  $2.20^{+0.15}_{-0.11}$ & $2.19^{+0.13}_{-0.11}$ &  $2.20^{+0.14}_{-0.11}$\\
\bottomrule
\end{tabular}
\tablefoot{
Here, $\mathcal{U} D_{\rm L}$ (Mpc) indicates a uniform prior for luminosity distance, $M_{\rm Chirp}$ ($\rm M_{\odot}$) is the chirp mass, $q$, the binary mass ratio, $\tilde\Lambda$, mass-weighted tidal deformability, and $M_{\rm TOV}$ ($\rm M_{\odot}$) is the maximum neutron star mass. }
\label{tab:All_Binary_Results}
\end{table*} 

\begin{figure}
	\centering
    \includegraphics[width=\linewidth]{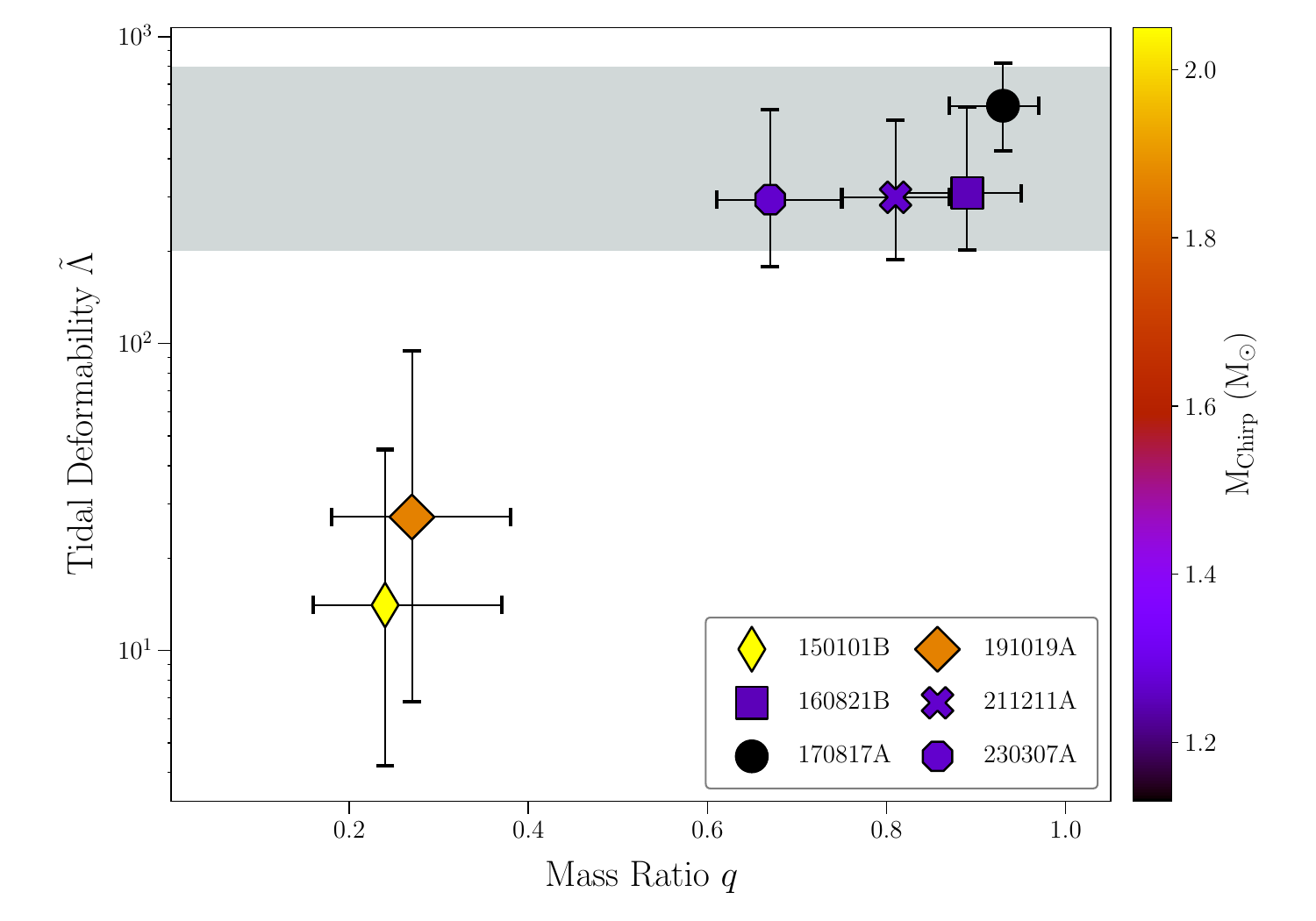}  
	\caption{Distribution of the mass-weighted tidal deformability, $\tilde\Lambda$, and the binary mass ratio, $q$. The shaded region $200\lesssim \tilde{\Lambda} \lesssim800$ highlights the upper limit for GW170817 observations \citep{Abbott2017PhRvL} and the lower limit is obtained from \citet{Altiparmak2022ApJ}.}
    \label{Fig:TidalMassRatio}  
\end{figure}

\begin{figure}
	\centering
	\includegraphics[width=\linewidth]{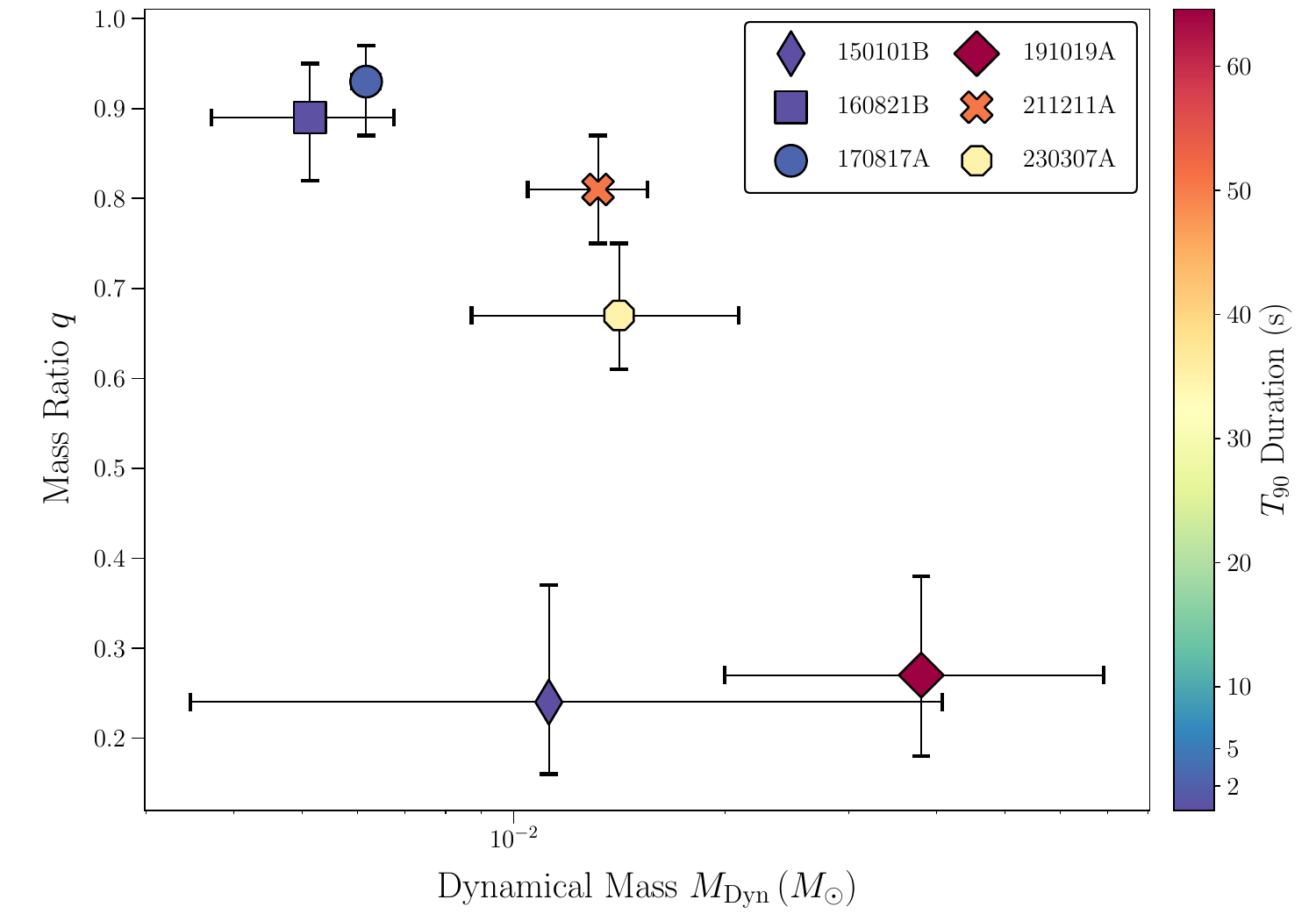}
	\caption{Distribution of the binary mass ratio $q$ and the dynamical ejected mass, $M_{\rm dyn}$ ($M_{\odot}$), for the short and long merger-driven GRBs.}
	\label{Fig:MassRatio_DynMass}
\end{figure}

\begin{figure}
	\centering
	\includegraphics[width=\linewidth]{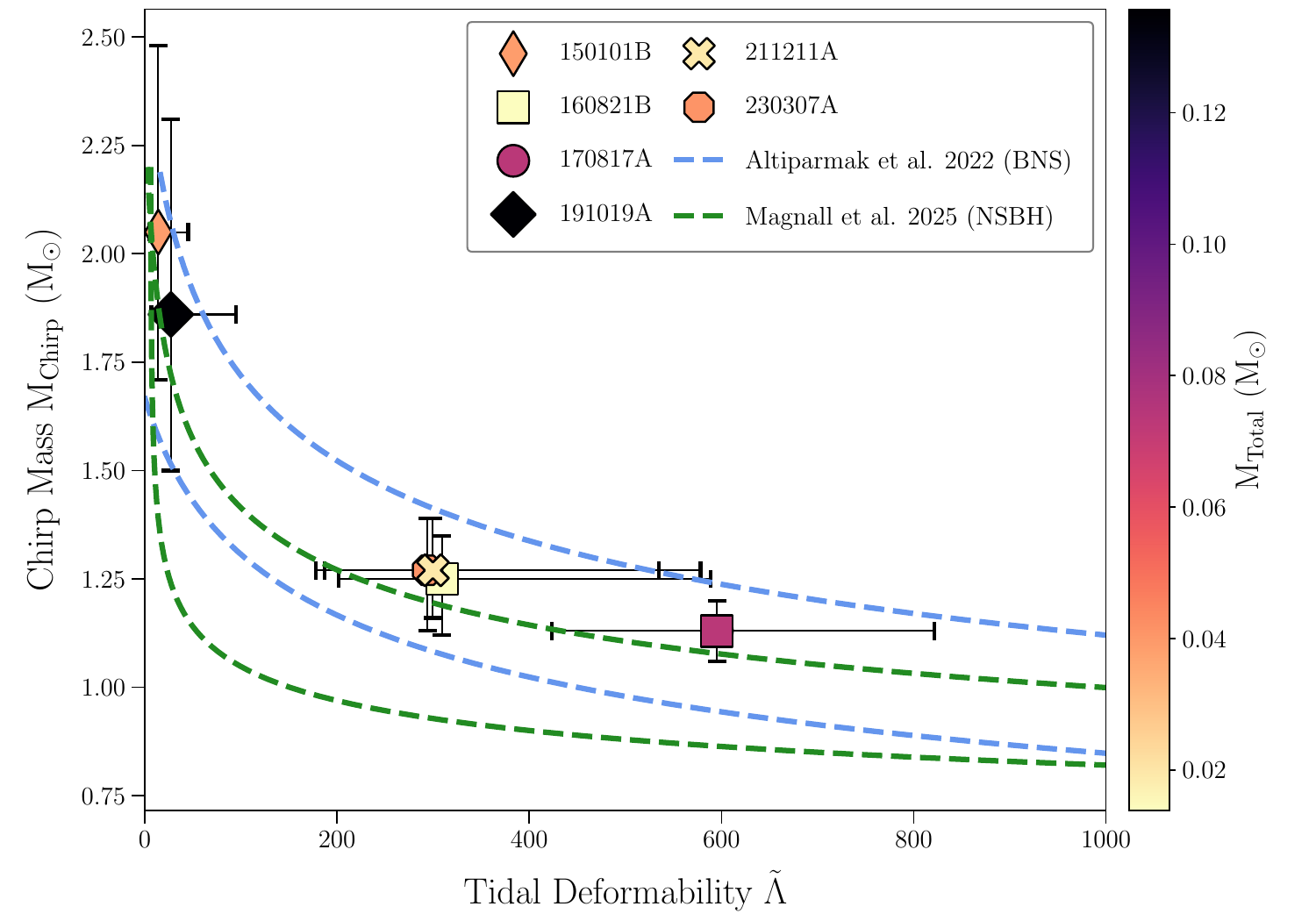}
	\caption{Chirp mass, $M_{\rm Chirp}$ ($M_{\odot}$), and mass-weighted tidal deformability, $\tilde\Lambda$, from our analysis plotted over the analytical limits for BNS mergers shown in blue \citep{Altiparmak2022ApJ} and for an NSBH merger, shown in green \citep{Magnall2025ApJ}. 
    }
	\label{Fig:TidalChirpRelationship}
\end{figure}
 
\subsection{Progenitor properties}\label{DiscussionProgenitor}
In \figref{Fig:ModelSelectionAndBayes}, we present all the models evaluated in this work and identify the preferred model as the one with the highest Bayes factor. We were able to conclusively establish the nature of the progenitor as BNS for GRB 160821B, GRB 170817A, GRB 211211A, and GRB 230307A. In the case of GRB 150101B and GRB 191019A, our analysis suggests a preference for an NSBH progenitor due to the highest Bayes factor, although a BNS origin is still viable for these two events. 

Comparing BNS and NSBH (see \figref{Fig:TidalMassRatio}), we observed that BNS mergers are characterized by large median values of tidal deformability $\langle \tilde\Lambda\rangle = 304\pm80$ and smaller $\langle\mathcal{M}_{\rm \,Chirp}\rangle=1.23 \pm0.06 M_{\odot}$, while NSBH mergers show significantly lower $\langle\tilde\Lambda\rangle = 20\pm4$ and higher $\langle\mathcal{M}_{\rm \,Chirp}\rangle=1.3 \pm2 M_{\odot}$ due the primary being a black hole. The binary mass ratio for BNS cases spans the range of $q\sim0.6-0.9$, while NSBH events have $q\sim0.2$ again indicative of a massive primary component.

In \figref{Fig:MassRatio_DynMass}, we show the binary mass ratio, $q$, and the dynamically ejected mass, where the data points are colored according to the burst duration of $T_{90}$ (sec). We observe that the mass ratio decreases as $M_{\rm dyn}$ increases. This implies that if the merging binaries have highly unequal masses, with the primary body being heavier than the secondary, a larger dynamical ejecta could be produced. This behavior could be attributed to the binary merger phase, during which the tidal forces sweep away a significant amount of mass from the outer layers of neutron stars, generating the dynamical ejecta. Indeed, mergers with unequal component masses (i.e., low $q$) have been shown to undergo a pronounced tidal disruption and produce large dynamical masses \citep{Shibata2003PhRvD,Rezzolla2010CQGra,Radice2016MNRAS,Bovard2017PhRvD,Barbieri2020EPJA,Papenfort2022MNRAS}. Such a dependence between $q$ with $M_{\rm Total}$ or $M_{\rm wind}$ has not clearly emerged from our analysis. In particular, from our analysis, we find that $M_{\rm Total}$ has a significant contribution from $M_{\rm wind}$ (see \secref{DynAndWindMassComparison}), where the latter is difficult to model in current simulations, which typically cover the dynamical timescales $\approx10-30$ milliseconds; whereas the wind ejecta is active at longer timescales and not fully understood at present \citep[see, e.g.,][]{Radice2018ApJ,Shibata2019,Foucart2023LRCA,Neuweiler2023,Collins2023MNRAS,Han2025PhRvD,Cook2025arXiv2508,Gutierrez2025arXiv2506}

Going on to focus on the BNS cases in our sample (\figref{Fig:MassRatio_DynMass}), we find that the two long-duration GRBs belonging to this progenitor class (211211A and 230307A) have a lower $q$ value compared to the short GRBs (160821B and 170817A). At the same time, the dynamical ejecta mass is larger for the long GRBs than for the short GRBs. As a result, the long GRBs and short GRBs originating from BNS mergers appear to occupy separate regions in the $q$--$M_{\rm dyn}$ plane, where short GRBs are characterized by $q$ and low $M_{\rm dyn}$ while long GRBs have low $q$ and high $M_{\rm dyn}$. This trend is observed when also considering NSBH cases, where a small $q$, large $M_{\rm dyn}$, and long burst duration has been found for GRB 191019A. The large uncertainties on the dynamical ejecta mass prevented us from drawing any strong conclusion for GRB\,150101B. In terms of the duration of the burst, a low $q$ would result in a  large ejected mass that could fallback onto the central remnant over longer timescales making the prompt duration longer, which is in agreement with the results reported by \citet{Musolino2024ApJ}. This is in line with our findings for the merger-driven long GRBs in the $q$-$M_{\rm dyn}$ plane, supporting a fallback origin of their long duration. 

Next, we compared our values of the chirp mass and tidal deformability, given in \figref{Fig:TidalChirpRelationship}, against the analytical relation derived from the nonmonotonic behavior of the speed of sound in neutron stars, obtained by \citet{Altiparmak2022ApJ} for BNS and by \citet{Magnall2025ApJ} for NSBH systems. \citet{Altiparmak2022ApJ} found that for BNS events $\tilde{\Lambda}$ and $M_{\rm Chirp}$, can be written as $\tilde{\Lambda}_{\rm min \, (max)} = a+b \,M_{\rm Chirp}^{\rm c}$, where $a = -50 (-20)$, $b = 500\, (1800)$, and $c = -4.5\,(-5.0)$; corresponding to "min" and "max" values. Meanwhile, in the case of NSBH, \citet{Magnall2025ApJ} provided an updated formulation for the same where $a = 6 (-5)$, $b = 296\, (1700)$, and $c = -6.6\,(-4.9)$. Firstly, our results on $\tilde{\Lambda}$ and $M_{\rm Chirp}$ are compatible with both analytical relations. Secondly, the expected trend is that the chirp mass will decrease as the tidal deformability increases. Since, in our analysis, such a dependence is not prescribed during inference, which means that we considered $\tilde{\Lambda}$ and $M_{\rm Chirp}$ independently, finding a similar trend strengthens the result from \citet{Altiparmak2022ApJ,Magnall2025ApJ}. Lastly, given that these two quantities can be measured with GW detectors, our analysis, which is purely based on EM observations and the one by \citet{Altiparmak2022ApJ,Magnall2025ApJ}, motivates further efforts in constraining the EOS with future multimessenger detections.

\subsection{Efficacy of simultaneous inference}\label{DiscussionInferenceEfficacy}
An analysis of eight GRBs (050709, 060614, 130603B, 160821B, 170817A, 200522A, 211211A, and 230307A) was recently carried out by \citet{Rastinejad2025}. However, their data analysis is different from that presented in this work. Indeed, rather than performing a simultaneous joint afterglow and KN inference, in their methodology, first the afterglow is fitted on the X-ray and radio data, while the optical and NIR observations are masked, and the afterglow model is extrapolated to the optical and NIR bands. Then the KN component is obtained by subtracting the afterglow model from the opt/NIR observations. Another difference with respect to our analysis is that in \citet{Rastinejad2025}, the viewing angle for AT2017gfo $\iota=22^{\circ}$ (and for all other GRBs, they assumed $\iota=0^{\circ}$) were fixed. In contrast, the viewing angle in our analysis (which is the same for both the afterglow and KN) was considered as a free parameter. Furthermore, it was recently illustrated that an independent inference of afterglow and KN, as the one followed by \citet{Rastinejad2025}, could lead to incorrect estimates of physical properties \citep{Wallace2025MNRAS}. They demonstrated that simultaneous inference, which is the method followed in this work, is more reliable.

In comparing our work with the results of \citet{Rastinejad2025}, we note that the total ejected mass of AT2017gfo is consistent. In this case, the KN emission could be easily disentangled from the afterglow and, thus, the systematic biases resulting from the different methods (i.e., simultaneous inference vs. independent inference) are minimal, although we note that the KN models are different. However, for GRB\,160821B and GRB\,230307A, we recovered slightly smaller values of the total ejected mass, while for GRB\,211211A, our value for $M_{\rm Total}$ is significantly lower (see \figref{Fig:EjectaMassesMtotal} and \secref{DiscussionEjectaMass}). 

\section{Conclusions}\label{ConclusionsSummary}
We performed a uniform and systematic Bayesian analysis to constrain both the afterglow and KN properties. Our methodology benefits from the simultaneous analysis of nonthermal (afterglow) and thermal (KN) emission, along with identical model assumptions for each GRB. Our analysis demonstrates that EM observations of merger-driven GRBs can be used to infer the progenitor properties, jet structure, and ejecta parameters. In the following, we summarize our main conclusions.

\begin{itemize}[itemsep=0.35em]
    \item We robustly identified the KN in all cases, except for GRB\,150101B. For this event, we cannot rule out that it might be dominated by afterglow-only emission. We cannot conclusively confirm the presence of a KN. 

    \item Taking into account all KNe in our sample, we found that the dynamically ejected mass is lower than the wind mass. This result is consistent with radiative transfer simulations that adopt a similar approach to ours, as well as with numerical-relativity simulations of compact binary mergers.

    \item Our analysis of the ejected wind mass and the collimation-corrected kinetic isotropic-equivalent energy of the jet shows that these two are dependent and scale together. This correlation, while statistically significant, has been identified for the first time in our work and future studies could unveil additional connections.
    
    \item We found that although most GRBs in our sample are consistent with BNS progenitors, GRB\,150101B and GRB\,191019A are compatible with a BNS scenario, while still exhibiting evidence that would suggest an NSBH origin.
    
    \item We present, for the first time, the binary properties of a sizable sample of GRB events with KNe. We have been able to consistently reproduce the binary properties of GW170817 and, thanks to the inclusion of other GRBs, we have populated the binary parameter space. 
    
    \item We found a potential trend among dynamical ejecta mass and binary mass ratio, as expected from numerical simulations where low mass ratios could produce strong tidal effects and result in a large dynamically ejected mass. 
    
    \item Our result regarding chirp mass and tidal deformability positively adheres to the analytical relation developed by \citet{Altiparmak2022ApJ} and \citet{Magnall2025ApJ}. Firstly, it strengthens their relationship obtained from analysis of the speed of sound and the EOS. Secondly, since these two quantities are measurable by GW observations, our analysis, based purely on EM observations, motivates the potential synergy in constraining EOS with future multimessenger detections.
\end{itemize}

\section{Future perspectives }
In the future, a similar analysis with a larger sample would be promising to constrain the observational and binary properties, providing further connections with neutron star physics and improving physical constraints. The modeling of EM counterparts can be improved by incorporating time- and frequency-dependent systematic uncertainties in Bayesian inference \citep{Hussenot-Desenonge2024MNRAS,Peng2024PhRvR,Jhawar2025PhRvD}. Updated KN and radiative transfer models \citep{Bulla2023MNRAS}, which account for a better treatment of microphysics and local evolution, can further improve the estimated ejecta parameters \citep[see also][]{Anand2023arXiv2307,Koehn2025arXiv}. The inference of binary properties can be further refined by leveraging the quasi-universal relations \citep{Koppel2019ApJ,Tootle2021ApJ}; prescribing the dependence of $\tilde{\Lambda}$ and $M_{\rm Chirp}$ \citep{Altiparmak2022ApJ,Magnall2025ApJ} directly in the inference framework instead of considering them independent. 

The advances in numerical and theoretical studies are of paramount significance in expanding and exploring the merger dynamics, the impact of the EOS, and the evolution of KN and ejecta physics, where observations of new KN and GW detections are essential. A new coincident detection of a GW merger and an EM counterpart would irrefutably provide an exceptional opportunity and wealth of information to further enhance our understanding. In the absence of GW detections, future observations with Vera Rubin Observatory and from GRBs detected by space-based observatories are essential to construct a larger sample and extend such analyses. The comparison of KN properties between events with and without a detected GRB will shed light on the effects of relativistic jets on the properties of ejected matter. 

\section*{Data availability}\label{DataAvailability}
The full \tabref{tab:ObservationsDataTable} with the observational data used in this
work is available at the CDS via \url{https://cdsarc.cds.unistra.fr/viz-bin/cat/J/A+A/710/A350}. The complete posterior distributions for \figref{Fig:AllKNPosterior_CornerSmall1} to \figref{Fig:AllGWPosterior_CornerSmall3} are available online at \href{https://doi.org/10.5281/zenodo.19708188}{https://doi.org/10.5281/zenodo.19708188}.

\begin{acknowledgements}
We thank the anonymous referee for helpful comments and suggestions. The authors acknowledge the work of our colleague and friend Dr. D.A. Kann, who prematurely passed away at the beginning of this project. The authors thank Prof. Dr. Luciano Rezzolla for his valuable support and insightful suggestions during the course of this research. The authors thank Dr. Christian Ecker and Dr. Konrad Topolski for their helpful discussions. P.S acknowledges the support of the State of Hesse within the Research Cluster ELEMENTS (Project ID 500/10.006) and the European Union’s Horizon 2020 Programme under the AHEAD2020 project (grant agreement n. 871158). We gratefully acknowledge financial support from INAF Mini Grants RSN4 (ID: 1.05.24.07.04). This research has used the SVO\footnote{\url{https://svo.cab.inta-csic.es}} Filter Profile Service "Carlos Rodrigo" \citep{Rodrigo2012ivoa,Rodrigo2020sea,Rodrigo2024A&A}, funded by MCIN/AEI/10.13039/501100011033/ through grant PID2023-146210NB-I00. M. B. acknowledges the Department of Physics and Earth Science of the University of Ferrara for the financial support through the FIRD 2024 and FIRD 2025 grants.
\end{acknowledgements}

\bibliographystyle{aa} 
\bibliography{aa58774-25}

\begin{appendix} 
\onecolumn
\section{Best-fitting light curves.}

\begin{figure}[h]
	\centering
	\includegraphics[width=0.46\linewidth]{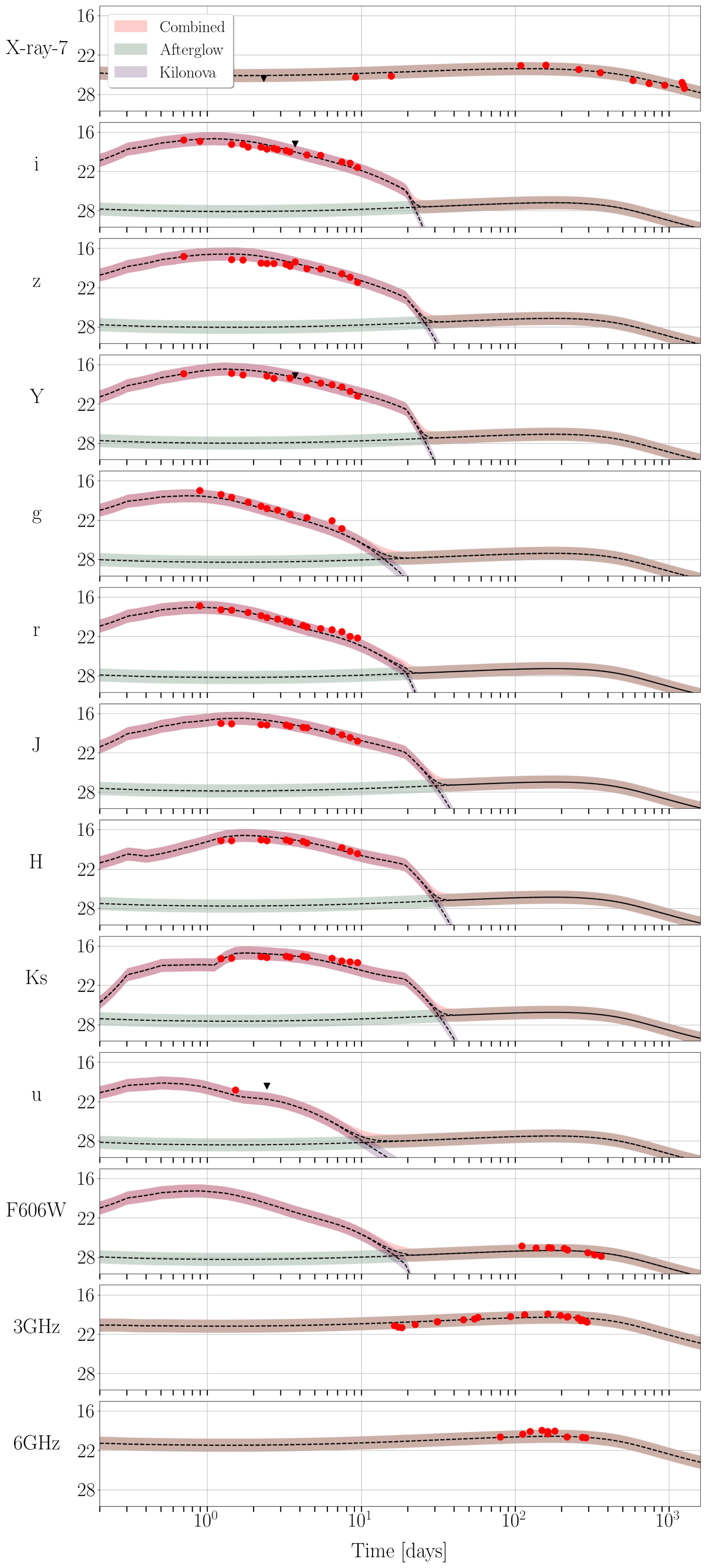}
	\includegraphics[width=0.46\linewidth]{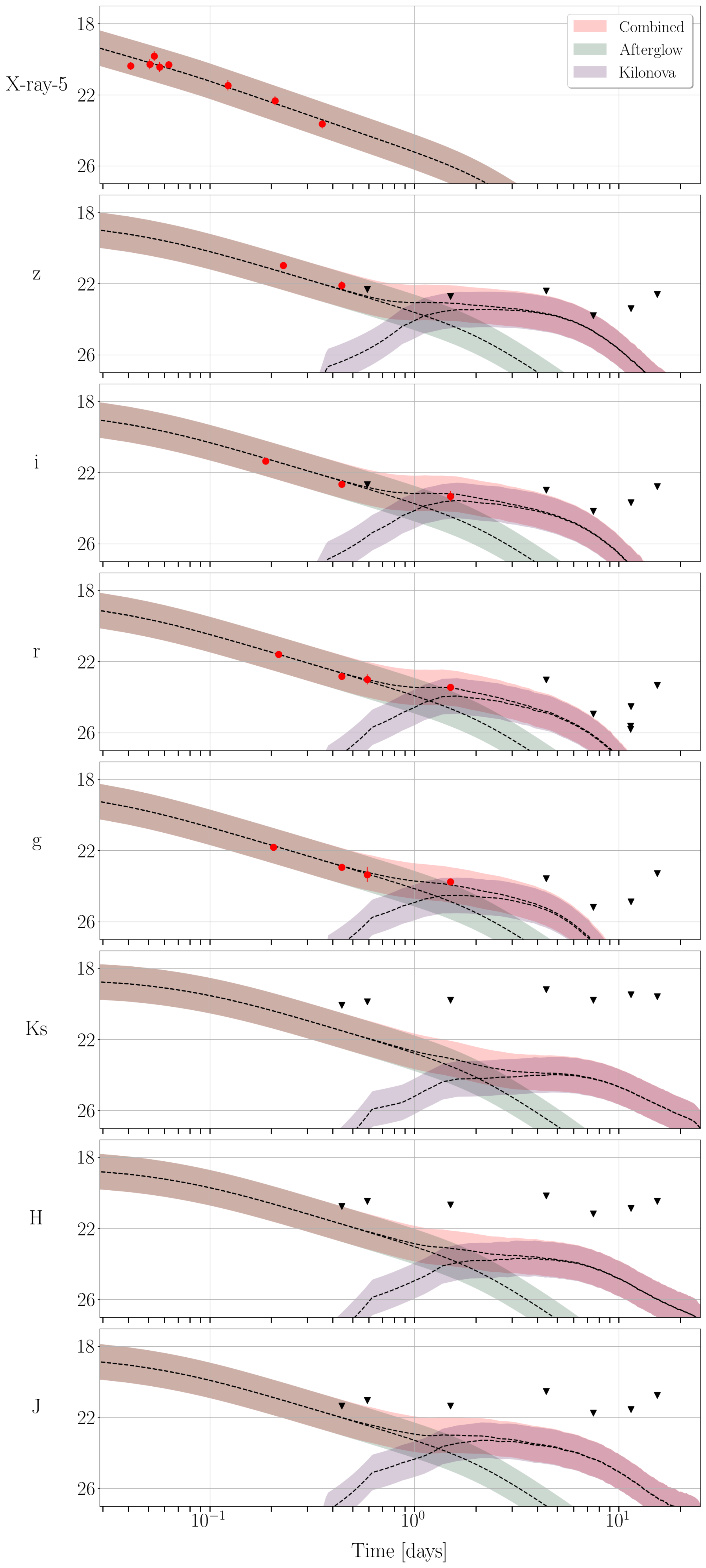}
	\caption{Same as \figref{Fig:150101BAnd160821BLightcurves}, the light curves for GRB\,170817A \& AT2017gfo (left) and GRB\,191019A (right).
    }
	\label{Fig:170817And191019ALightcurves}
\end{figure}

\begin{figure}[h]
	\centering
	\includegraphics[width=0.46\linewidth]{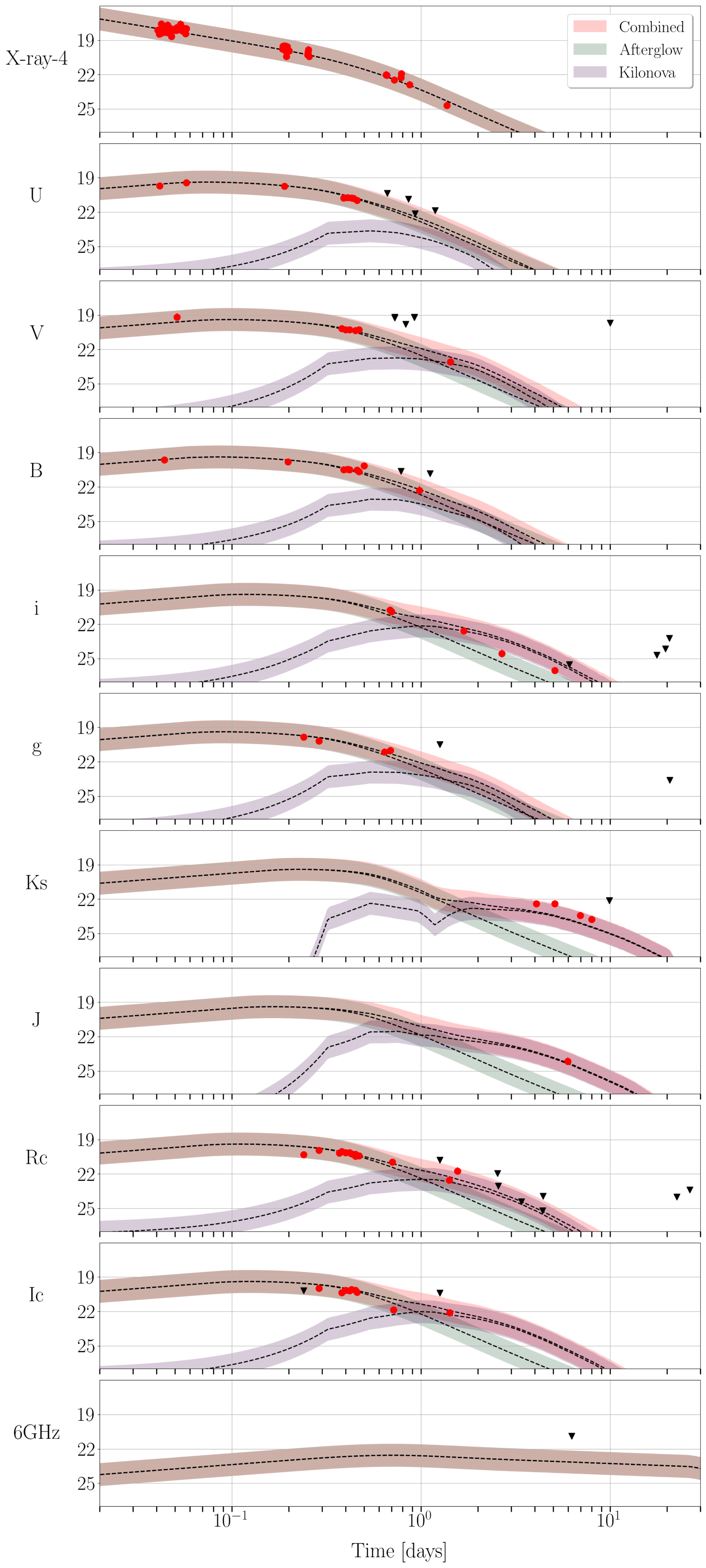}
	\includegraphics[width=0.46\linewidth]{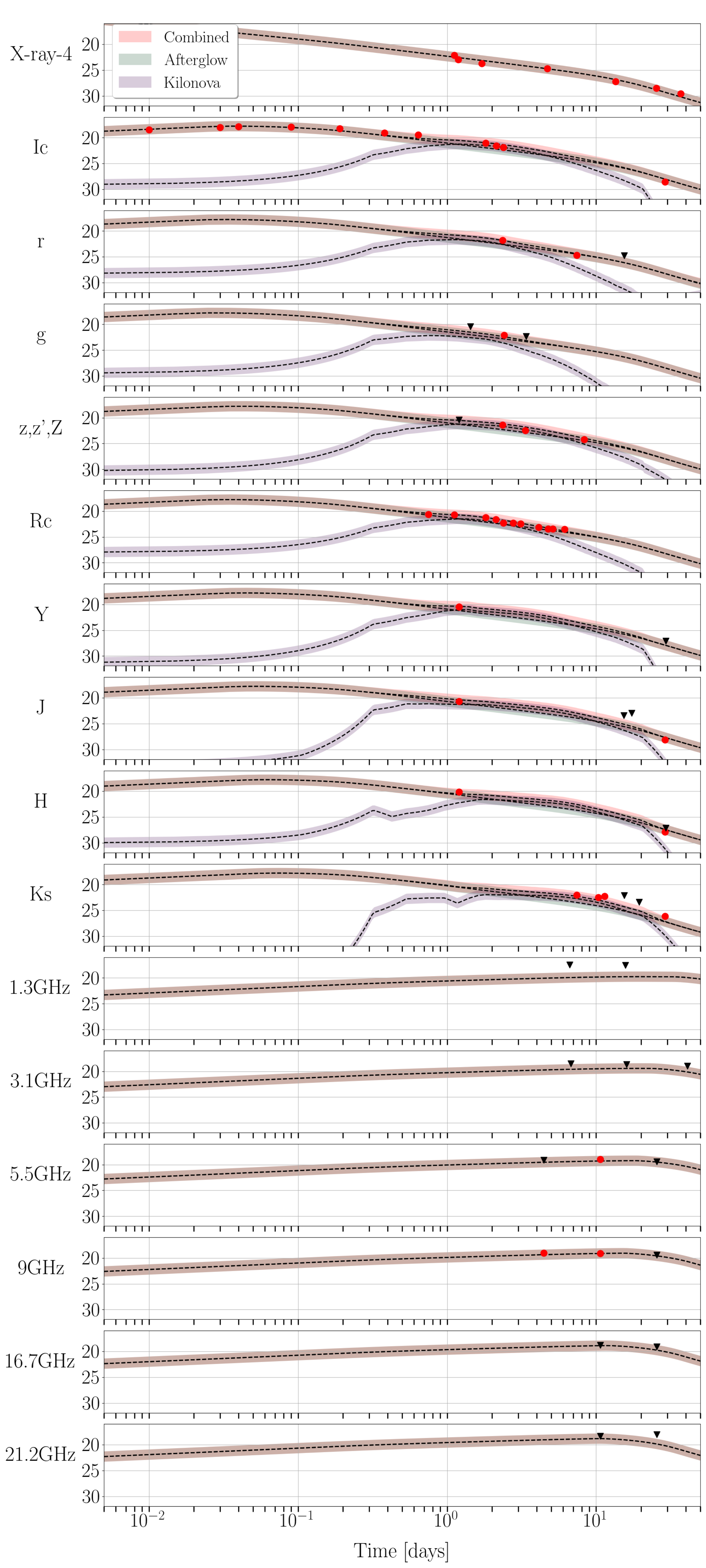}
	\caption{Same as \figref{Fig:150101BAnd160821BLightcurves}, the light curves for GRB\,211211A (left) and GRB\,230307A (right). 
    }
	\label{Fig:211211AAnd230307ALightcurves}
\end{figure}

\twocolumn
\onecolumn
\section{Observational data. }

\begin{longtable}{llcccrr}
\caption{Observations of all GRBs in the sample.}\\
\toprule
GRB & Days & Band/Filter & Mag (AB) & Err (AB)  & Telescope/Instrument & Reference \\
\toprule
150101B & 1.50 & X-ray-1keV & 28.46 & inf & Swift XRT & Ref1 \\
150101B &  1.66 & r & 23.01 & 0.17 & Magellan/Baade IMACS & Ref2 \\
150101B &  2.65 & r & 23.53 & 0.26 & Magellan/Baade IMACS & Ref2 \\
150101B & 2.67 & H & 21.40 & inf & VLT HAWK-I & Ref2 \\
150101B & 2.68 & K & 21.50 & inf & VLT HAWK-I Ref2 \\
\midrule				
...	& 			&	& &	 & 	&		\\
\midrule
150101B & 43.63 & H & 23.90 & inf & VLT HAWK-I & Ref2 \\
150101B & 48.39 & r & 24.60 & inf&  Magellan/Baade IMACS & Ref2 \\
150101B & 48.41 & r & 16.51 & inf & Magellan/Baade IMACS & Ref2 \\
150101B &  48.41 & g & 17.42 & inf & Magellan/Baade IMACS & Ref2 \\
150101B &  48.42 & i & 16.08 & inf & Magellan/Baade IMACS & Ref2 \\
150101B &  48.42 & z & 15.77 & inf & Magellan/Baade IMACS & Ref2 \\
\bottomrule
\label{tab:ObservationsDataTable}
\end{longtable}
\tablefoot{All data are in AB magnitudes, and corrected for Galactic extinction and intrinsic extinction from the host. The ``inf" value specifies a nondetection (upper limit). The reference keys are Ref1 \citep{Troja2018GRB150101B} and Ref2 \citep{Fong2016ApJ}. The full data table is available as online material via CDS (see \secref{DataAvailability}).}\\

\section{Priors for the best-fit models. }

\begin{table*}[h!]
\centering
\caption{Priors used for simultaneous inference for models summarized in \tabref{tab:All_KN_Results}.}
\renewcommand{\arraystretch}{1.9}
\begin{tabular}{@{}>{\raggedright\arraybackslash}m{2.4cm}*{6}{>{\centering\arraybackslash}m{2.2cm}}@{}}
\toprule
& 150101B  & 160821B & 170817A & 191019A & 211211A &  230307A   \\ 
& NSBH-TH & BNS-TH & BNS-GS & NSBH-TH & BNS-TH & BNS-TH   \\  
\midrule
$\log(M_{\rm dyn})$  $M_{\odot}$&  $\mathcal{U}$ (-3, -1) & $\mathcal{U}$ (-3, -1)  & $\mathcal{U}$ (-3, -1) & $\mathcal{U}$ (-3, -1) & $\mathcal{U}$ (-3, -1) & $\mathcal{U}$ (-3, -1) \\
$\log(M_{\rm wind})$  $M_{\odot}$ & $\mathcal{U}$ (-3, -0.5)  &  $\mathcal{U}$ (-3, -0.5) & $\mathcal{U}$ (-3, -0.5) & $\mathcal{U}$ (-3, -0.5) & $\mathcal{U}$ (-3, -0.5) & $\mathcal{U}$ (-3, -0.5) \\ 
$\Phi$ (deg) & 30 & $\mathcal{U}$ (15, 75) & $\mathcal{U}$ (15, 75) & 30 & $\mathcal{U}$ (15, 75) & $\mathcal{U}$ (15, 75)\\
\midrule
$\iota$ (rad)        & Sine (0.0, $\pi$/4)  & 	Sine (0.0, $\pi$/4) & Sine (0.20, 0.60) & Sine (0.0, $\pi$/8) & Sine (0.0, $\pi$/4)    & Sine (0.0, $\pi$/4) \\ 		
$\log(E_0)$  (erg)  &  $\mathcal{U}$ (48, 54)  & $\mathcal{U}$ (49, 53) & $\mathcal{U}$ (49, 54) & $\mathcal{U}$ (49, 53) & $\mathcal{U}$ (49, 54)   & $\mathcal{U}$ (47, 55)  \\ 
$\log(n_0)$  (cm$^{-3}$)  &  $\mathcal{U}$ (-4, 2)   & $\mathcal{U}$ (-8, 4) & $\mathcal{U}$ (-4, 2) & $\mathcal{U}$ (-3, 7) & $\mathcal{U}$ (-6, 2)  & $\mathcal{U}$ (-7, 4)\\
$\theta_{\rm c}$  (rad) &  $\mathcal{U}$ (0.01, $\pi$/10)  &  $\mathcal{U}$ (0.01, $\pi$/10) & $\mathcal{U}$ (0.02, 0.15) & $\mathcal{U}$ (0.01, $\pi$/10) & $\mathcal{U}$ (0.01, $\pi$/10) & $\mathcal{U}$ (0.01, $\pi$/10) \\ 
$\theta_{\rm w}$ (rad)  &   -   & - & $\mathcal{U}$ (0.6, 0.99) & - &   - & - \\ 
$p$              &  $\mathcal{U}$ (2.01, 3.0)  & $\mathcal{U}$ (2.01, 3.0) & $\mathcal{U}$ (2.01, 3.0) & $\mathcal{U}$ (2.01, 3.0) & $\mathcal{U}$ (2.01, 3.0) & $\mathcal{U}$ (2.01, 3.0) \\ 
$\log(\epsilon_e)$ &   $\mathcal{U}$ (-3, 0)  & $\mathcal{U}$ (-5, 0) & $\mathcal{U}$ (-5, 0) & $\mathcal{U}$ (-5, 0) & $\mathcal{U}$ (-5, 0) & $\mathcal{U}$ (-5, 0)\\ 
$\log(\epsilon_B)$ &  $\mathcal{U}$ (-5, 0)   & $\mathcal{U}$ (-10, 0) & $\mathcal{U}$ (-10, 0) & $\mathcal{U}$ (-10, 0) & $\mathcal{U}$ (-6, 0) & $\mathcal{U}$ (-10, 0)\\  
\bottomrule
\end{tabular}
\tablefoot{
{The angle of inclination $\mathcal{U}\,\iota$ is uniform on a sphere. In the case of the TH jet structure, $\theta_{\rm w}$ is not a model parameter and $\Phi$ is fixed = 30 (deg) for all NSBH models (see \secref{ModelingEM}).}
}
\label{tab:All_KN_Priors}
\end{table*}

\section{Best-fit parameters for all the evaluated models for each GRB. }

\begin{table*}[t]
    \centering
    \caption{Results of all evaluated models for GRB\,150101B. }
    \renewcommand{\arraystretch}{1.9}
    \begin{tabular}{@{}>{\raggedright\arraybackslash}m{2.4cm}*{7}{>{\centering\arraybackslash}m{2.2cm}}@{}}
    \toprule
     & GS & TH & BNS-GS & BNS-TH &   NSBH-GS & NSBH-TH \\  
    \midrule
    $\log(M_{\rm dyn})$ ($M_{\odot}$) &  - & - & $-2.35^{+0.42}_{-0.35}$ & $-2.29^{+0.39}_{-0.36}$ &  $-2.03^{+0.62}_{-0.45}$ & $-1.95^{+0.56}_{-0.51}$\\ 
    $\log(M_{\rm wind})$ ($M_{\odot}$) & -  & -  & $-1.85^{+0.39}_{-0.45}$ & $-1.73^{+0.34}_{-0.39}$ & $-1.71^{+0.39}_{-0.41}$ & $-1.56^{+0.36}_{-0.38}$\\ 
    $\Phi$  (deg) &  - & - & $48.23^{+15.04}_{-17.08}$ & $41.21^{+18.32}_{-16.12}$ &  30 & 30 \\
    \midrule
    $\iota$  (rad)      & $0.11^{+0.12}_{-0.07}$  & $0.11^{+0.08}_{-0.06}$ & $0.17^{+0.10}_{-0.09}$ & $0.20^{+0.11}_{-0.11}$        & $0.14^{+0.11}_{-0.08}$  & $0.20^{+0.11}_{-0.13}$ \\ 
    $\log(E_0)$ (erg)   &  $52.86^{+0.66}_{-0.86}$ & $52.68^{+0.73}_{-0.74}$ & $52.86^{+0.66}_{-0.85}$ & $52.72^{+0.70}_{-0.75}$   & $52.69^{+0.69}_{-0.83}$ & $52.36^{+0.84}_{-0.71}$ \\ 
    $\log(n_0)$ (cm$^{-3}$)   &  $-2.83^{+1.06}_{-0.67}$   & $-2.99^{+0.91}_{-0.58}$ & $-2.79^{+1.02}_{-0.70}$ & $-2.68^{+1.02}_{-0.70}$ &    $-2.85^{+1.00}_{-0.65}$  & $-2.83^{+1.13}_{-0.67}$\\  
    $\theta_{\rm c}$ (rad) &  $0.22^{+0.05}_{-0.06}$  &  $0.22^{+0.05}_{-0.06}$ & $0.22^{+0.05}_{-0.06}$ & $0.21^{+0.05}_{-0.07}$ &   $0.23^{+0.05}_{-0.06}$ & $0.22^{+0.05}_{-0.07}$ \\ 
    $\theta_{\rm w}$ (rad) &  $0.46^{+0.18}_{-0.18}$   & - & $0.47^{+0.16}_{-0.17}$ & - &   $0.50^{+0.16}_{-0.20}$ &-\\ 
    $p$              &  $2.24^{+0.10}_{-0.07}$  & $2.21^{+0.09}_{-0.06}$ & $2.20^{+0.08}_{-0.06}$ & $2.22^{+0.09}_{-0.07}$ &  $2.18^{+0.08}_{-0.05}$ & $2.19^{+0.08}_{-0.05}$\\ 
    $\log(\epsilon_e)$ &   $-1.27^{+0.68}_{-0.78}$  & $-1.51^{+0.80}_{-0.59}$ & $-1.38^{+0.73}_{-0.75}$ & $-1.17^{+0.61}_{-0.63}$ &  $-1.36^{+0.74}_{-0.74}$ & $-0.93^{+0.54}_{-1.02}$\\    
    $\log(\epsilon_B)$ &  $-3.52^{+1.11}_{-0.82}$   & $-3.30^{+0.92}_{-0.89}$ & $-3.40^{+0.94}_{-0.84}$ & $-3.55^{+0.83}_{-0.78}$ &  $-3.62^{+0.97}_{-0.74}$ & $-3.28^{+1.12}_{-0.99}$\\ 				 
    \midrule
    Evidence ($\mathcal{Z}$)  & $-11.58\pm0.03$  & $-11.64\pm 0.03$  & $-11.53\pm 0.04$  & $-11.45\pm0.04$ &  $-11.67\pm0.04$ & $-11.02\pm0.04$\\ 
    $\ln (\mathcal{B}^{\rm Test}_{\rm{Ref}})$  & -0.57  & -0.62 & -0.51 & -0.44 & -0.65 & 0.0    \\ 
    \toprule
    \end{tabular}
	\label{tab:150101B_AllModelsResults}
\end{table*}

\begin{table*}[t]
\centering
\caption{Results of all evaluated models for GRB\,160821B.}
\renewcommand{\arraystretch}{1.9}
    \begin{tabular}{@{}>{\raggedright\arraybackslash}m{2.4cm}*{7}{>{\centering\arraybackslash}m{2.2cm}}@{}}
	\toprule
	    &  GS  &   TH & BNS-GS &   BNS-TH & NSBH-GS &   NSBH-TH \\  
    \midrule
	$\log(M_{\rm dyn})$  ($M_{\odot}$)&-&- & $-2.38^{+0.26}_{-0.26}$ &   $-2.29^{+0.12}_{-0.14}$ &   $-1.74^{+0.15}_{-0.13}$ & $-1.92^{+0.11}_{-0.07}$  \\ 
	$\log(M_{\rm wind})$ ($M_{\odot}$) &-&- & $-1.77^{+0.11}_{-0.11}$    & $-2.06^{+0.11}_{-0.13}$ &   $-2.62^{+0.17}_{-0.16}$ & $-2.38^{+0.14}_{-0.17}$\\ 
	$\Phi$ (deg)&-&-&  $62.95^{+7.63}_{-8.19}$  & $70.71^{+6.94}_{-7.57}$ & 30 & 30 \\ 
	\midrule
	$\iota$  (rad)   & $0.26^{+0.06}_{-0.05}$ &    $0.14^{+0.08}_{-0.04}$      & $0.09^{+0.05}_{-0.04}$ &  $0.30^{+0.05}_{-0.07}$  &  $0.45^{+0.05}_{-0.06}$ &  $0.34^{+0.07}_{-0.07}$   \\ 
    $\log(E_0)$ (erg) & $51.10^{+0.23}_{-0.36}$ &   $51.04^{+0.16}_{-0.21}$   & $51.76^{+0.34}_{-0.51}$ &  $50.52^{+0.50}_{-0.25}$ &  $50.66^{+0.25}_{-0.16}$ & $50.93^{+0.45}_{-0.35}$   \\ 
	$\log(n_0)$ (cm$^{-3}$) & $-1.28^{+0.60}_{-0.79}$   &  $-3.39^{+1.10}_{-0.60}$   &   $-6.41^{+0.77}_{-0.84}$  & $-1.67^{+0.79}_{-1.00}$ &  $-0.71^{+0.82}_{-0.65}$ &  $-1.09^{+0.93}_{-1.04}$ \\  
	
	$\theta_{\rm c}$ (rad)  & $0.19^{+0.05}_{-0.04}$   &  $0.10^{+0.07}_{-0.03}$&  $0.20^{+0.05}_{-0.07}$   & $0.22^{+0.05}_{-0.09}$  &  $0.27^{+0.03}_{-0.04}$ &  $0.24^{+0.04}_{-0.04}$  \\ 
	$\theta_{\rm w}$ (rad)  & $0.34^{+0.15}_{-0.17}$   & -& $0.33^{+0.25}_{-0.18}$  & - &  $0.36^{+0.05}_{-0.06}$ & - \\ 
	$p$     & $2.67^{+0.06}_{-0.09}$ &    $2.66^{+0.06}_{-0.06}$ &   $2.16^{+0.04}_{-0.05}$  & $2.34^{+0.15}_{-0.15}$ &  $2.43^{+0.05}_{-0.06}$ & $2.20^{+0.06}_{-0.05}$ \\ 
	
	$\log(\epsilon_e)$ & $-0.13^{+0.07}_{-0.19}$ &  $-0.43^{+0.17}_{-0.13}$&   $-1.00^{+0.27}_{-0.20}$  & $-0.33^{+0.15}_{-0.18}$ &   $-0.17^{+0.10}_{-0.12}$ & $-0.65^{+0.24}_{-0.21}$  \\  
	$\log(\epsilon_B)$ & $-3.35^{+1.31}_{-0.39}$   & $-1.58^{+0.43}_{-0.87}$&   $-1.38^{+0.66}_{-1.12}$  & $-2.23^{+0.51}_{-0.65}$ &  $-3.11^{+0.55}_{-0.64}$ & $-2.75^{+0.68}_{-0.90}$ \\  
	\midrule
	Evidence ($\mathcal{Z}$)  & -30.06$\pm$0.53 &   -27.32$\pm$0.55 & -30.92$\pm$0.48  & -15.70$\pm0.03$ & -42.51$\pm0.60$ & -39.54$\pm0.53$   \\ 
	$\ln (\mathcal{B}^{\rm Test}_{\rm{Ref}})$ &  -26.12  & -11.62  & -3.05  & 0.0    &  -27.19 &  -23.84  \\ 
	\toprule
\end{tabular}
\label{tab:160821B_AllModelsResults}
\end{table*}

\begin{table*}[t]	
	\centering
    \caption{Results of all evaluated models for GRB\,170817A and AT207gfo.  }
	\renewcommand{\arraystretch}{1.9}
    \begin{tabular}{@{}>{\raggedright\arraybackslash}m{2.4cm}*{7}{>{\centering\arraybackslash}m{2.2cm}}@{}}
    \toprule
	    & GS & TH & BNS-GS & BNS-TH & NSBH-GS & NSBH-TH   \\
    \midrule
	$\log(M_{\rm dyn})$ ($M_\odot$)  &  - & - &   $-2.21^{+0.02}_{-0.02}$  & $-2.41^{+0.02}_{-0.02}$ & $-2.44^{+0.02}_{-0.01}$ &  $-2.52^{+0.02}_{-0.02}$  \\
	$\log(M_{\rm wind})$  ($M_\odot$) &  -  & - & $-1.17^{+0.01}_{-0.01}$ &  $-1.36^{+0.02}_{-0.02}$   &  $-1.41^{+0.02}_{-0.01}$  & $-1.27^{+0.02}_{-0.02}$ \\
	$\Phi$ (deg) & -  & -  & $70.41^{+1.05}_{-1.46}$ & $18.59^{+2.13}_{-1.73}$   & 30  & 30 \\
	\midrule
    $\iota$  (rad) &   $0.31^{+0.04}_{-0.04}$ &  $0.21^{+0.04}_{-0.03}$ & $0.57^{+0.01}_{-0.01}$ & $0.61^{+0.03}_{-0.03}$   & $0.59^{+0.02}_{-0.02}$   & $0.46^{+0.02}_{-0.02}$  \\
	$\log(E_0)$ (erg) &  $51.45^{+0.31}_{-0.21}$ & $51.21^{+0.52}_{-0.31}$ & $52.23^{+0.10}_{-0.09}$ & $53.18^{+0.17}_{-0.20}$   & $52.65^{+0.09}_{-0.09}$  &  $52.14^{+0.17}_{-0.11}$ \\
	$\log(n_0)$ (cm$^{-3}$) & $ 1.32^{+0.35}_{-0.29}$ & $1.24^{+0.14}_{-0.18}$ & $-2.96^{+0.10}_{-0.08}$ & $ -0.39^{+0.27}_{-0.28}$    &  $ -2.29^{+0.08}_{-0.11}$    & $-3.11^{+0.19}_{-0.16}$ \\
	$\theta_{\rm c}$  (rad) &  $0.139^{+0.005}_{-0.004}$ & $0.27^{+0.02}_{-0.05}$ & $0.138^{+0.002}_{-0.002}$  & $0.29^{+0.01}_{-0.01}$ &   $0.135^{+0.002}_{-0.002}$   & $0.27^{+0.02}_{-0.01}$  \\
    $\theta_{\rm w}$  (rad) &  $0.65^{+0.03}_{-0.03}$ & - &  $0.64^{+0.01}_{-0.01}$ & -  & $0.636^{+0.004}_{-0.004}$   & - \\
	$p$   &  $2.02^{+0.01}_{-0.01}$ & $2.03^{+0.01}_{-0.01}$  & $2.14^{+0.01}_{-0.01}$  &  $2.09^{+0.01}_{-0.01}$   & $2.14^{+0.01}_{-0.01}$  & $2.12^{+0.01}_{-0.02}$ \\
	$\log(\epsilon_{\rm e})$  &  $-1.43^{+0.27}_{-0.26}$ & $-1.74^{+0.28}_{-0.35}$ & $-1.85^{+0.10}_{-0.10}$ & $-2.21^{+0.29}_{-0.22}$  & $-1.85^{+0.12}_{-0.15}$ &  $-2.19^{+0.14}_{-0.19}$ \\
	$\log(\epsilon_{\rm B})$   & $-2.26^{+0.24}_{-0.37}$  & $-2.28^{+0.20}_{-0.22}$ & $-2.25^{+0.08}_{-0.12}$ & $-5.75^{+0.24}_{-0.19}$     & $-3.32^{+0.26}_{-0.21}$  &  $-2.45^{+0.16}_{-0.32}$ \\
	\midrule
    Evidence ($\mathcal{Z}$) & -286.11  & -287.43 & -119.41 & -136.07  & -161.72    & -152.50 \\
	$\ln (\mathcal{B}^{\rm Test}_{\rm{Ref}})$   & -166.70   &  -168.02 & 0.0  & -16.66  & -42.31    & -33.09 \\
	\toprule
	\end{tabular}

\label{tab:170817_AllModelsResults}
\end{table*}

\begin{table*}[t]	
	\centering
    \caption{Results of GRB\,191019A, adapted from \citet{Stratta2025ApJ} with new results by considering a TH jet structure. }
	\renewcommand{\arraystretch}{1.9}
    \begin{tabular}{@{}>{\raggedright\arraybackslash}m{2.4cm}*{7}{>{\centering\arraybackslash}m{2.2cm}}@{}}
	\toprule
	    & GS & TH & BNS-GS  & BNS-TH & NSBH-GS & NSBH-TH   \\
    \midrule
	$\log(M_{\rm dyn})$ ($M_\odot$)  &  - & - &  $-1.74^{+0.30}_{-0.38}$  & $-1.83^{+0.23}_{-0.27}$ & $-1.60^{+0.25}_{-0.22}$ & $-1.55^{+0.28}_{-0.20}$   \\
	$\log(M_{\rm wind})$  ($M_\odot$) &  -  & - & $-1.28^{+0.46}_{-0.47}$ &   $-1.17^{+0.28}_{-0.26}$ &  $-1.03^{+0.27}_{-0.39}$  & $-0.83^{+0.18}_{-0.23}$ \\
	$\Phi$ (deg) & -  & -  & $40.13^{+11.88}_{-10.30}$ &   $44.56^{+11.19}_{-13.04}$ & 30  & 30 \\
	\midrule
    $\iota$  (rad) &   $0.26^{+0.05}_{-0.05}$ & $0.12^{+0.04}_{-0.08}$ & $0.05^{+0.13}_{-0.03}$  &  $0.08^{+0.08}_{-0.04}$  & $0.17^{+0.08}_{-0.14}$   & $0.06^{+0.04}_{-0.03}$ \\
	$\log(E_0)$ (erg) &  $50.44^{+0.15}_{-0.14}$ & $50.80^{+0.45}_{-0.42}$ & $50.55^{+0.30}_{-0.21}$ &  $50.56^{+0.33}_{-0.23}$  & $50.58^{+0.25}_{-0.22}$  & $50.47^{+0.51}_{-0.22}$ \\
	$\log(n_0)$ (cm$^{-3}$) & $ -0.27^{+0.21}_{-0.32}$ & $-2.91^{+2.67}_{-1.37}$ & $-2.14^{+0.70}_{-0.80}$ &   $-2.22^{+0.64}_{-1.00}$  &  $ -1.99^{+0.86}_{-0.89}$    & $ -2.05^{+0.66}_{-1.39}$ \\
	$\theta_{\rm c}$  (rad) &  $0.23^{+0.04}_{-0.04}$ & $0.21^{+0.07}_{-0.04}$ &  $0.21^{+0.05}_{-0.10}$ &    $0.13^{+0.12}_{-0.04}$  &  $0.22^{+0.05}_{-0.10}$   & $0.19^{+0.05}_{-0.09}$ \\
    $\theta_{\rm w}$  (rad) &  $0.57^{+0.15}_{-0.25}$ & - &  $0.39^{+0.20}_{-0.18}$ &  - & $0.50^{+0.12}_{-0.26}$   & -\\
	$p$   &  $2.23^{+0.16}_{-0.18}$ & $2.65^{+0.11}_{-0.43}$ & $2.72^{+0.06}_{-0.06}$ &   $2.79^{+0.05}_{-0.07}$  & $2.73^{+0.06}_{-0.11}$  & $ 2.75^{+0.07}_{-0.07}$  \\
	$\log(\epsilon_{\rm e})$  & $-0.3$ & $-0.3$ & $-0.3$ & $-0.3$ & $-0.3$ & $-0.3$ \\
	$\log(\epsilon_{\rm B})$   & $-2.0$  & $-2.0$ & $-2.0$ & $-2.0$ & $-2.0$ & $-2.0$   \\
	\midrule
    Evidence ($\mathcal{Z}$)  &  $-19.2$ & $-21.09$ & $-13.5$ &  $-12.58$ & $-12.4$    & $-11.64$  \\
	$\ln (\mathcal{B}^{\rm Test}_{\rm{Ref}})$   &  -7.56  & -9.45 &  -1.86 &  -0.94 &  -0.76  & 0.0 \\
	\toprule
	\end{tabular}

\label{tab:191019A_AllModelsResults}
\end{table*}

\begin{table*}[t]
	\centering
    \caption{Results of all evaluated models for GRB\,211211A. }
	\renewcommand{\arraystretch}{1.9}
    \begin{tabular}{@{}>{\raggedright\arraybackslash}m{2.4cm}*{7}{>{\centering\arraybackslash}m{2.2cm}}@{}}
	\toprule
	    & GS  &  TH &  BNS-GS &  BNS-TH & NSBH-GS  & NSBH-TH  \\  
	\midrule
	$\log(M_{\rm dyn})$ ($M_\odot$) &- &- &  $-1.68^{+0.08}_{-0.09}$ & $-1.88^{+0.07}_{-0.10}$ & $-1.87^{+0.09}_{-0.09}$     & $-2.00^{+0.01}_{-0.01}$  \\ 	
	$\log(M_{\rm wind})$ ($M_\odot$) &- &-&   $-2.45^{+0.06}_{-0.06}$  & $-2.20^{+0.07}_{-0.09}$  & $-2.22^{+0.11}_{-0.08}$     & $-1.99^{+0.03}_{-0.03}$     \\ 
	$\Phi$ (deg) &- &-&  $60.98^{+7.17}_{-9.45}$  & $71.66^{+1.98}_{-4.98}$  & 30   & 30    \\ 
	\midrule
	$\iota$  (rad)     & $0.07^{+0.04}_{-0.02}$  & $0.06^{+0.01}_{-0.01}$ &   $0.02^{+0.01}_{-0.01}$ & $0.005^{+0.001}_{-0.001}$  & $0.13^{+0.05}_{-0.07}$    &  $0.005^{+0.001}_{-0.001}$   \\ 
    $\log(E_0)$ (erg)  & $52.01^{+0.19}_{-0.42}$  &  $49.96^{+0.08}_{-0.06}$  &  $52.98^{+0.31}_{-0.20}$ & $51.54^{+0.22}_{-0.23}$  & $51.20^{+0.50}_{-0.19}$    & $51.31^{+0.08}_{-0.12}$     \\ 
	$\log(n_0)$ (cm$^{-3}$)   & $0.15^{+0.70}_{-0.30}$  &  $-2.28^{+0.17}_{-0.18}$  & $1.87^{+0.45}_{-0.79}$ & $-6.25^{+0.33}_{-0.34}$ & $2.95^{+0.24}_{-0.27}$     & $-6.35^{+0.28}_{-0.31}$  \\ 
	$\theta_{\rm c}$ (rad) & $0.16^{+0.02}_{-0.02}$ &    $0.15^{+0.01}_{-0.01}$ &  $0.14^{+0.02}_{-0.05}$  & $0.017^{+0.003}_{-0.002}$ & $0.24^{+0.02}_{-0.03}$     & $0.018^{+0.002}_{-0.002}$    \\ 
	$\theta_{\rm w}$ (rad) & $0.16^{+0.09}_{-0.03}$ &  - &  $0.15^{+0.03}_{-0.02}$ & - &  $0.50^{+0.15}_{-0.07}$    & -    \\ 
	$p$     & $2.12^{+0.04}_{-0.02}$ &  $2.32^{+0.01}_{-0.02}$   & $2.16^{+0.07}_{-0.04}$ & $2.39^{+0.02}_{-0.03}$  & $2.19^{+0.04}_{-0.03}$   & $2.37^{+0.02}_{-0.03}$    \\ 
	$\log(\epsilon_e)$ & $-1.68^{+0.39}_{-0.19}$ & $-0.07^{+0.03}_{-0.05}$& $-2.45^{+0.32}_{-0.19}$ & $-0.10^{+0.05}_{-0.07}$ & $-0.68^{+0.22}_{-0.19}$    & $-0.23^{+0.05}_{-0.04}$    \\ 
	$\log(\epsilon_B)$ &  $-3.39^{+0.23}_{-0.38}$   &  $-1.31^{+0.14}_{-0.14}$  &  $-5.02^{+0.27}_{-0.27}$ & $-1.40^{+0.37}_{-0.35}$   & $-5.06^{+0.20}_{-1.27}$     & $-0.79^{+0.12}_{-0.11}$     \\ 
	\midrule
	Evidence ($\mathcal{Z}$)   & $-130.87{\pm 0.12}$  &   $-126.12{\pm 0.14}$  &  $-100.23{\pm 0.14}$ & $-89.28{\pm 0.07 }$  & $-137.23{\pm 0.14}$    & $-107.18{\pm 0.16}$   \\  
	$\ln (\mathcal{B}^{\rm Test}_{\rm{Ref}})$  & -30.15  &  -31.39  & -5.50   & 0.0  &  -42.50    & -12.45  \\ 
	\bottomrule
	\end{tabular}
	
	\label{tab:211211A_AllModelsResults}
\end{table*}

\begin{table*}[t]
    \centering
    \caption{Results of all evaluated models for GRB\,230307A.}
    \renewcommand{\arraystretch}{1.9}
    \begin{tabular}{@{}>{\raggedright\arraybackslash}m{2.4cm}*{7}{>{\centering\arraybackslash}m{2.2cm}}@{}}
    \toprule
         & GS & TH & BNS-GS  & BNS-TH &   NSBH-GS & NSBH-TH \\  
    \midrule
    $\log(M_{\rm dyn})$ ($M_\odot$)  &  - & - & $-1.72^{+0.15}_{-0.23}$ & $-1.85^{+0.17}_{-0.21}$ &  $-1.98^{+0.50}_{-0.09}$ & $-1.97^{+0.16}_{-0.07}$\\ 
    $\log(M_{\rm wind})$  ($M_\odot$) & -  & -  & $-1.56^{+0.17}_{-0.19}$ & $-1.57^{+0.19}_{-0.30}$ & $-1.91^{+0.19}_{-0.38}$ & $-2.23^{+0.12}_{-0.08}$\\ 
    $\Phi$ (deg) &  - & - & $68.48^{+3.83}_{-8.09}$ & $63.74^{+6.92}_{-10.20}$ & 30 & 30 \\ 
    \midrule
    $\iota$ (rad)       & $0.07^{+0.03}_{-0.03}$  & $0.03^{+0.01}_{-0.01}$ & $0.09^{+0.02}_{-0.02}$ & $0.06^{+0.03}_{-0.02}$        & $0.06^{+0.03}_{-0.02}$  & $0.04^{+0.02}_{-0.01}$ \\ 
    $\log(E_0)$ (erg)    &  $50.48^{+0.32}_{-0.42}$ & $51.79^{+0.36}_{-0.45}$ & $50.71^{+0.42}_{-0.26}$ & $51.22^{+0.56}_{-0.53}$   & $51.30^{+0.35}_{-0.65}$ & $51.41^{+0.43}_{-0.34}$ \\ 
    $\log(n_0)$  (cm$^{-3}$)  &  $-3.45^{+0.44}_{-0.31}$   & $-5.32^{+0.68}_{-1.01}$ & $-3.10^{+0.31}_{-0.39}$ & $-4.25^{+0.96}_{-0.77}$ &    $-3.80^{+0.59}_{-0.66}$  & $-4.96^{+0.68}_{-0.55}$\\  
    $\theta_{\rm c}$ (rad) &  $0.13^{+0.03}_{-0.02}$  &  $0.06^{+0.02}_{-0.02}$ & $0.12^{+0.02}_{-0.02}$ & $0.09^{+0.04}_{-0.03}$ &   $0.08^{+0.07}_{-0.02}$ & $0.07^{+0.02}_{-0.01}$ \\ 
    $\theta_{\rm w}$ (rad) &  $0.26^{+0.18}_{-0.11}$   & - & $0.36^{+0.14}_{-0.20}$ & - &   $0.21^{+0.10}_{-0.08}$ &-\\ 
    $p$              &  $2.49^{+0.10}_{-0.14}$  & $2.65^{+0.04}_{-0.04}$ & $2.48^{+0.05}_{-0.05}$ & $2.50^{+0.05}_{-0.05}$ &  $2.48^{+0.05}_{-0.06}$ & $2.55^{+0.04}_{-0.05}$\\ 
    $\log(\epsilon_e)$ &   $-0.22^{+0.10}_{-0.05}$  & $-0.38^{+0.16}_{-0.19}$ & $-0.15^{+0.07}_{-0.08}$ & $-0.33^{+0.20}_{-0.14}$ &  $-0.21^{+0.11}_{-0.15}$ & $-0.40^{+0.14}_{-0.16}$\\ 
    $\log(\epsilon_B)$ &  $-0.98^{+0.37}_{-0.47}$   & $-1.84^{+0.79}_{-0.72}$ & $-1.89^{+0.43}_{-0.71}$ & $-1.92^{+0.57}_{-0.68}$ &  $-2.26^{+0.87}_{-0.74}$ & $-1.72^{+0.81}_{-0.69}$\\ 
    \midrule
    Evidence ($\mathcal{Z}$)  & $-58.27\pm 0.04$  & $-54.77\pm 0.04$  & $-54.09\pm 0.04$  & $-50.74\pm0.04$ &  $-61.10\pm0.04$ & $-58.25\pm0.04$\\ 
    $\ln (\mathcal{B}^{\rm Test}_{\rm{Ref}})$  & -7.53  & -4.03 & -3.35 & 0.0  & -10.36 & -7.51  \\ 
\toprule
\end{tabular}
\label{tab:230307A_AllModelsResults}
\end{table*}

\onecolumn
\section{Posterior distributions for the best-fit models.  }
\begin{figure*}[h]
\centering
\begin{minipage}[t]{0.47\textheight}
    \centering
    \includegraphics[width=0.90\linewidth]{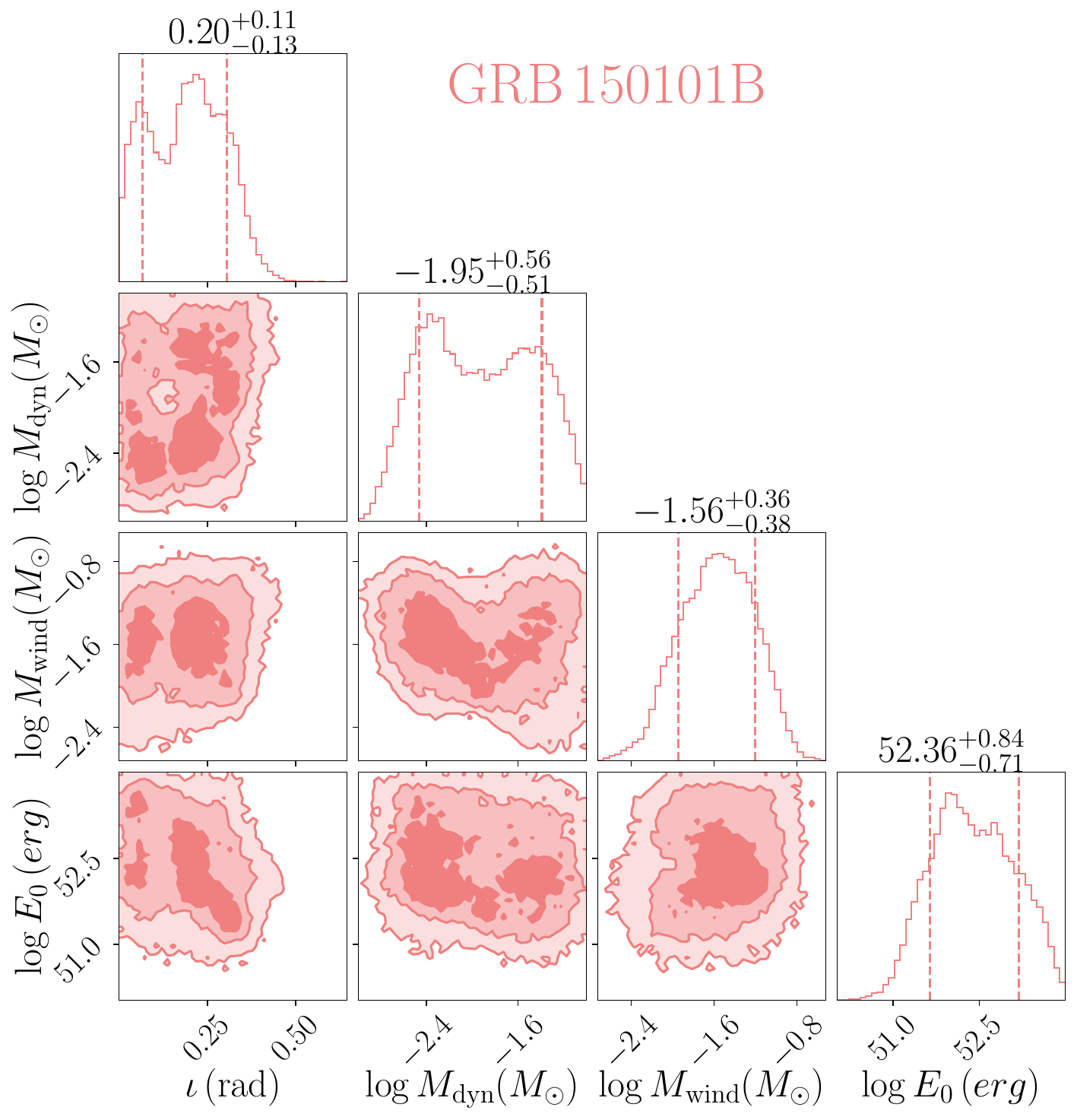}
\end{minipage}
\begin{minipage}[t]{0.47\textheight}
    \centering
    \includegraphics[width=0.90\linewidth]{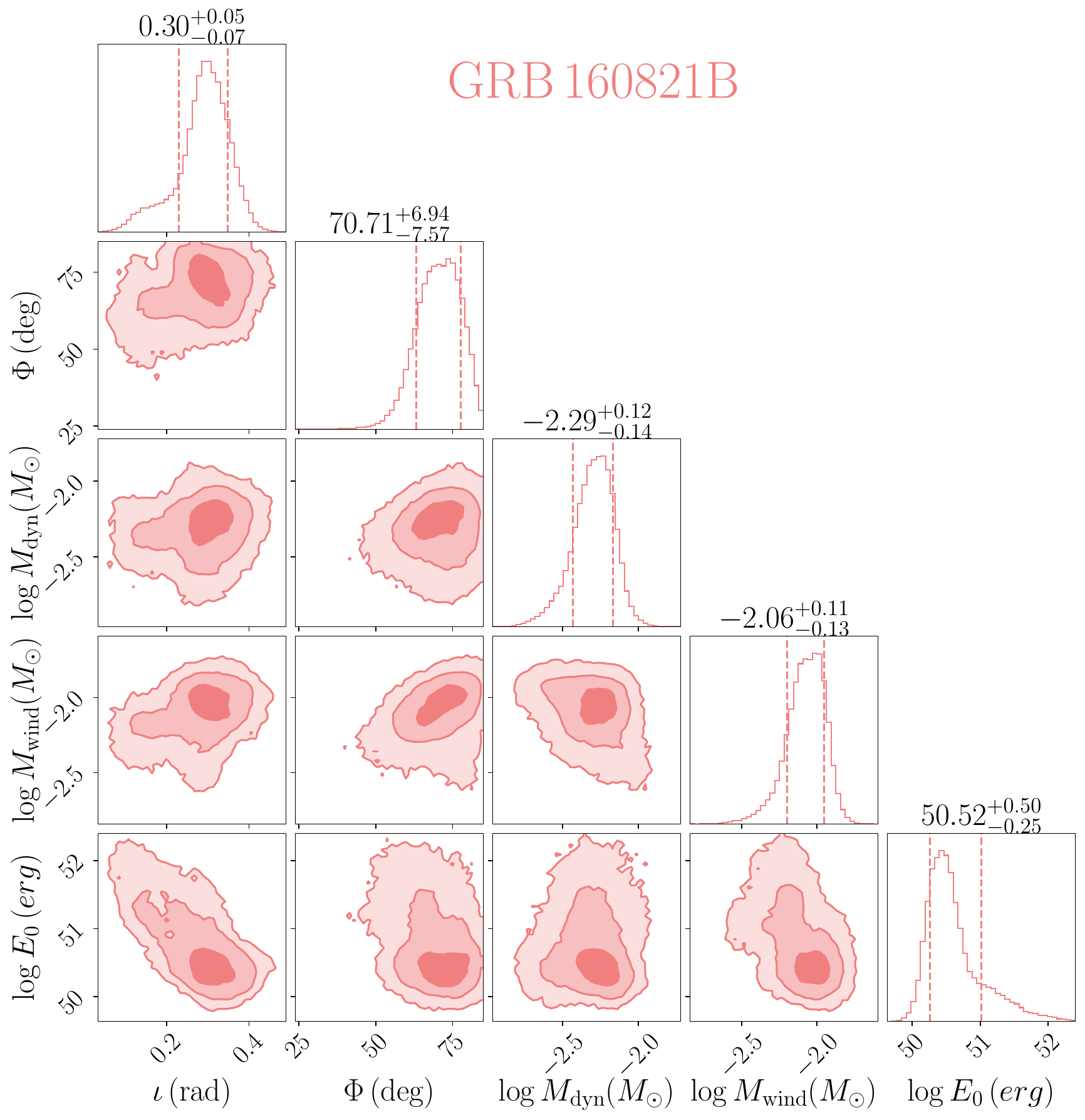}
\end{minipage}

\caption{Posterior distribution of the parameters in \tabref{tab:All_KN_Results}, shown as corner plots. Different shadings mark the 39.3\%, 86.5\%, and 97.9\% confidence intervals. The 68\% confidence interval is indicated with dashed lines, and the median values are shown above each panel for the 1D posterior probability distributions. The full posterior distributions are available here: \href{https://doi.org/10.5281/zenodo.19708188}{https://doi.org/10.5281/zenodo.19708188}. }
\label{Fig:AllKNPosterior_CornerSmall1}
\end{figure*}

\begin{figure*}[h]
\centering
\begin{minipage}[t]{0.47\textheight}
    \centering
    \includegraphics[width=0.94\linewidth]{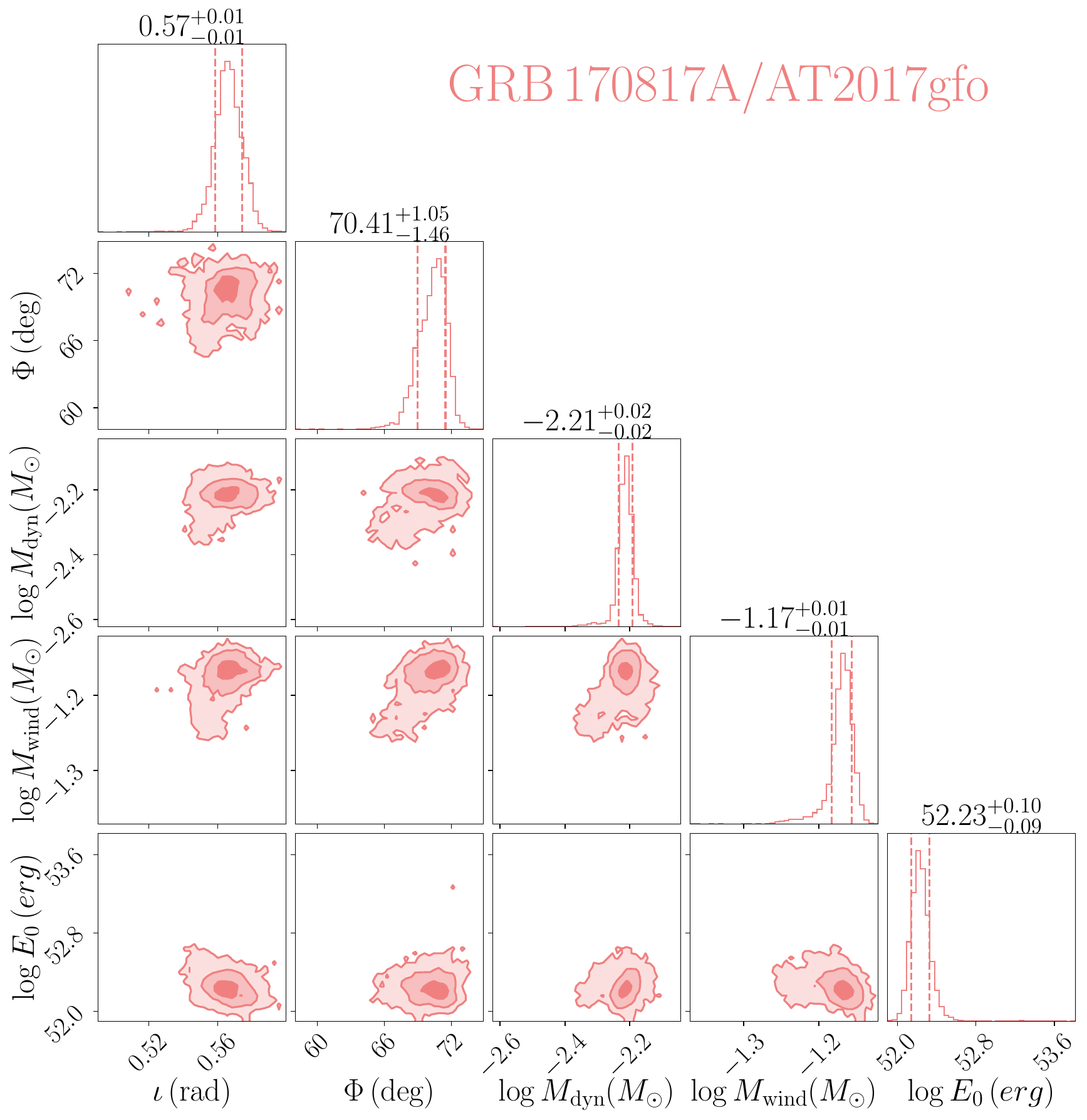}
\end{minipage}
\begin{minipage}[t]{0.47\textheight}
    \centering
    \includegraphics[width=0.94\linewidth]{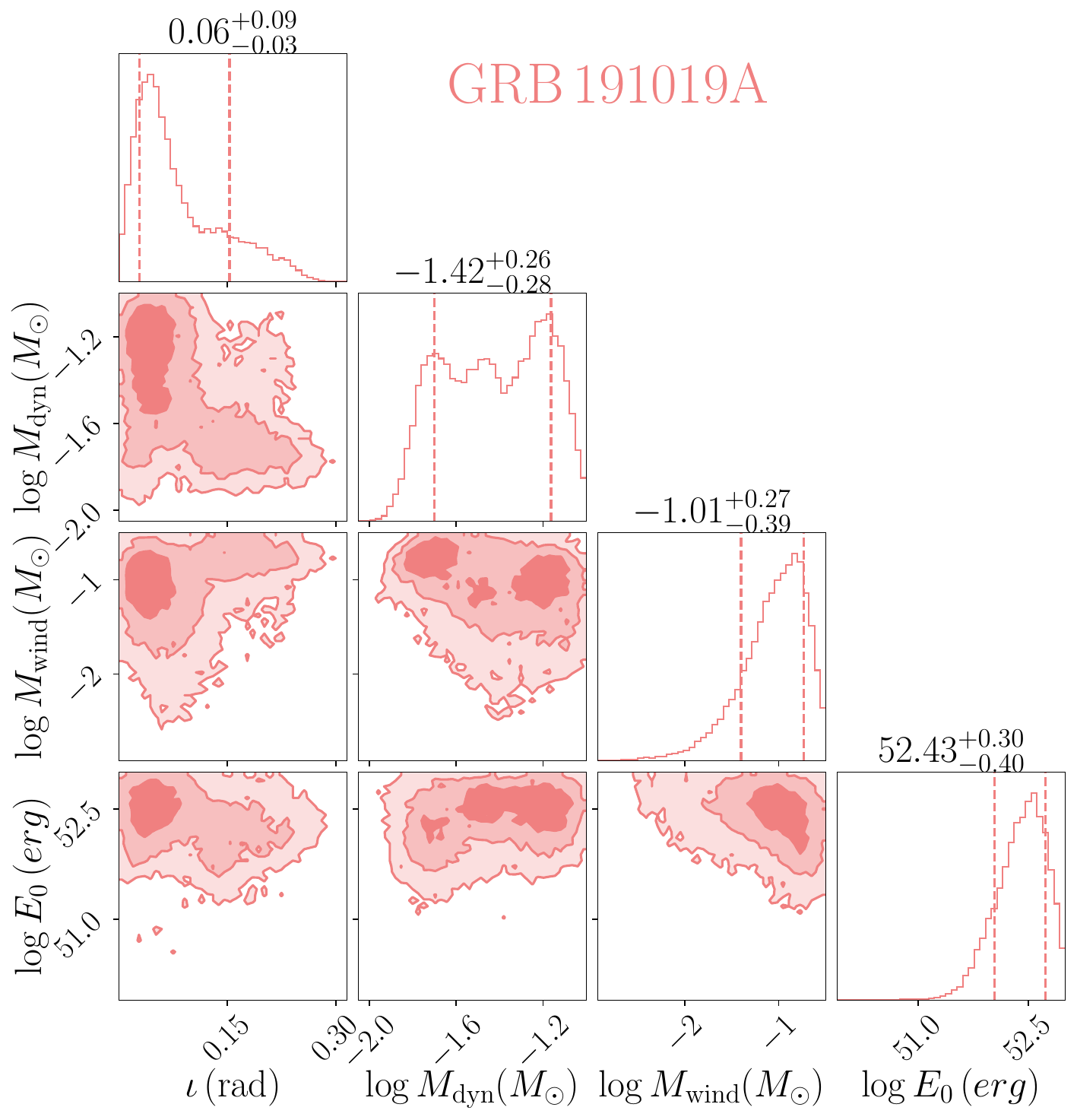}
\end{minipage}

\caption{Same as \figref{Fig:AllKNPosterior_CornerSmall1}. The full posterior distributions are available here: \href{https://doi.org/10.5281/zenodo.19708188}{https://doi.org/10.5281/zenodo.19708188}. }
\label{Fig:AllKNPosterior_CornerSmall2}
\end{figure*}

\begin{figure*}[h]
\centering
\begin{minipage}[t]{0.47\textheight}
    \centering
    \includegraphics[width=0.94\linewidth]{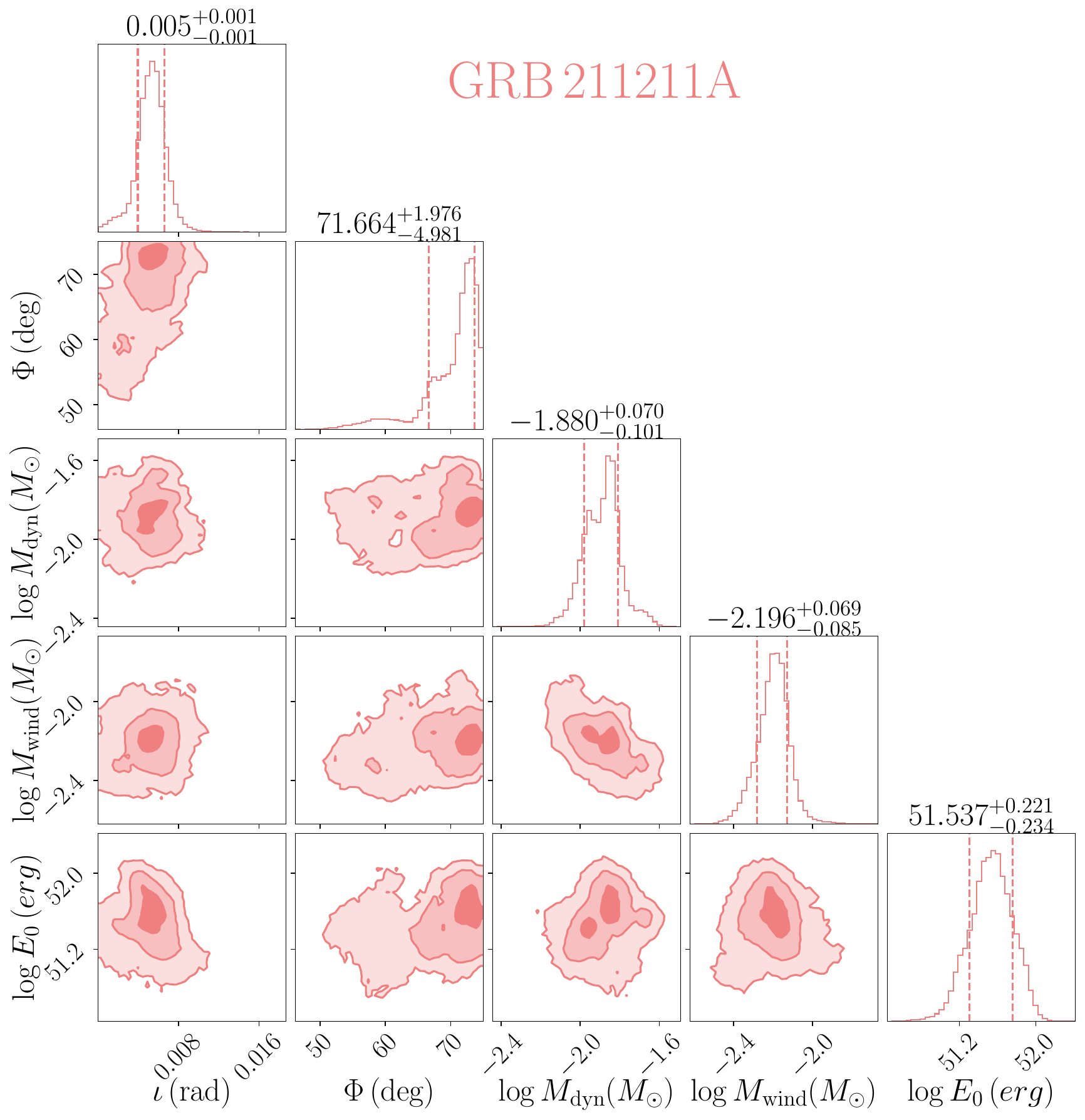}
\end{minipage}
\begin{minipage}[t]{0.47\textheight}
    \centering
    \includegraphics[width=0.94\linewidth]{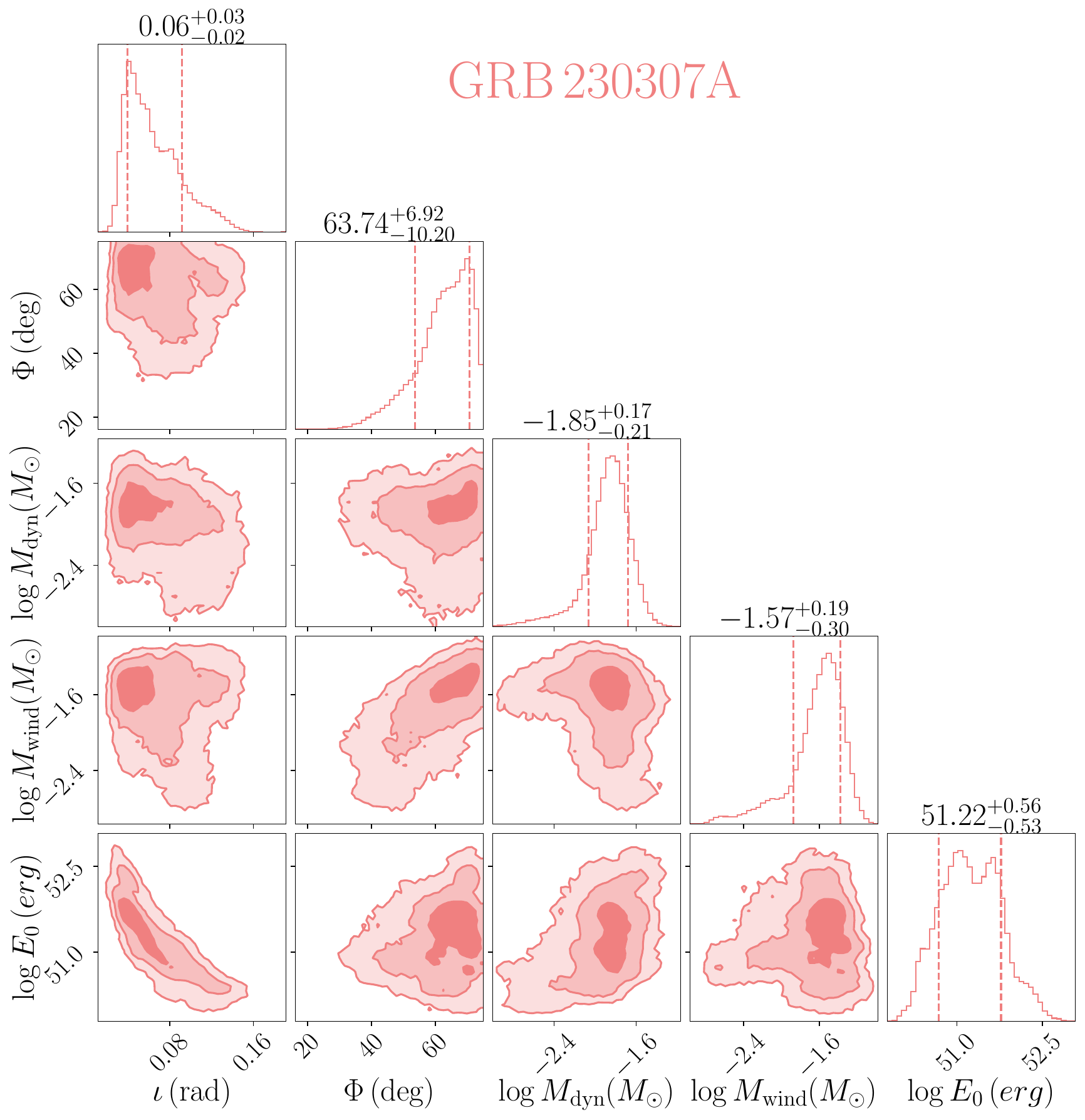}
\end{minipage}

\caption{Same as \figref{Fig:AllKNPosterior_CornerSmall1}. The full posterior distributions are available here: \href{https://doi.org/10.5281/zenodo.19708188}{https://doi.org/10.5281/zenodo.19708188}. }
\label{Fig:AllKNPosterior_CornerSmall3}
\end{figure*}

\begin{figure*}
\centering
\begin{minipage}[t]{0.53\textheight}
    \centering
    \includegraphics[width=0.85\linewidth]{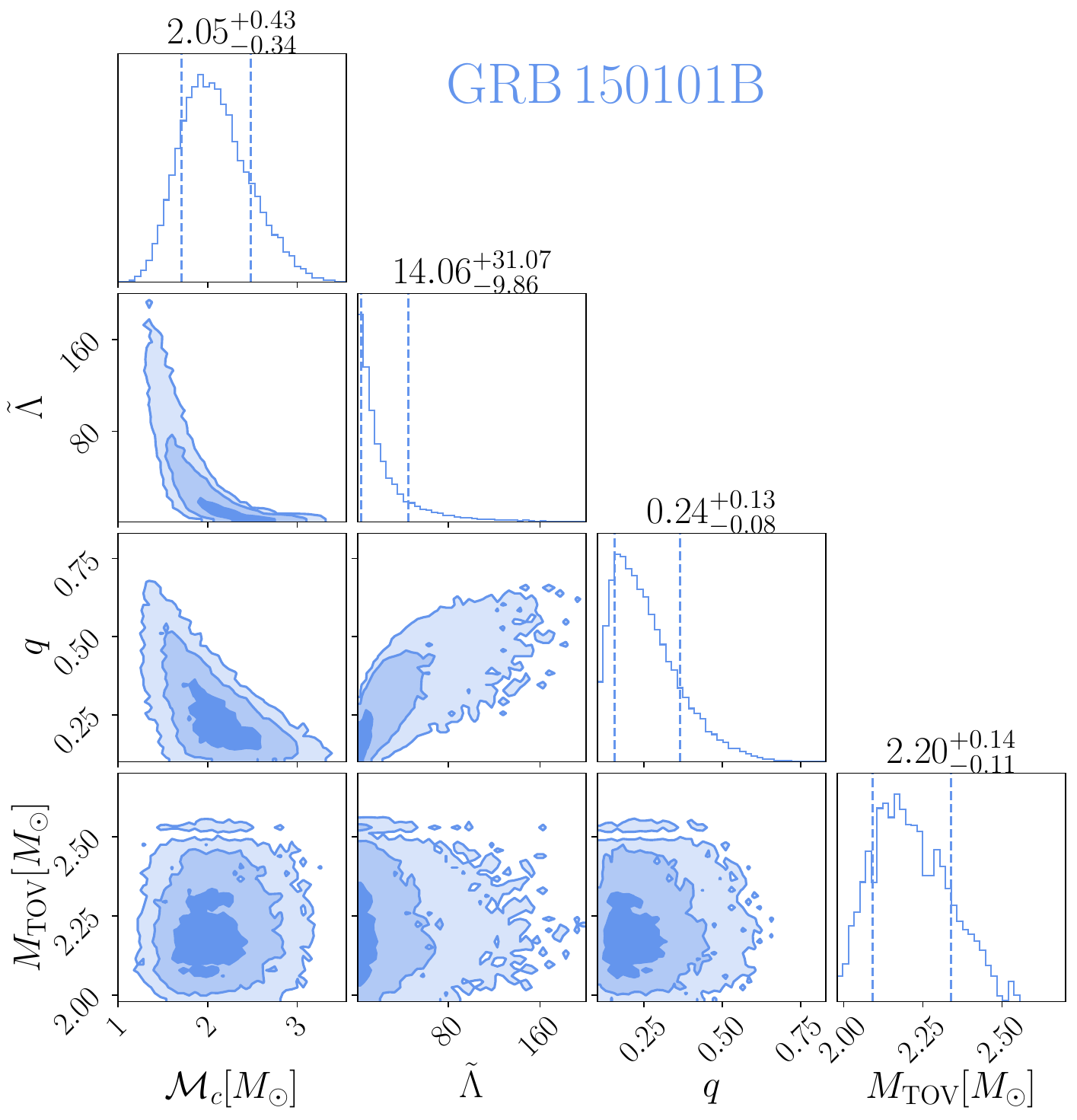}
\end{minipage}
\begin{minipage}[t]{0.53\textheight}
    \centering
    \includegraphics[width=0.85\linewidth]{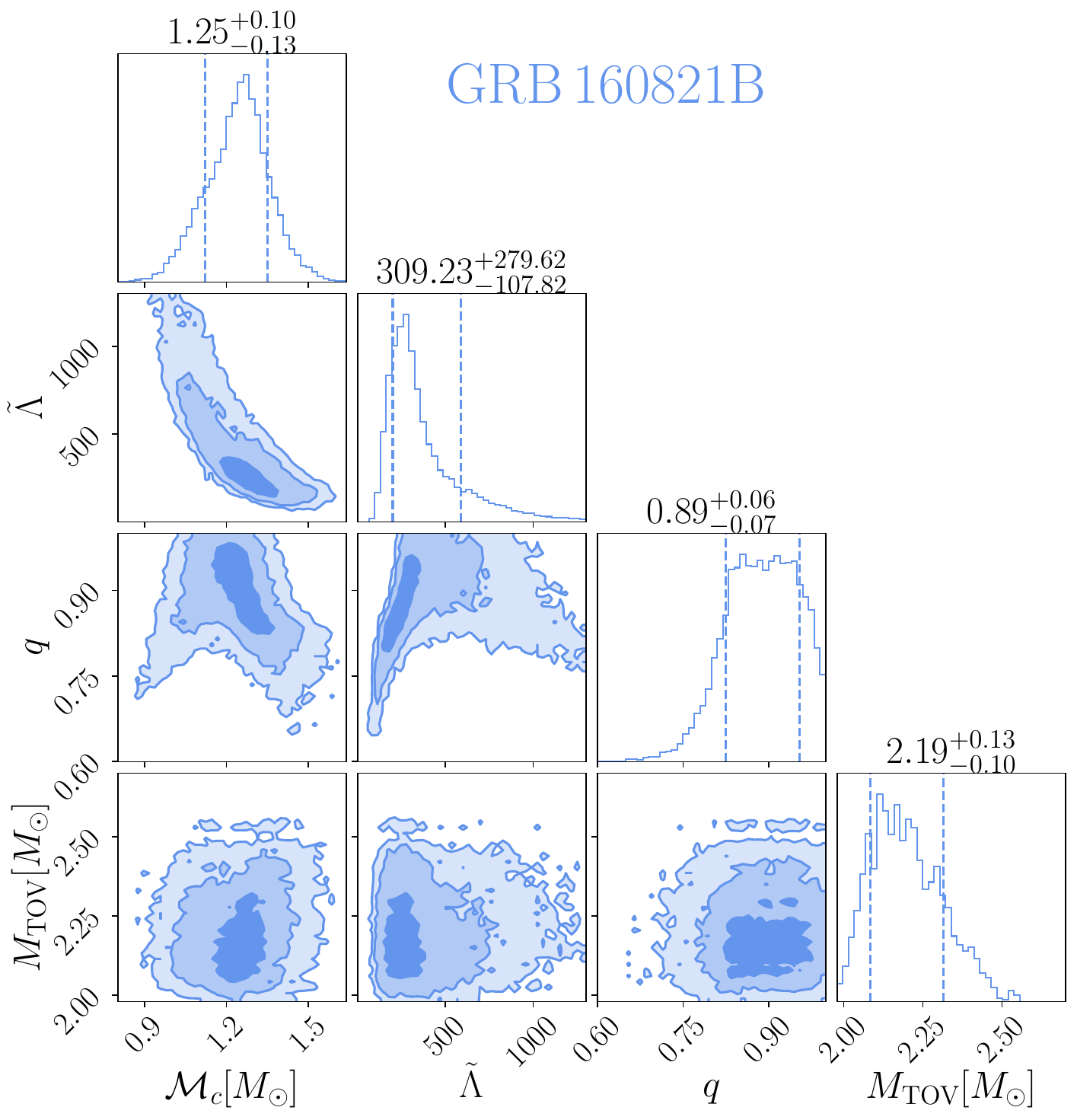}
\end{minipage}

\caption{Posterior distribution of the binary progenitor parameters, corresponding to the models in \tabref{tab:All_KN_Results} and values in \tabref{tab:All_Binary_Results} displayed as corner plots. Different shadings mark the 39.3\%, 86.5\%, and 97.9\% confidence intervals. The 68\% confidence interval is indicated with dashed lines, and the median values are shown above each panel for the 1D posterior probability distributions. The full posterior distributions are available here: \href{https://doi.org/10.5281/zenodo.19708188}{https://doi.org/10.5281/zenodo.19708188}. }
\label{Fig:AllGWPosterior_CornerSmall1}
\end{figure*}

\begin{figure*}
\centering
\begin{minipage}[t]{0.53\textheight}
    \centering
    \includegraphics[width=0.85\linewidth]{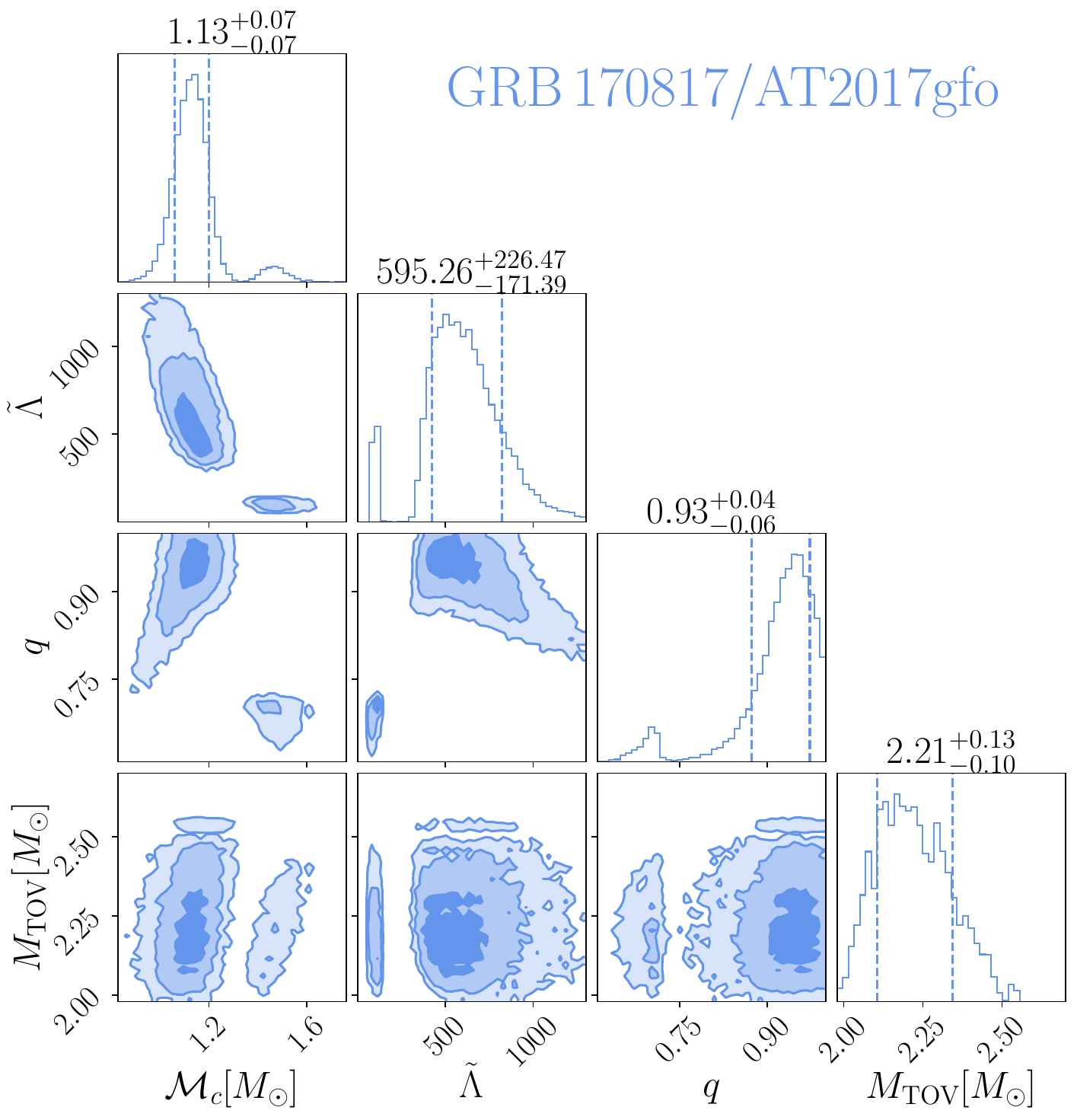}
\end{minipage}
\begin{minipage}[t]{0.53\textheight}
    \centering
    \includegraphics[width=0.85\linewidth]{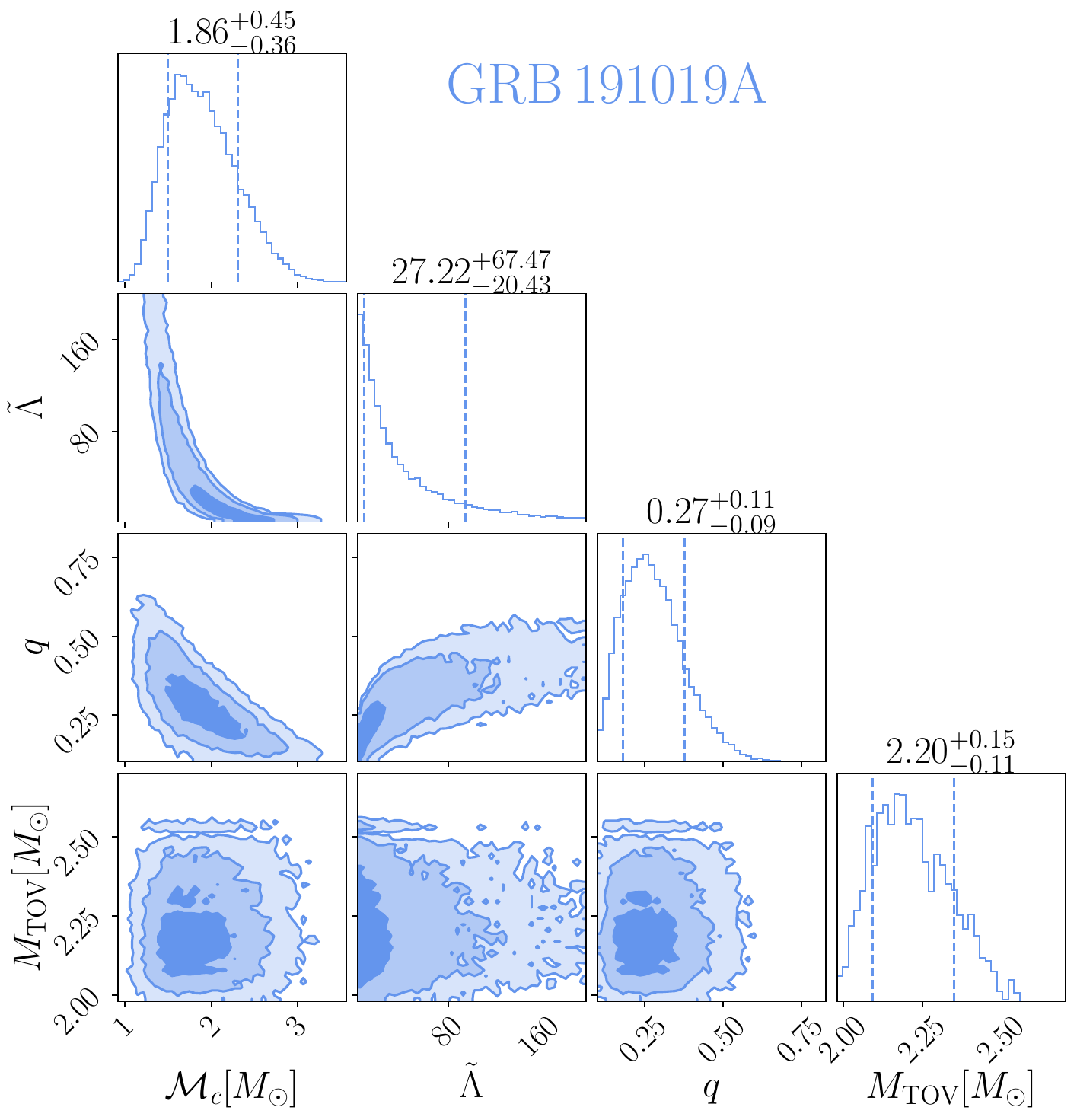}
\end{minipage}

\caption{Same as \figref{Fig:AllGWPosterior_CornerSmall1}. The full posterior distributions are available here: \href{https://doi.org/10.5281/zenodo.19708188}{https://doi.org/10.5281/zenodo.19708188}. }
\label{Fig:AllGWPosterior_CornerSmall2}
\end{figure*}

\begin{figure*}
\centering
\begin{minipage}[t]{0.53\textheight}
    \centering
    \includegraphics[width=0.85\linewidth]{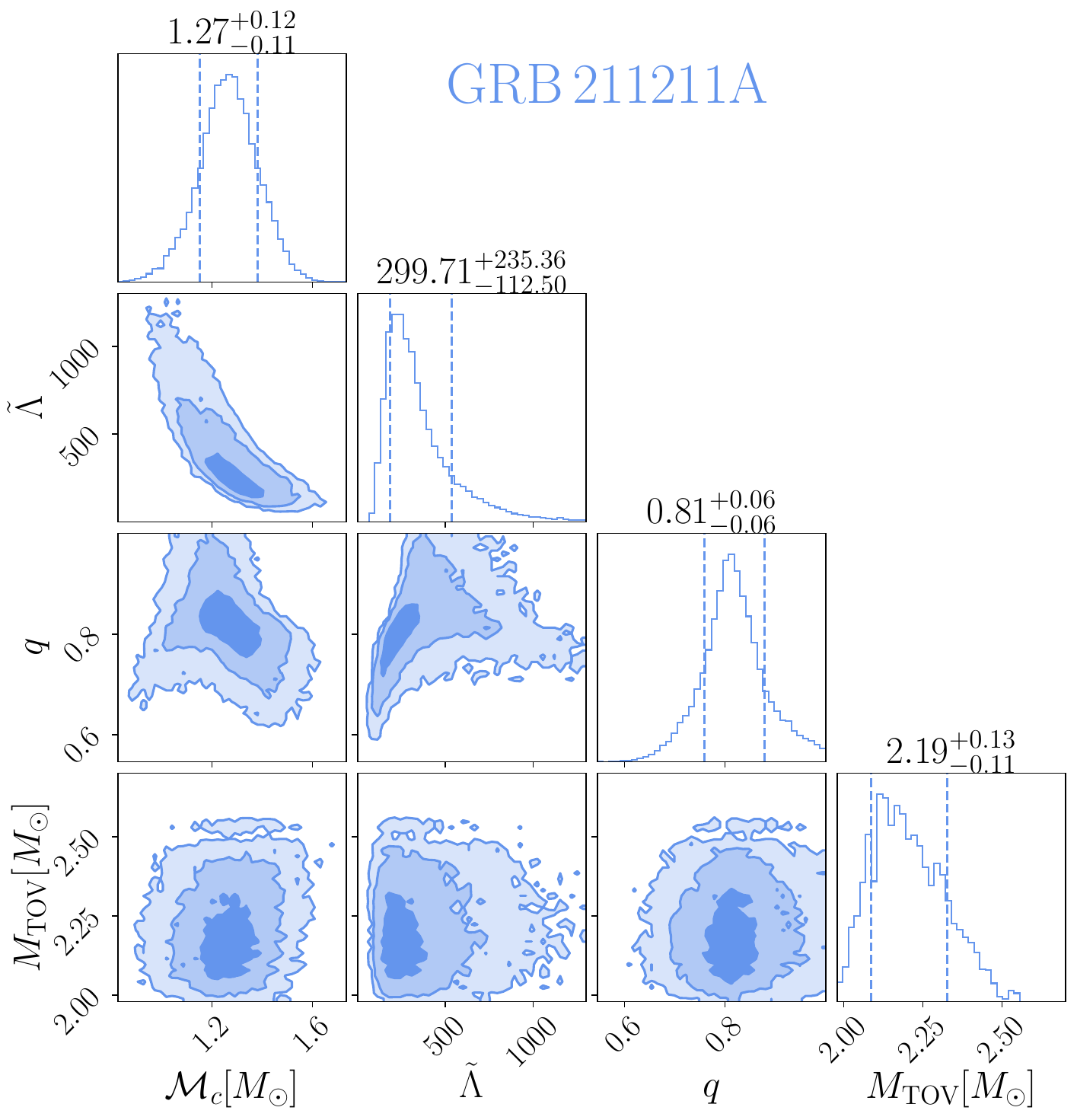}
\end{minipage}
\begin{minipage}[t]{0.53\textheight}
    \centering
    \includegraphics[width=0.85\linewidth]{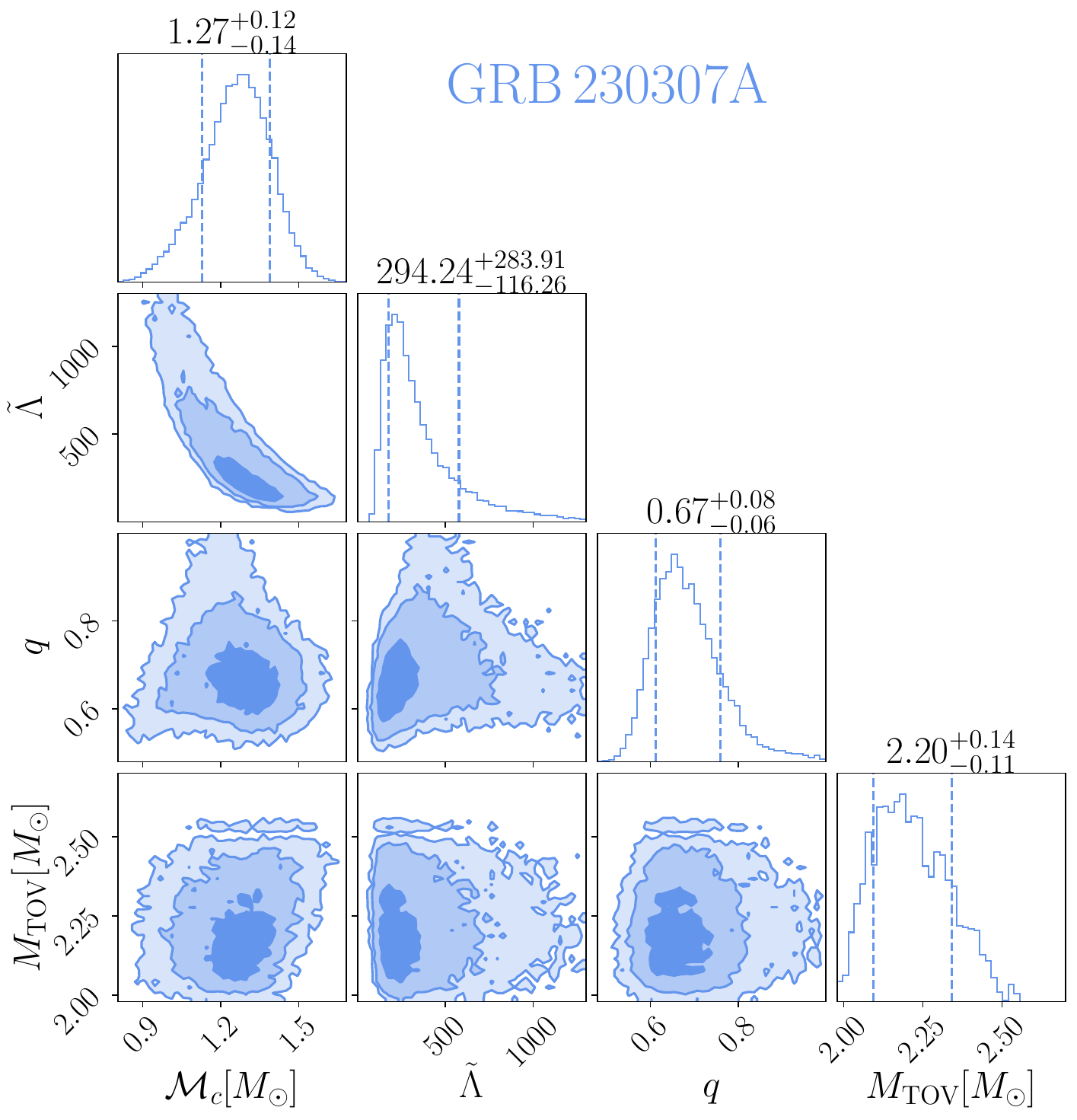}
\end{minipage}

\caption{Same as \figref{Fig:AllGWPosterior_CornerSmall1}. The full posterior distributions are available here: \href{https://doi.org/10.5281/zenodo.19708188}{https://doi.org/10.5281/zenodo.19708188}. }
\label{Fig:AllGWPosterior_CornerSmall3}
\end{figure*}

\end{appendix}

\end{document}